\documentclass[aps,pre,amsmath,amssymb,groupedaddress]{revtex4}
\usepackage{graphicx,color}
 \graphicspath{{./}{./figures/}}
\usepackage{ams math}
\usepackage{enumerate}
\usepackage{mathbbol}
\usepackage{amsfonts}
\usepackage{natbib}

\usepackage{bm}

\usepackage{color}

\newcommand{\fpi}{{(4\pi)^{D/2}}}
\newcommand{\as}{_{\alpha}}
\newcommand{\bs}{_{\beta}}
\newcommand{\cs}{_{\gamma}}
\newcommand{\ds}{_{\delta}}
\newcommand{\ab}{_{\alpha\beta}}
\newcommand{\ac}{_{\alpha\gamma}}
\newcommand{\ad}{_{\alpha\delta}}
\newcommand{\bc}{_{\beta\gamma}}
\newcommand{\bd}{_{\beta\delta}}
\newcommand{\cd}{_{\gamma\delta}}
\newcommand{\aas}{_{\alpha\alpha}}
\newcommand{\bbs}{_{\beta\beta}}
\newcommand{\kk}{_{\kappa\kappa}}
\newcommand{\kd}{_{\kappa\Delta}}
\newcommand{\dd}{_{\Delta\Delta}}
\newcommand{\kap}{_{\kappa}}
\newcommand{\Del}{_{\Delta}}
\newcommand{\g}{{\Gamma}}
\newcommand{\intd}{{\int d^Dx}}

\newcommand{\n}{{\vec n}}
\newcommand{\f}{{\vec f}}
\newcommand{\h}{{\vec h}}

\newcommand{\T}{{k_B T}}

\newcommand{\F}{\hbox{$\cal F$}}
\newcommand{\pb}{{\bf p}}
\newcommand{\q}{{\bf q}}
\newcommand{\kb}{{\bf k}}
\newcommand{\ks}{{\bf k}}
\newcommand{\rv}{{\vec r}}
\newcommand{\tv}{{\vec t}}
\newcommand{\x}{{\bf x}}

\newcommand{\eps}{\epsilon}

\newcommand{\pt}{{\partial}}

\newcommand{\nn}{\nonumber}

\newcommand{\half}{\frac{1}{2}}

\newcommand{\be}{\begin{equation}}
\newcommand{\ee}{\end{equation}}
\newcommand{\bea}{\begin{eqnarray}}
\newcommand{\eea}{\end{eqnarray}}
\newcommand{\beq}{\begin{eqnarray}}
\newcommand{\eeq}{\end{eqnarray}}

\newcommand{\p}{{\bf p}}

\newcommand{\Gc}{{\cal G}}

\begin{document}

\title{Anomalous elasticity, fluctuations and disorder in elastic 
membranes}

\author{Pierre Le Doussal}
\affiliation{CNRS-Laboratoire de Physique Th\'eorique de l'Ecole
  Normale Sup\'erieure, 24 rue Lhomond, 75231 Paris Cedex, France}
\email{ledou@lpt.ens.fr}
\author{Leo Radzihovsky} 
\affiliation{Department of Physics,
  University of Colorado, Boulder, CO 80309}
\affiliation{Kavli Institute for Theoretical Physics, University of California, Santa Barbara, CA 93106}
\email{radzihov@colorado.edu}

\begin{abstract}
  Motivated by freely suspended graphene and polymerized membranes in
  soft and biological matter we present a detailed study of a
  tensionless elastic sheet in the presence of thermal fluctuations
  and quenched disorder. The manuscript is based on an extensive draft
  dating back to 1993, that was circulated privately. It presents the
  general theoretical framework and calculational details of numerous
  results, partial forms of which have been published in brief Letters
  \cite{LRprl,LRrapid}.  The experimental realization atom-thin
  graphene sheets \cite{Geim2004} has driven a resurgence in this
  fascinating subject, making our dated predictions and their detailed
  derivations timely. To this end we analyze the statistical mechanics
  of a generalized $D$-dimensional elastic ``membrane'' embedded in
  $d$ dimensions using a self-consistent screening approximation
  (SCSA), that has proved to be unprecedentedly accurate in this
  system, {\em exact} in three complementary limits: (i)
  $d\rightarrow\infty$, (ii) $D\rightarrow 4$, and (iii) $D=d$.
  Focusing on the critical ``flat'' phase, for a homogeneous
  two-dimensional ($D=2$) membrane embedded in three dimensions
  ($d=3$), we predict its universal roughness exponent $\zeta =
  0.590$, length-scale dependent elastic moduli exponents $\eta =
  0.821$ and $\eta_u = 0.358$, and an anomalous Poisson ratio, $\sigma
  = -1/3$. In the presence of random uncorrelated heterogeneity the
  membrane exhibits a glassy wrinkled ground state, characterized by
  $\zeta' = 0.775, \eta' = 0.449,\eta'_u = 1.101$ and a Poisson ratio
  $\sigma'= -1/3$. Motivated by a number of physical realizations
  (charged impurities, disclinations and dislocations) we also study
  power-law correlated quenched disorder that leads to a variety of
  distinct glassy wrinkled phases. Finally, neglecting self-avoiding
  interaction we demonstrate that at high temperature a ``phantom''
  sheet undergoes a continuous crumpling transition, characterized by
  a radius of gyration exponent, $\nu =0.732$ and $\eta = 0.535$. Many
  of these universal predictions have received considerable support
  from simulations. We hope that this detailed presentation of the
  SCSA theory will be useful to further theoretical developments and
  corresponding experimental investigations on freely suspended
  graphene.

 \end{abstract}

\maketitle

\tableofcontents

\section{Introduction}
\label{Intro} 

\subsection{Preamble and modern graphene motivation}
\label{preamble} 

In this manuscript we present a detailed statistical mechanics study
of a tensionless elastic sheet, as realized by graphene and
polymerized membranes in soft and biological matter, in the presence
of thermal fluctuations and quenched disorder. The manuscript is based
on a draft dating back to 1993, that was circulated privately. Recent
experimental realization of a freely suspended graphene
\cite{suspendGrapheneNature2007} following the 2004 pioneering works
of Geim and Novoselov \cite{Geim2004}, has led to a renaissance of
this subject \cite{GeimMacDonald,reviewRMPGraphene} making our dated
predictions and their detailed derivations timely.  Some of the
predictions have appeared in our earlier Letters \cite{LRprl,LRrapid}
in a rather terse form, challenging researchers to reproduce and
utilize our results in the modern graphene context. This motivated us
to complete this detailed manuscript.

As we will discuss in detail, the key qualitative ingredient of a
low-tension elastic sheet is that (in contrast to its
tension-controlled counterpart), at an arbitrary low temperature and
weak disorder its out-of-plane fluctuations $h_{rms}\sim L^\zeta$ grow
unboundedly with its in-plane linear size $L$, where $\zeta$ is the
universal roughness exponent of the ``flat'' phase of the fluctuating
membrane. As a result of such strong thermally and disorder-driven
fluctuations in the flat phase (first emphasized for the former by
Nelson and Peliti\cite{NP}), nonlinear elasticity is always
qualitatively important, {\em nonperturbatively} so, beyond a
nonlinear length scale $\xi_{nl}(T)$. On longer scales the fluctuating
flat sheet is {\em critical} at zero tension without any fine-tuning
-- a ``critical {\em phase}'' -- and is characterized by universal
power-law correlations, manifested by its anomalous elasticity driven
by thermal fluctuations and/or quenched disorder. Namely, as predicted
by theory \cite{AL,GDLP,LRprl}, it is characterized by universal
length-scale-dependent (enhanced) bending rigidity $\kappa(\q)\sim
q^{-\eta}$, and in-plane (softened) elastic moduli
$\mu(\q)\sim\lambda(\q) \sim q^{\eta_u}$, with a universal negative
Poisson ratio of $-1/3$\cite{LRprl}. All these and other properties of
such a tensionless membrane, e.g., a spectacular absence of a linear
elastic response with power-law universal nonlinear elasticity, then
follow and are controlled by an infrared stable fixed point. Although
in the presence of tension these universal singularities will be
cut off by the corresponding tension-dependent length scale, they
can, nevertheless, be experimentally observable in graphene for a
sufficiently low tension. 

In our work \cite{LRprl,LRrapid}, detailed here, we derived and
explored these properties extensively in the presence of thermal
fluctuations and random quenched disorder using a field-theoretic
method that we developed in the context of membranes, the
Self-Consistent Screening Approximation (SCSA), leading to accurate
predictions for the exponents.  As can be seen in
Fig. \eqref{etaExponentsFig} these predictions were confirmed in
numerical simulations already in the 90's.

Graphene, as the most prominent modern realization of an atom thin
elastic sheet, is subject to strong thermal fluctuations, and the
nonlinear length $\xi_{nl}(T)$ is reduced to the scale of its
nanometer lattice constant.  The renormalization of the Young modulus,
$K_0(\q) \sim q^{\eta_u}$, predicted by the theory, has been recently
observed experimentally on graphene sheets using the defect density as
a control parameter for the relevant wavevector $q$
\cite{simulationsGraphene,experimentElasticModuli} (see also
\cite{experimentElasticModuli2}).  The best fitting value $\eta_u
\approx 0.36$ found in this experiment is in close agreement with the
SCSA prediction \cite{LRprl}, $\eta_u = 0.358$.  Note that, contrarily
to the claim there \cite{experimentElasticModuli}, the
$\epsilon$-expansion \cite{AL}, which predicts a nearly vanishing
value $\eta_u=2 - \frac{24}{25} \epsilon \approx 0.08$, strongly
underestimates the renormalization of $K_0$ by thermal out-of-plane
fluctuations, which is much more accurately captured by the SCSA
method employed here. A similar renormalization was also observed in
ab-initio atomistic Monte Carlo (MC) simulations, based on a realistic
inter-atomic potential, for graphene at room temperature
\cite{simulationsGraphene}, with exponent values close to the SCSA
predictions (see also \cite{Troster}). 

Here we do not attempt any review of the subject but only quote a few
recent developments (among many).  After its first development and
application to polymerized membranes\cite{LRprl,LRrapid}, the SCSA
method was pushed to the next order (including higher order diagrams
in $1/d_c$ expansion) by Gazit \cite{Gazit} who found $\eta=0.789$,
$\eta_u=0.422$ and $\zeta=0.605$. The relative stability of these
values with the order of expansion may explain why the simpler (first
order) SCSA is so accurate. The non perturbative RG (NPRG) method was
applied in Ref.\onlinecite{Mouhanna1} who found $\eta=0.849$. A
quantum extension of the SCSA was developed in \cite{LedouRipples} to
study ripples in graphene (see also \cite{MouhannaQuantum} and
\cite{PacoRipples}).  Indeed experiments show that the suspended
graphene is not perfectly flat but exhibits static ripples
\cite{suspendGrapheneNature2007}. Understanding of the nature of these
ripples, their interplay with thermal fluctuations and disorder, and
of the statistical properties of the effective gauge fields that they
induce for electrons \cite{LedouGaugefields} remains unsatisfactory.
Finally, the crumpling transition was further studied in a number of
works, using NPRG \cite{Mouhanna1} and numerical studies, e.g., see
\cite{CrumplingPerforatedBowick} and reference therein, to quote a
few.

While there has been much activity and contact between theory and
experiments on the study of thermal fluctuations in graphene, the
effect of quenched disorder that arises from a variety of graphene
lattice defects (e.g. vacancies and interstitials, adatoms,
impurities, dislocations, grain boundaries, ripples and electron
puddles)
\cite{DislocationsInGraphene,GrainboundariesGraphene,AdatomsInteractions,
suspendGrapheneNature2007,ControledRipples},
remains significantly less explored. This is despite of the fact that
much of the general theory of fluctuating disordered membranes
(detailed here) has long been developed dating back to early 90's
\cite{NRlett,RNpra,NRpra,ML,MLG,LRprl,LRrapid,RLjphys,NS}. In recent
developments the SCSA, and related RG methods, were further used in
studying the effect of quenched disorder in graphene, e.g., in
\cite{Mirlin} and for a peculiar model of long-range disorder in 3D
printed artificial membranes \cite{Kosmrlj}.  The role of {\it
  long-range disorder}, which may result from some of the defects
listed above, remains also relatively unexplored in the context of
graphene.  Here we study it in great details and show that long-range
disorder leads to a rich variety of flat glassy phases and to crumpled
states, depending on disorder's strength and range. We hope that our
pedagogical exposition and detailed derivations in the present paper
will help fill the gap between theory, numerics and experiments on the
effect of disorder on graphene and motivate further studies.

\subsection{Motivation and background}

\subsubsection{Membranes}
Fluctuating membranes has been a subject of much research over the
last three decades \cite{Jerusalem}. On the theoretical side the
interest is motivated by the opportunity to study the interplay
between geometry, statistical mechanics and field theory. At the same
time these two- and higher-dimensional generalizations of flexible
one-dimensional manifolds are ubiquitous in nature and laboratory, and
much of the theoretical research activity has been stimulated by the
many experimental realizations. Biological membranes as e.g., walls of
living cells, are one of the early most extensively studied
realizations, which consist of amphiphilic lipid bilayer often with
protein networks permeating the membrane and giving it its structural
integrity and elasticity. In laboratory much of the effort has been
directed to utilize the self-assembling nature of various kind of
amphiphilic lipid and surfactant molecules.  These molecules when put
into an aqueous solution self-assemble into structures of various
shapes and topologies, often tens of microns in extent. Some of the
studied assemblies include open bilayers and monolayers, spherical,
cylindrical and tori topologies, and multi-layered lyotropic
microstructures \cite{Jerusalem}.

Generically, these self-assembling membranes are two-dimensional
fluids characterized by a vanishing in-plane shear modulus, with
constituent molecules free to diffuse within the sheet. These {\em
  fluid} membranes are therefore somewhat fragile, and in the presence
of fluctuations will often break up into smaller parts or undergo
topological transformations.  As detailed in
Ref.\onlinecite{Jerusalem} their thermal fluctuations and resulting
properties are quite distinct from tensionless {\em elastic} sheets
that is our focus here.

The self-assembling membranes have many potential technological
applications such as the use of the phospholipid vesicles and tubules
for drug encapsulation and delivery, blood substitutes, perfumes, and
antifouling paints \cite{Schnur}.  Because these applications require
stable membranes, many of the experimental efforts have been directed
toward cross-polymerizing liquid membranes. Using multipli-bonded
phospholipids such as diacetylenic lipids, UV irradiation can be used
to activate the bonds and create a {\em polymerized} membrane with
bonds virtually unbreakable on the scale of thermal energies at
room-temperature. These two- and higher-dimensional realizations of
linear polymers have been the subject of extensive studies dating back
to late 1980's\cite{Jerusalem} and are the focus of our manuscript.

There are many other naturally occurring realizations of polymerized
membranes.  The inner surface of red blood cells contains the
fishnet-like biopolymer spectrin network which can be extracted and
studied in isolation from the lipid bilayer \cite{Branton,Schmidt}.
Red blood cells themselves, with the spectrin attached to the lipid
cell wall, can also be described by membrane theories, as emphasized
by Lipowsky and Girardet \cite{LG}.  Leibler~\cite{Leibler} has
suggested that simple ``paracrystals'' of proteins like tropomyosin
provide another example of a biological tethered surface.  Inorganic
examples of tethered surfaces include graphite oxide sheets in an
appropriate solvent \cite{Hwa} and the ``rag'' sheet-like structures
found in MoS$_2$ \cite{Chian}. And of course, as discussed above,
laboratory realizations of single-atom thin graphene sheets have
rekindled a renaissance in the study of electronic and mechanical
properties of fluctuating elastic sheets
\cite{GeimMacDonald,reviewRMPGraphene}. For other experimental
realizations, see Ref.~\onlinecite{Jerusalem}.

\subsubsection{Elastic membrane and its critical ``flat'' phase}

Elastic polymerized membranes, two-dimensional generalizations of
linear polymer chains \cite{KKN,Jerusalem}, have thus been a focus of
a large number of theoretical studies over the course of three decades
starting in mid 1980's.  Analogously to one-dimensional polymers,
membranes are expected to be crumpled at high temperatures \cite{KKN},
though for most physical realizations a membrane crumpling scale (that
scales exponentially with a ratio of the bending rigidity to
temperature) is astronomically long.  On the other hand at low
temperature based on simple considerations one may expect that a
tensionless membrane will undergo a transition to a ``flat phase'' in
which it is on average flat but with strong fluctuations about a
spontaneously selected plane \cite{PKN}.

On the other hand, at a simple-minded, mean-field level a membrane is
quite analogous to a ferromagnetic spin system, with the local normals
to the membrane playing the role of spins. The ``crumpled'' and the
flat states of a membrane are then the analog of the paramagnetic and
ferromagnetic phases, respectively.  On the other hand, it is expected
that {\em two}-dimensional systems with {\em short-range} interactions
are forbidden to undergo a transition in which a continuous symmetry
is broken spontaneously \cite{Hohenberg,MerminWagner,Coleman}, as
happens when the $O(3)$ rotational symmetry of the crumpled phase is
spontaneously reduced to $O(2)$ symmetry in the flat phase. It is the
low-energy spin-wave fluctuations, that are responsible for the
destruction of a two-dimensional ordered phase in a conventional
$O(N)$ spin system \cite{Polyakov}. In fact, this expectation is
realized in liquid membranes and in linear polymers, whose flat phase
is strictly speaking (though typically difficult to observe) destroyed
by thermal conformational fluctuations at any finite temperature in a
thermodynamic limit.

In stark qualitative contrast, the flat phase of a {\em polymerized}
membrane was predicted \cite{NP} to be {\em stable} at low
temperatures, a highly nontrivial observation in thermodynamic limit,
that is seemingly in conflict with Hohenberg-Mermin-Wagner
theorems. In polymerized membranes, through the shear modulus coupling
of the in-plane and out-of-plane deformations, fluctuations that try
to (and do so in a fluid membrane) destabilize the flat phase, instead
infinitely stiffen its bending rigidity via a statistical
``corrugation'' effect \cite{NP,Jerusalem}.  The resulting anomalous
elasticity then in fact stabilizes the flat phase against these very
same fluctuations, in a spectacular phenomenon of order-from-disorder.

This striking phenomenon was first demonstrated using a simple
one-loop self-consistent theory that {\it assumed} a
non-renormalization of in-plane elastic moduli, leading to a roughness
exponent $\zeta = 1/2$ and $\eta = 1,\eta_u = 0$ \cite{NP}.  Later
detailed renormalization group calculations \cite{AL}, controlled by
an $\epsilon=4-D$ expansion, predicted renormalized elastic constants
$\lambda(\q) \sim \mu(\q) \sim q^{\eta_u}$, $\eta_u>0$, with
${\eta_u=4-D-2 \eta}$, a relation imposed exactly by the underlying
rotational invariance.  This study found $\eta=12\epsilon/(24+d_c)$
($d_c=d-D$, the codimension of the manifold), leading to $\zeta\approx
1/2$ for physical membranes \cite{AL}.  A complementary
$1/d$-expansion similarly confirmed the stability and anomalous
elasticity of the flat phase and predicted $\zeta\approx 2/d$ for a
two-dimensional polymerized membrane \cite{GDLP}.

\subsubsection{Crumpling transition and the crumpled phase}

Similar to conventional polymers, the crumpled phase of a membrane is
characterized by a radius of gyration $R_G$, which describes the
average size of the crumpled membrane inside the $3$-dimensional
embedding space. The radius of gyration that characterizes the
crumpled phase is predicted to scale as a power of the internal size
of the membrane, $R_G(L)\sim L^\nu$
\cite{KardarNelsonCrumpledPRL,ALcrumpled,DuplantierPRL}.  Flory type
arguments, which are based on dimensional analysis predict
$\nu=(D+2)/(d+2)$. This approach had lead to accurate predictions of
$\nu$ for one-dimensional polymers, but it is not clear how accurate
its prediction is for membranes and higher dimensional manifolds.

However, the high-temperature crumpled phase has turned out to be
elusive to numerical and experimental realization, because the
self-avoiding interaction of a membrane tend to stabilizes the flat
phase. At first, a high-temperature crumpled phase has only been seen
in computer simulations of the so-called ``phantom'' membranes, in the
absence of self-avoiding interaction, with interactions between
nearest neighbor monomers only \cite{KantorNelson,ARP,AN}.  The
crumpled phase however has subsequently been seen in Monte Carlo
simulations of self-avoiding tethered surfaces modeled by impenetrable
flexible plaquette.  It has also been demonstrated to exist as an
intermediate phase between a collapsed and the flat phases of the
membrane with finely-tuned attractive interactions \cite{PlishkeBoal}.
Experimentally investigated graphite oxide sheets \cite{Hwa}, MoS$_2$
structures \cite{Chian} and red blood cell ``ghosts'' \cite{Schmidt}
were also observed to exist in the crumpled phase.

The theory of the crumpling transition \cite{PKN, GDLP} started with
considering phantom membranes.
%
%
%
%Analogously, as we discuss in detail in the rest of the paper,
%the flat phase is described by the average size of the height
%deviations from the average perfectly flat configuration. Powerful
%theoretical techniques have been developed to treat the crumpled
%phase, flat phase and the crumpling transition. 
%Fluctuations can be
%treated by considering $D$-dimensional polymerized manifolds embedded
%in $d$-dimensions, and then carrying out expansions in $\epsilon=4-D$
%or $1/d$ to treat the physically relevant case $D=2$ and $d=3$.
In mean-field approximation the crumpling transition between the flat
and crumpled phases is second-order.  However, there are several open
problems with the theory. First the renormalization group calculation
of Paczuski {\it et al.} \cite{PKN} did not find a perturbative
critical point for the physical case of $d = 3 < d_c = 219$. Although
they interpreted these runaway flows as a fluctuation-driven
first-order transition, the explicit demonstration of the order of the
transition has remained open.  Other related technical issues will be
discussed at the end of Section \eqref{sec:crumpling}.  Secondly, the
role of self-avoidance at the crumpling transition remains to be
elucidated. Indeed, if $2 D - \nu_{\rm c} d>0$ self-avoidance should
be relevant, where $\nu_{\rm c}$ is the radius of gyration exponent at
the transition.
%\cite{LeDouRadzCrumplingUnpublished}.

\subsubsection{Quenched disorder heterogeneity in an elastic membrane}

Motivated by a variety of physical realizations, subsequent studies of
elastic membranes considered the effects of quenched internal disorder
in addition to thermal fluctuations \cite{NRlett,RNpra}.  Various
kinds of local random heterogeneities are an almost inevitable feature
of real membranes and graphene. Examples of disorder include holes or
tears in the polymerized network, variations in the local coordination
number and impurities in the form of functional proteins and lipids of
odd size incorporated at random into the biological membrane.  Partial
and random polymerization of self-assembled microstructures will also
inadvertently lead to defects in the form of vacancies, interstitials,
dislocations and disclinations in a polymerized membrane and graphene
sheets \cite{NRpra}.

In such elastic membranes most of the defects leading to disorder will
relax on much longer time scale than the conformational degrees of
freedom of the membrane and can therefore be treated as static, i.e.,
quenched. In contrast, the annealed (dynamic) disorder can be shown to
be unimportant, since it leads to only nonsingular renormalization of
the elastic coefficient.  Quenched disorder in a membrane leads to a
local, random curving and stretching of the sheet even in its zero
temperature ground state. Disorder that curves the membrane locally
breaks the up-down symmetry of the sheet, while the stretching
disorder respects this inversion symmetry \cite{RNpra}.  As was first
described theoretically by Nelson and Radzihovsky \cite{RNpra}, in the
continuum description of a membrane the important effects of disorder
can be summarized by a quenched extrinsic curvature and random stress
disorders, which are analogous to the random field and random exchange
models of the magnetic spin systems, respectively.  Vacancies and
interstitials introduce local uncorrelated random strains and
therefore lead to disorder that has short-range spatial
correlations. On the other hand, disclinations, dislocations and
grain-boundaries resulting from partially polymerized liquid membranes
will lead to local stresses that are power-law correlated and
therefore can be modeled by strain and curvature disorder, but with
power-law correlations \cite{NRpra}.

In the crumpled phase of phantom membranes the disorder leads to a
swelling of the membrane but does not modify the size exponent $\nu$
\cite{RNpra}.  Simple arguments also suggest that disorder will modify
the crumpling transition but careful analysis remains to be
performed. At nonzero temperature, the effect of disorder on the flat
phase were studied extensively and more carefully.  Based on
renormalization group calculations and expansions in dimensionality it
was found that at long wavelengths, short-range weak stress-only
disorder does not lead to significant modifications of the properties
of pure membranes \cite{RNpra}.  At vanishingly small temperatures the
disorder becomes important and its effective strength grows at long
length scales. At zero temperature the short-range curvature disorder
dominates and leads to a new ground state whose roughness is found to
scale as $L^{\zeta'}$, analogously to the finite temperature
roughness, but now in the statistically rough ground state
\cite{MLG,ML}. This zero-temperature ``crinkled'' fixed point, was
found to be marginally unstable to temperature.  The
$\epsilon$-expansion of \cite{ML} predicted a roughness exponent
$\zeta'=1 - \frac{3}{14} \epsilon \approx 0.571$. The numerical
simulation found $\zeta'=0.81 \pm 0.02$ \cite{MLG}.  As we detail in
current manuscript, we extended the SCSA to disordered membranes, and
found the zero temperature crinkled phase solution with the
ground-state roughness exponent $\zeta'=0.775$ much closer to the
numerical simulation value of Ref.\onlinecite{MLG}.

The early theoretical consideration of disordered polymerized
membranes was motivated by beautiful experiments on polymerizable
lipid vesicles subjected to UV irradiation \cite{Bensimon}. Upon
partial polymerization the vesicles were observed to undergo what
appeared to be a reversible, first-order wrinkling phase transition
either spontaneously or upon cooling (depending on experimental
conditions). During the transition a spherically symmetric fluctuating
vesicle would undergo a shrinking and wrinkling transformation to a
raisin glassy structure. 

Since, as described above, the short-range disorder was shown to be
irrelevant at finite temperatures, the generation and effects of
long-range disorder were naturally considered.  A simple microscopic
model was used to demonstrate that long-range disorder in the form of
networks of grain-boundaries and unscreened disclinations will
naturally arise in partially polymerized membranes \cite{NRpra}.  It
was found that unlike short-range disorder, the resulting long-range
strains and extrinsic curvature disorders with power-law correlations,
with exponents larger than a critical value, lead to an instability of
the membrane's flat phase\cite{NRpra,LRrapid}. The strength and onset
of the instability was also estimated from simple extensions of
Harris-type arguments. The nature of the resulting phase was
investigated theoretically in \cite{RLjphys} and was described as a
"crumpled glass" with spin glass like order in the membrane
normals. Although the detailed nature of this phase remains open the
disorder-activated instability predicted by the theory appeared to be
a good candidate to explain the wrinkling transition observed in the
experiments.

\subsection{Self-consistent screening approximation (SCSA)}

\begin{figure}[htbp]
        \centering
        \includegraphics[width=0.7\textwidth,scale=1]{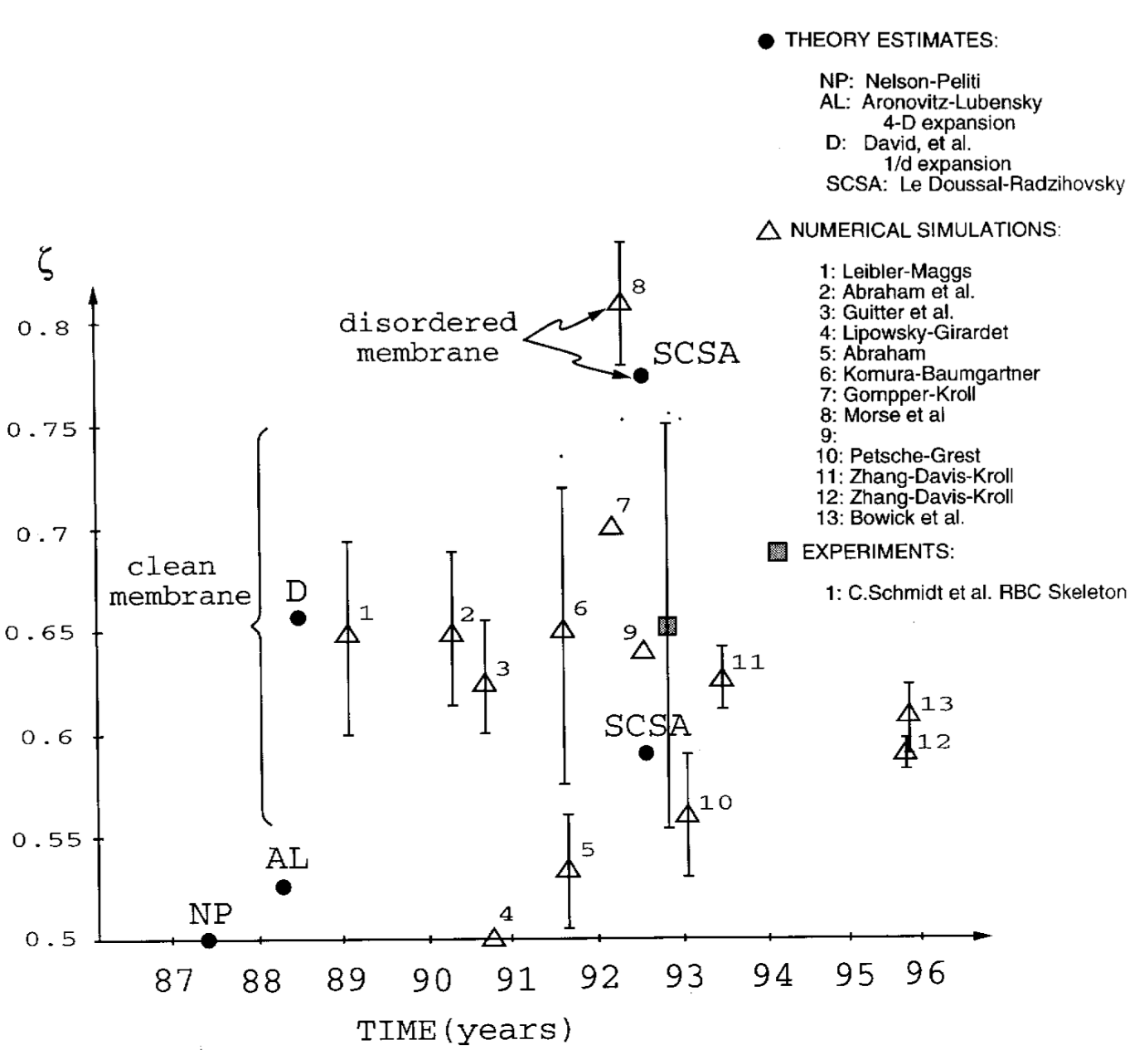}
        \caption{\label{etaExponentsFig} Universal roughness exponent
          $\zeta$ of a tensionless elastic membrane (with physical
          dimensions $D=2$, $d_c=1$), as a function of the year, in
          the 1987-1996 range. $\zeta$ is also directly related to the
          anomalous (length-scale-dependent) elasticity exponents
          $\eta=2(1-\zeta)$ and $\eta_u=4 \zeta-2$.  Black dots:
          theoretical predictions (NP \cite{NP}, AL \cite{AL}, D
          \cite{GDLP}, SCSA \cite{LRprl}).  Triangles: numerical
          simulations (with error bars) (1\cite{LeiblerMaggs},
          2\cite{AN}, 3\cite{GuitterMC}, 4\cite{LipowskyGirardet},
          5\cite{Abraham}, 6\cite{KomuraBaumgartner},
          7\cite{GompperKroll}, 8\cite{MLG}, 
          10\cite{PetscheGrest}, 11\cite{ZhangDavisKroll},
          12\cite{ZhangDavisKroll2}, 13\cite{BowickFlat}. Square:
          experiment on red blood cell
          cytoskeleton\cite{Schmidt}. Note that the SCSA prediction
          for the Poisson ratio $\sigma=-1/3$ was confirmed in
          numerical simulations in \cite{PoissonBowick1} who obtained
          $\sigma=-0.32(4)$.}
\end{figure}

Theoretical investigations have led to much progress in qualitative
understanding of polymerized membranes, especially of the flat phase
where the challenging self-avoiding interaction is believed to be
unimportant.  Unfortunately, the quantitative predictions of the
$\epsilon$- and $1/d$-expansions are far from the physical dimensions
of a two-dimensional membrane embedded in 3d, explaining the lack of
accuracy of these approaches. The original attempt at a
self-consistent approach failed to obtain an independent set of
equations to determine $\eta$ and $\eta_u$, reproducing the rotational
symmetry based relation between these exponents
\cite{NP}. Furthermore, for heterogeneous membranes controlled by
quenched disorder even the qualitative questions of the nature of the
instability and the resulting phase remained a mystery.

This motivated us in 1992 to study homogeneous \cite{LRprl} and
heterogeneous \cite{LRrapid} membranes with a Self-Consistent
Screening Approximation (SCSA), generalized from the studies of $O(N)$
magnets \cite{Bray}. As we detail below, this analysis allowed us to
perform calculations directly for membranes of physical dimensions
(and their arbitrary $D,d$ generalizations) and uniquely determines
the relevant exponents.  For disorder-free thermal membranes we
calculated the roughness exponent and found $\zeta=0.590$, which fell
in the middle of all the known predictions from numerical simulations
at the time. The comparison with numerics is summarized in
Fig.\ref{etaExponentsFig}.  In addition the SCSA prediction for the
Poisson ratio, $\sigma = -1/3$, was later confirmed by numerical
simulations (see caption of Fig.\ref{etaExponentsFig}). Furthermore,
as detailed below, we also extended the SCSA method to describe the
crumpling transition in thermal, homogeneous membranes.

In contrast to its original application to the $O(N)$ model, we found
that the SCSA is unprecedentedly accurate when applied to the problem
of polymerized membranes. As we demonstrated, although uncontrolled
(no small systematic expansion parameter), this method encompasses
both the $\epsilon$ and $1/d$ approximations.  Namely, it is exact to
first order in $\epsilon$ for arbitrary $d$, exact for $d = D$, and by
construction, it is exact in the limit of $d\rightarrow\infty$.  Our
results thus interpolate between two extreme dimensionalities, and are
therefore expected to be substantially more accurate for physical
membranes, than the RG perturbative expansions in dimensionalities.

Encouraged by the quantitative success of the SCSA, as we detail
below, we extended it to membranes in the presence of short- and
long-range disorder \cite{LRprl,LRrapid}. We found a rich phase
diagram that describes flat phases of a homogeneous and of a
short-range-disordered membrane, previously studied, as well as new
wrinkled glassy phases towards which the ordinary flat phase is
unstable when long-range disorder is turned on. The SCSA theory thus
captures the nature of the new disordered glassy phases and makes
quantitatively accurate predictions about their statistical
conformations and effective mechanical properties.

\subsection{Outline of the manuscript}

In the present manuscript we present the previously unpublished
calculational details of our studies of polymerized membranes from the
mid 1990s. As discussed in Section \ref{preamble} these results became
even more relevant after 2004 with the synthesis of atom-thin graphene
sheets. In the next Section \ref{sec:generalizedmodel}, we review the
main properties of the continuum model of an elastic membrane, and
describe the mean-field theory, thermal fluctuations and the model for
a disordered membrane. In Section \ref{PureMembrane} we present the
details of the SCSA method and apply it to the calculation of the
roughness $\zeta$ and anomalous elasticity $\eta,\eta_u$ exponents for
a homogeneous thermally fluctuating membrane in its flat phase. In
Section \ref{sec:crumpling} we extend the SCSA method to the study of
the crumpling transition of a phantom membrane, from which the lower
critical dimension for the finite temperature transition can be
calculated.  From the lower critical dimension the roughness exponent
for the flat phase can be extracted and we find that it is in complete
agreement with one found in Section \ref{sec:generalizedmodel}.  In
Section\ref{MembraneDisorder} we derive the effective model for the
flat phase in presence of disorder in terms of replicated out-of-plane
height fields. We use this model to extend the SCSA equations to the
case of the disordered membrane. In Section \ref{sec:analysis} we
analyze the SCSA equations for short-range disorder, and obtain the
disorder related exponents $\zeta',\eta',\eta_u'$.  Finally, in
Section \ref{sec:LRdis} we analyze the SCSA equations for the
heterogeneous membrane in presence of long-range disorder, and derive
the rich phase diagram containing a variety of glassy wrinkled phases,
determined by the range and strength of quenched disorder. We analyzed
in detail the properties of these glassy phases. 

\section{Generalized model of a polymerized membrane} \label{sec:generalizedmodel} 
\subsection{Homogeneous membrane: mean-field theory}
As will become clear from our analysis, it useful to study a
higher-dimensional generalization of a physical two-dimensional
membrane. Without much extra difficulty we will study an elastic
$D$-dimensional manifold of interconnected particles fluctuating in
the embedding $d$-dimensional space. Physically important degrees of
freedom are the positions of the constituent particles.  At long
length-scale, continuum description that we will be interested in
here, the discrete network, on scales longer than the typical mesh
length $a$, can be described as a continuous manifold with the
internal points labeled by a $D$-dimensional coordinate $\x$ with
components $x\as$, $\alpha=1,\cdots D$, and their location in the
embedding space specified by a $d$-dimensional vector $\rv(\x)$, with
components $r_i(\x)$, $i=1,\cdots d$.  We consider the case of
positive codimension $d_c=d-D>0$.  Underlying symmetries of the
problem impose constraints on the possible local Landau-Ginzburg
Hamiltonians that one can write to describe the long wavelength
properties of the membrane. The rotational and translational
invariances of the embedding space require the effective Hamiltonian
to be a local, analytic expansion in the scalar product of a local
tangent vector-field $\tv\as(\x)=\pt\as\rv(\x)$.  For a homogeneous
membrane these invariances also exist in the internal space.  They
restrict the Hamiltonian to a scalar with respect to rotations of the
$\x$ coordinate system and require constant coefficients in the
Landau-Ginzburg expansion \cite{Landau,commentTubulesGeneralization}.

Focussing our attention on membranes with only local interactions both
in the embedding and internal spaces and keeping only terms that are
most dominant at long scales, we obtain the Hamiltonian previously
used in a number of studies of polymerized
membranes\cite{KKN,PKN,KantorNelson,ARP}
\begin{equation}
\label{hamEFF}
{\cal F}[\vec{r}] =\intd\left[{1\over2}\;\kappa(\partial^2
\vec{r})^{2} + {1\over2} \tau(\partial_{\alpha}\vec{r})^{2}
+ u(\partial_{\alpha}\vec{r}\cdot\partial_{\beta}\vec{r})^{2} +
\tilde{v}(\partial_{\alpha}\vec{r}\cdot\partial_{\alpha}\vec{r})^{2}\right]\;,
\end{equation}
where the sums over the repeated indices are implied.  Upon
identifying the tangent vector $\tv\as(\x)$ or equivalently the local
normal as the order parameter, the above Hamiltonian takes the form of
a $\phi^4$-theory typically used in critical phenomena to describe
second-order phase transitions. We therefore expect that at least in
mean-field theory the membrane will undergo a continuous phase
transition.

By minimizing the above Hamiltonian with respect to $\vec r$, we find
that for positive values of the reduced temperature $\tau$
($\tau=T-T_c^{MF} > 0$) the order parameter $\tv\as$ vanishes,
describing the crumpled phase of the membrane characterized by
randomly distributed local normals with zero average.  On the other
hand for $T<T_c^{MF}$ the bending energy (first term) dominates and
the energy is minimized by aligning the local normals. The resulting
ordered flat phase spontaneously breaks the full $O(d)\times O(D)$
rotational symmetry of the membrane down to $O(d)\times O(D)/
O(D)$. In the flat phase the average tangent is nonzero and in
mean-field theory is given by,
\begin{equation}
\label{MFTtangent}
\vec r({\bf x}) = t (x_1,\cdots x_D,0) \quad , \quad
|\tv\as\,^{MF}| =t={1\over2}
\left[{-\tau \over u+D\tilde{v}}\right]^{1/2}\;.
\end{equation}
$t$ is the scalar order parameter of the mean-field theory which
vanishes in the high-temperature crumpled phase. In the flat phase it
measures the shrinkage factor of the membrane with respect to its
internal size $L$ due to its out-of-plane thermal undulations. We
observe that the usual mean-field order parameter exponent $\beta=1/2$
has emerged from Eq. \eqref{MFTtangent}.  In the flat phase the
quartic $(u,\tilde v)$ interactions become essential to render the
Hamiltonian positive.

It is enlightening to examine the physical origin as well as the
geometrical structure of the various terms appearing in the effective
Hamiltonian, Eq. \eqref{hamEFF}.  The first term represents the bending
rigidity of the membrane which arises from the energetic and entropic
interactions, e.g., between amphiphilic rod-like lipids in the case of
biological membranes, or from carbon-carbon bond stiffness in
graphene. This interaction can be derived from a lattice model of
locally interacting normals $\n_{{\bf x}}$, similar to the Heisenberg
model of spin systems
\begin{equation}\label{hamBi}
{\cal H}_{\rm bend} =- \tilde \kappa\sum_{<{{\bf x}} {\bf x}'>}\n_{{\bf x}}\cdot
\n_{{\bf x}'}\;.
\end{equation}
Ignoring a constant term, generalizing to $D$-dimensions and taking
the continuum limit, we obtain,
\begin{equation}\label{hamBiii}{\cal H}_{\rm bend} 
= {1\over2}\kappa\intd(\partial_{\alpha}\n({\bf x}))^2,
\end{equation}
where $\kappa \sim \tilde \kappa$ is the bending rigidity modulus.
Since the normal is related to the tangent $\pt\as\rv$ by a simple
rotation, we obtain the bending rigidity term of Eq.\eqref{hamEFF}.
Geometrically this contribution is simply the square of the extrinsic
curvature \cite{GaussCurv,Jerusalem}, originally proposed by Frolich
to describe the oblate shape of red blood cells.

The rest of the terms in the Landau-Ginzburg expansion are the elastic
contributions to the membrane energy arising from an expansion of the
local nearest neighbor tethering interaction,
\begin{equation}\label{hamEi}{\cal H}_{\rm el} = \sum_{{\bf x}, {\bf e}} V(\rv({\bf x}+{\bf e}) - \rv({\bf x})) \end{equation}
in powers of the local tangent $\pt\as\rv$ in the continuum limit.

To study the flat phase of the membrane it is convenient to rewrite
the effective Hamiltonian Eq.\eqref{hamEFF} in a slightly different
form. By completing the square in the elastic interaction we obtain,
\begin{equation}
\label{Fflat}
{\cal F}[\rv] =\intd\left[{1\over2}
\;\kappa(\partial^2\rv)^{2} + \mu(u_{\alpha\beta})^2 + 
{1\over2}\lambda(u_{\alpha\alpha})^2\right],
\end{equation}
where $\mu$ and $\lambda$ are known as the Lam\'e coefficients and are
related to the previous coupling constants, $\mu=4ut^4$,
$\lambda=8\tilde{v}t^4$.  $u\ab$ is the strain tensor related to the
membrane metric tensor $\pt\as\rv\cdot\pt\bs\rv$ inherited from the
embedding,
\begin{equation}
\label{strainPure}
u\ab={1\over2 t^2}(\pt\as\rv\cdot\pt\bs\rv- t^2\delta\ab).
\end{equation}
The strain tensor measures the local deformation of the membrane
relative to the metric of the lowest energy configuration, which for
homogeneous (disorder-free) membranes is the dilated flat metric
$t^2\delta\ab$.  In this form it becomes obvious that the total energy
is minimized by minimizing the curvature and the elastic terms
independently. The curvature term is minimized by any constant tangent
vector, while the elastic terms vanish when the metric is precisely
the preferred flat metric of the homogeneous membrane,
$\pt_a\rv\cdot\pt\bs\rv= t^2\delta\ab$, with the obvious solution,
\begin{equation}
\label{rGround}
\rv(\x)= t \left(x_1, x_2, \ldots , x_D, 0,
0,\ldots,0\right)\;.
\end{equation}

The effects of the fluctuations in the flat phase can be studied by
expanding the general conformation of the membrane around this
perfectly flat ground state. A general deformation of the membrane
that assumes no overhangs is described in the Monge representation,
\begin{equation}\label{Monge}\rv(\x)= t \left((x\as + u\as(\x)){\hat e}\as + \h(\x)\right)\;, \end{equation}
where $u\as(\x)$ are the in-plane $D$ phonon displacements, $\h(\x)$
are the $d_c$ out-of-plane membrane undulation modes \cite{ModeCount},
and the ${\hat e}_\alpha$ form a $D$-dimensional basis of in plane
unit vectors.  In this representation the strain tensor becomes,
\begin{eqnarray}\label{strainPureFlat}u\ab & = & \frac{1}{2} \left(  \pt\as u\bs +\pt\bs u\as+\pt\as\h\cdot\pt\bs\h\;
+  \partial_\alpha u_\gamma\partial_\beta u_\gamma\right) \label{st1},  \\
& \simeq & \frac{1}{2} \left(  \pt\as u\bs +\pt\bs u\as+\pt\as\h\cdot\pt\bs\h  \right) \;, \label{st2}
\end{eqnarray}
where the phonon non-linearity (last term) has been neglected when
going from the first line to the second, since it is quite clearly
subdominant to $\partial_\alpha u_\beta$ at long wavelengths
\cite{NP}.  Aside from the bending rigidity term the effective
Hamiltonian takes on the standard form of elastic energy of an
isotropic solid \cite{Landau,commentTubulesGeneralization}
\begin{equation}\label{Fflatii}
{\cal F}[\h,u\as] =\intd\left[{1\over2}
\;\kappa(\partial^2\h)^{2} + \mu(u_{\alpha\beta})^2 + 
{1\over2}\lambda(u_{\alpha\alpha})^2\right]. 
\end{equation}
Note that in the broken symmetry flat phase the underlying
embedding-space rotational invariance of the model \eqref{hamEFF},
manifest in \eqref{strainPure}, is encoded in the non-linear form of
$u_{\alpha \beta}$.  For instance, this nonlinear form ensures that
the (apparent) distortion $u^0_1(x_1,x_2) = x_1 (\cos \theta-1)$,
$h^0(x_1,x_2) = x_1\sin \theta$, corresponding to a rigid rotation of
a flat membrane (in e.g., the $1-3$ plane) consistently gives a
vanishing strain, $u^0_{\alpha \beta}=0$.

%For instance, for a physical membrane ($D=2$,
%$d=3$) rotated in the $x_1,x_3$ plane the rotation $u_1(x_1,x_2) \to
%(\cos \theta-1) x_1 + u_1 \cos \theta - h \sin \theta$, $h(x_1,x_2)
%\to (x_1 + u_1) \sin \theta + h \cos \theta$ leaves $u_{\alpha \beta}$
%in \eqref{st1} invariant exactly, and its approximation \eqref{st2}
%(used below) invariant up to order $O(\theta^2)$.

\subsection{Homogeneous membrane: thermal fluctuations}

We now consider the effect of thermal fluctuations around the
mean-field flat phase, defined by the in-plane phonons $u_\alpha$ and
out-of-plane flexural modes $\vec h$. The statistical averages of any
observable $O[\h,u\as]$ are defined from the Gibbs measure 
\bea
\langle O[\h,u\as] \rangle = {\cal Z}^{-1} \int {\cal D} u_\alpha
{\cal D}\vec h \, O[\h,u\as] \, e^{- {\cal F}[\h,u\as]/\T}. 
\eea
As we already described in the Introduction, the fluctuations in the
flat phase have been studied by the methods of renormalization group
together with $\eps$- and $1/d_c$-expansions \cite{AL,GDLP}, as well
as with the SCSA described below \cite{LRprl}, with the key result
that the flat phase is stable to thermal fluctuations up to the
crumpling transition temperature $T_c$.  Surprisingly, the
fluctuations actually tend to stabilize the flat phase \cite{NP} and
lead to the elastic properties that differ remarkably from that of
classical theory of local elasticity \cite{Landau}. As we will show
below, in the presence of these fluctuations the membrane is described
by a universal (fixed point) renormalized Hamiltonian of the same form
as Eq.\eqref{Fflatii}, but with length scale-dependent renormalized
elastic moduli, with the bending rigidity diverging as a power law of
length scale,
\be
\kappa_R(\kb)\sim k^{-\eta}\label{kappa_eta},
\ee
and vanishing Lam\'e coefficients
\be \mu_R(\ks)\sim\lambda_R(\ks)\sim k^{\eta_u}.
\ee
The underlying rotational invariance in the embedding space, that
requires that coarse-graining (RG) preserves the full nonlinear form
of the strain tensor, Eq.\eqref{st2}, implies the exact relation
\be \eta_u = 4 - D - 2\eta.
\label{exponentrelation} 
\ee
Thus there is only a single independent universal exponent
characterizing the large scale fluctuations of a homogeneous membrane.
A related property is the roughness of the membrane due to thermal
fluctuations, defined by the root-mean-square fluctuations of the
height difference of two distant points on the membrane,
\begin{eqnarray}\label{Rough} 
\langle(\h(\x)-\h(0))^2\rangle&=& 2 d_c T
\int{d^D k\over (2\pi)^D}{1-e^{i\ks\cdot\x}\over\kappa_R(\ks) k^4}\; 
\sim x^{2\zeta}\;, 
\quad  \quad \langle h_i(\ks) h_j(-\ks)\rangle 
 = \frac{T}{\kappa_R(\ks) k^4}\delta_{ij} \sim \frac{1}{k^{4-\eta}}\;, 
\end{eqnarray}
where the Fourier transform is defined in the standard way
$\h(\ks)=\intd\h(\x){\rm e}^{-i\ks\cdot\x}$, we use units in which the
total membrane area is $1$ (i.e., $(2\pi)^D\delta^{(D)}(\kb-\kb) =
1$), and $\zeta$ is the out-of-plane roughness exponent, related to
$\eta$ as
\begin{equation}
\label{ZetaIdentity}
\zeta=(4-D-\eta)/2\;. 
\end{equation}
The in-plane phonon fluctuations also acquire an anomalous dimension,
and at large scale have a universal power-law roughness, 
\begin{eqnarray}\label{Rough2} 
\langle(u_\alpha(\x)-u_\alpha(0))^2\rangle \sim x^{2 \zeta_u}\;, 
\quad  \quad \langle u_\alpha(\ks) u_\beta(-\ks)\rangle \simeq 
\frac{T}{ \mu_R({\bf k}) k^2} P^T_{\alpha \beta}({\bf k})
+ \frac{T}{( 2 \mu_R({\bf k}) + \lambda_R({\bf k}) ) k^2}
 P^L_{\alpha \beta}({\bf k})\sim \frac{1}{k^{2+\eta_u}} \;, 
\end{eqnarray}
where $\zeta_u$ is the in-plane phonon roughness exponent, related to
$\eta_u$ and therefore to the transverse roughness exponent $\zeta$
according to
\begin{equation}\label{Zeta_uIdentity}\zeta_u=(2-D+\eta_u)/2\; = 2 \zeta - 1. \end{equation}
It is clear from Eq.\eqref{Rough} that for $\zeta < 1$, while
transverse thermal height fluctuations diverge with system size,
$h_{rms}\sim L^\zeta$, they remain smaller than membrane's in-plane
extent, $L$, and thus the rotational broken symmetry ``flat'' phase
(though qualitatively quite distinct from its literally flat
mean-field $T=0$ form) is stable to low-temperature thermal
fluctuations. Based on Eq.\eqref{ZetaIdentity}, this stability is
satisfied in the case of $D=2$ for $\eta > 0$, driven by the anomalous
enhancement of the membrane's bending rigidity, \eqref{kappa_eta} by
the very same thermal fluctuations that try (but fail at low
temperature) to destabilize the flat phase. This order-from-disorder
phenomenon thus circumvents the naive application of the
Hohenberg-Mermin-Wagner theorem\cite{MerminWagner,Hohenberg,Coleman},
that would otherwise argue for an instability of the two-dimensional
flat phase to arbitrarily weak thermal fluctuations.

In the subsequent section we will derive these results, computing the
anomalous exponents, by using the SCSA method which encompasses both
the $\eps$- and $1/d_c$-expansion results as special limits and as we
will argue is therefore significantly more accurate than all the
previous dimensional expansions.  In the next subsection we describe
how the model of a homogeneous membrane of this section can be
generalized to treat the effects of quenched internal disorder that
appears to be relevant to understanding a variety of experimental
systems, as described in the Introduction.

\subsection{Model of a heterogeneous membrane: quenched disorder}
\label{sec:disM} 
The model for a homogeneous membrane can be extended to treat the
effects of disorder, as was first done by Nelson and Radzihovsky
\cite{NRlett}.  They showed that the effect of weak disorder in the
high temperature phase is simply to swell the size of the crumpled
membrane, without modifying the radius of gyration exponent $\nu$. On
the other hand based on $1/d$-expansion we subsequently argued that
strong disorder leads to a transition to crumpled glassy phase in
which the membrane is frozen into an isotropic spin-glass like
configuration \cite{RLjphys}.  Here, however, we concentrate on weak
quenched disorder and study its effects only in the flat phase.

To understand how disorder modifies the membrane model we return to
the effective Hamiltonian of Eq.\eqref{Fflat} and first examine the
elastic and bending effects.  The simplest contribution to disorder
will come from a larger or smaller molecule embedded inside the
regular crystalline matrix, or vacancies and interstitials in
graphene, that will lead to a local compression or dilation.  The
strain tensor measures the local deformation of the membrane relative
to the metric of the lowest energy configuration.  For a homogeneous
membrane the reference metric $g^o\ab({\bf x})$ is (aside from a
global scaling factor) a flat metric, $\delta\ab$.  However, in the
presence of disorder, the reference metric $g^o\ab({\bf x})$ will be
modified to reflect the local deformation of the membrane due to the
presence of impurities.  In the continuum limit this will lead to a
rough membrane even at $T=0$ and with the ground state described by a
non-trivial disorder-dependent conformation. This can be modeled by
generalizing the expression for the strain $u^o\ab$,
Eq.\eqref{strainPure} to 
\begin{eqnarray}\label{strainDisorder}
u\ab(\x) &=& {1\over2 t_o^2}(g\ab(\x) -g^o\ab(\x))\;.
% \\
%               &= & {1\over2 t_o^2}(\pt_\alpha\rv\cdot\pt_\beta\rv -
%      \pt_\alpha\rv_o\cdot\pt_\beta\rv_o)\;. 
\end{eqnarray}  
One particularly simple example is to choose
$g^o\ab(\x)=\pt\as\rv_o(\x)\cdot\pt\bs\rv_o(\x)$, where $\rv_o(\x)$ is
a given, random configuration. The elastic energy is not frustrated as
it can be perfectly minimized by having $\rv(\x)$ conform to
$\rv_o(\x)$. This is the membrane analog of the Mattis glass model for
a non-frustrated random magnet \cite{Mattis} (for a recent realization
of this model see \cite{Kosmrlj}).  However, this is clearly not the
most general case of disorder, since $g^o\ab(\x)$ can be chosen a
priori to be an arbitrary local symmetric tensor, in which case the
ground state will generically be frustrated.

The deformation will be small if the size of the impurity molecules is
not too different from the size of the host molecules. This allows us
to model the effects of the random impurities by taking $g^o\ab(\x)$
to be,
\begin{equation}\label{metr}g^o\ab(\x) = t_o^2\left[\delta\ab +
    2c\ab(\x)\right]\;,
\end{equation}
where $c\ab(\x)$ is proportional to a local stress due to local
heterogeneity in impurity concentration.

Inserting the new strain tensor \eqref{strainDisorder} inside
Eq.\eqref{Fflat} we see that $g^o\ab(\x)$ and therefore $c\ab(\x)$
couple quadratically to the order parameter $\partial_\alpha \vec
r$. Hence this type of disorder preserves the inversion symmetry of
the membrane and therefore can only model the physics of defects
symmetrically positioned with respect to the plane of the membrane.

Another, qualitatively distinct type of disorder, which breaks this
up-down symmetry is modeled by introducing a local quenched random
extrinsic curvature $\f(\x)$ into the bending rigidity term of
Eq.\eqref{Fflatii} \cite{ML},
\begin{equation}\label{CurveEnergy}{\cal F}^B_d[\h,u\as] =\intd{1\over2}
\;\kappa(\partial^2\h-\f(\x))^{2}\;. \end{equation}
Such disorder type is generically induced by, e.g., adatoms in
graphene, or asymmetrical proteins embedded in a biological membrane.

Upon expanding the resulting full effective Hamiltonian and dropping
the unimportant terms that do not involve the membrane conformational
fields, we obtain,
\begin{eqnarray}\label{FDisorder}
{\cal F}_d[\h,u\as]&=&\intd\left[{1\over2}\;\kappa(\partial^2\h)^{2} + 
\mu(u_{\alpha\beta})^2 + {1\over2}\lambda(u_{\alpha\alpha})^2\right.\left.-\kappa\partial^2\h\cdot\f(\x)-2\mu u\ab c\ab(\x) - 
\lambda u\aas c\bbs(\x)\right]\;.
\end{eqnarray}
As expected $c\ab(\x)$ disorder acts like the external local random
stress due to defects embedded into the membrane lattice. We also
observe that the random curvature $\f(\x)$ and random stress
$c\ab(\x)$ disorders symmetry-wise are the analogs of
%
%\cite{IM,RandomBond}
%
the random field \cite{IM}, and the random $T_c$ (random bond)
\cite{RandomBond} types of disorders in random magnets, previously
extensively studied.  In retrospect therefore, the disorder operators
in Eq. \eqref{FDisorder} could have been written down immediately
simply from symmetry considerations, and noting that the randomness in
the elastic constants leads to irrelevant operators \cite{RNpra,ML}
(as long as they remain positive).

For simplicity we take $c\ab(\x)$ and $\f(\x)$ to be quenched Gaussian
random fields with zero mean and correlations, \cite{CommentDist}
\begin{eqnarray}\label{DisorderCorr}\overline{c\ab(\x) c\cd(\x^\prime)}&=&
\hat \Delta_{\lambda}(\x-\x^\prime) \delta\ab \delta\cd
+ \hat \Delta_{\mu}(\x-\x^\prime)\left (\delta\ac\delta\bd +
\delta\ad\delta\bc\right)\;,\\
\overline{f_i(\x) f_j(\x\prime)}&=&\delta_{i j}
\hat \Delta_{\kappa}(\x-\x^\prime)\;,
\end{eqnarray}
where overbars denote averages over disorder realizations, and the hat
on the correlators is purely for later notational convenience. For
uncorrelated disorder coming for example from randomly positioned
impurities, $\hat \Delta_{\lambda}(\x-\x^\prime)$, $\hat
\Delta_{\mu}(\x-\x^\prime)$ and $\hat \Delta_{\kappa}(\x-\x^\prime)$
can be taken to be short-ranged, proportional to $\delta^{(D)}
(\x-\x^\prime)$, that is understood to be cutoff at short scales by
the disorder correlation length.  However, a more general type of
disorder, that might arise from unscreened disclination charges,
grain-boundaries or possibly frozen-in tilt order will lead to
long-range correlated disorder with power-law correlations
\cite{NRpra}. We will extensively analyze both of these types of
disorders in subsequent sections.

The two point statistics of fluctuations of an observable $\phi$ in a
finite temperature heterogeneous system are determined by a combination
of thermal fluctuations and quenched disorder as $\overline{ \langle
  \phi \phi \rangle} = \overline{(\phi - \langle \phi \rangle)^2} +
\overline{ \langle \phi \rangle^2}$, respectively, where the second
term is the sample-to-sample fluctuations of the heterogeneous ground
state, while the first connected correlator,
$\overline{\langle\phi\phi \rangle}_{conn}\equiv\overline{(\phi -
  \langle \phi \rangle)^2}$ quantifies thermal fluctuations around
this nontrivial background.  Accordingly, the roughness of a membrane
with quenched disorder in the flat phase is characterized by the
correlation function of the difference between out-of-plane height
fluctuations at two different points, averaged over realizations of
disorder
\begin{equation}
\label{Rough3}
\overline{\langle\left(\h(\x)-\h(0)\right)^2\rangle}=
\overline{\langle\left(\h(\x)-\h(0)\right)^2\rangle}_{conn}+
\overline{\langle\h(\x)-\h(0)\rangle^2}\;. 
\end{equation}
In above equation the ``connected'' part measures the roughness due to
thermal membrane fluctuations about a disorder specific, thermally
averaged background configuration $\langle\h(x)\rangle$. The second
contribution quantifies static spatial correlations in membrane
roughness purely due to the presence of disorder. At large length
scales we expect (and will show) that the two parts of the correlation
functions $\delta h_c^{rms}(\x)$ and $\delta h^{rms}(\x)$ scale
independently with length $x=|\x|$, with two possibly different roughness exponents
\begin{eqnarray}\label{Roughii}
\delta h_c^{rms}(x)^2&=&\overline{\langle(\h(\x)-\h(0))^2\rangle}_{conn}\approx
A_c x^{2\zeta}\;, \\
\delta h^{rms}(x)^2 &=& \overline{(\langle\h(\x)\rangle-\langle\h(0)\rangle)^2}
\approx A \, x^{2\zeta'}\;.
\end{eqnarray}
Aside from a possible buckling transition which might spontaneously
break the up-down ($\h(\x)\rightarrow -\h(\x)$) symmetry of the
membrane (which is not likely to occur in the absence of external
forces or disorder), for a disorder-free membrane, i.e., for
$\f(\x)=c\ab(\x)=0$, the thermally averaged height vanishes,
$\langle\h(\x)\rangle=0$. We therefore expect the amplitude $A$ of the
disconnected component to vanish for a homogeneous (disorder-free)
membrane, while the amplitude $A_c$ of the connected part to vanish as
$T\rightarrow 0$.  We observe then that the flat phase roughness in
general is characterized by two independent roughness exponents
$\zeta$ and $\zeta'$.

It is convenient to define the height-height correlation functions in
Fourier space
\begin{eqnarray}\label{RoughFourier}
\overline{\langle(h_i(\ks)-\langle h_i(\ks)\rangle)(h_j(\ks')-
\langle h_j(\ks')\rangle)\rangle}&\sim& |\ks|^{\eta-4}\; (2 \pi)^D \delta({\bf k}+{\bf k}'), \\
\overline{\langle h_i(\ks)\rangle\langle h_j(\ks')\rangle}&
\sim& |\ks|^{\eta'-4}\; (2 \pi)^D \delta({\bf k}+{\bf k}'), 
\end{eqnarray}
where the anomalous exponents $\eta$ and $\eta'$ describe singular
renormalization of the bending rigidity $\kappa(\kb)$,
Eq.(\ref{kappa_eta}) and disorder correlator $\Delta_\kappa(\kb)\sim
k^{-\eta'}$, respectively (see below). They are related to the
roughness exponents according to 
\bea
\zeta = \frac{4 - D - \eta}{2}\ , \quad \zeta'= \frac{4 - D - \eta'
  }{2}. 
\label{zetazetaprime}
\eea

One can also study the roughness of the in-plane phonon
  deformations. As for the out-of plane deformations there is, quite
  generally, a thermal (connected) and a disorder parts to these
  fluctuations. The former has the same form as that for the
  homogeneous membrane in \eqref{Rough2}, though with disorder
  potentially modifying and controlling the universal anomalous
  exponents. The latter, characterizing the ground state
  zero-temperature phonon roughness, in Fourier space is given by
\begin{eqnarray}\label{uuDisconnectedFourier}
\overline{\langle u_\alpha(\ks)\rangle\langle u_\beta(-\ks)\rangle}
&\sim&\frac{4\hat{\Delta}^R_\mu(\ks)}{k^2} P^T_{\alpha\beta}({\bf k})
+ \frac{\hat{\Delta}^R_L(\ks)}{k^2}P^L_{\alpha \beta}({\bf k})
\sim \frac{1}{k^{2+\eta'_u}}\;,
\end{eqnarray}
where the anomalous exponent $\eta'_u$ describes a singular
renormalization of the stress disorder variances, with
%$\hat{\Delta}^R_{\mu,\lambda}(\kb)\sim k^{-\eta'_u}$, and
%$\hat{\Delta}_L\equiv \left[4\mu^2(\hat\Delta_\lambda+2\hat\Delta_\mu)
%  +4\mu\lambda(D\hat\Delta_\lambda+2\hat\Delta_\mu)
%  +\lambda^2D(D\hat\Delta_\lambda+2\hat\Delta_\mu)\right]/(2\mu+\lambda)^2$.
%  This can be written shorter as 
$\hat{\Delta}_L\equiv \left[4\mu^2(\hat\Delta_\lambda+2\hat\Delta_\mu)
  +(4\mu+D \lambda) \lambda (D\hat\Delta_\lambda+2\hat\Delta_\mu)
\right]/(2\mu+\lambda)^2$.  In \eqref{uuDisconnectedFourier} the
superscript (and superscript) $R$ indicates that all couplings must be
replaced by their renormalized ($k$-dependent) values - see below for
an equivalent explicit expression.  In coordinate space in-plane
phonon roughness is given by generalization of \eqref{Rough2}
\bea && 
\overline{ \langle(u_\alpha(\x)-u_\alpha(0))^2\rangle_{\text conn}}
 \sim x^{2 \zeta_u}\;, \quad \zeta_u = \frac{2-D+\eta_u}{2}\;, \\
&& \overline{ \langle u_\alpha(\x)-u_\alpha(0) \rangle \langle
  u_\alpha(\x)-u_\alpha(0) \rangle} \sim x^{2 \zeta'_u}\;, \quad
\zeta'_u = \frac{2-D+\eta'_u}{2},
\eea 
where $\eta'_u=4-D - 2 \eta'$. Anticipating the following sections, it
is useful to give the phonon propagator in terms of the independent
renormalized couplings $\tilde \mu(\ks)$, $\tilde b(\ks)$, $\tilde
\Delta_\mu(\ks)$, $\tilde \Delta_b(\ks)$ since these are the ones
which obey, within SCSA, multiplicative renormalization (see
\eqref{couplings2}-\eqref{couplings2d}).  Note that the tilde label
indicates exactly the same thing as the $R$ subscript, i.e., it
denotes renormalized couplings. Let us start by rewriting
\eqref{Rough2} by expressing $\tilde \lambda(\ks)$ as a function of
$\tilde b(\ks)$ and $\tilde \mu(\ks)$. In presence of disorder this
gives the following formula for the disorder average of the connected
(i.e., the thermal part) of the fluctuations
\begin{eqnarray}\label{Rough23} 
\overline{ \langle u_\alpha(\ks) u_\beta(-\ks)\rangle}_{\rm conn} \simeq \frac{T P^T_{\alpha \beta}({\bf k})}{ \tilde \mu({\bf k}) k^2} 
+ 
\left( \frac{D}{\tilde \mu({\bf k})} - \frac{\tilde b({\bf k})}{\tilde \mu({\bf k})^2}  \right) \frac{T P^L_{\alpha \beta}({\bf k})}{2 (D-1) k^2}\;.
\end{eqnarray}
To obtain the corresponding formula for the disorder part of the
fluctuations, i.e., the off-diagonal component of the replica phonon
propagator, one can simply replace each term in this expression by its
corresponding replica matrix, as parameterized in \eqref{MNcoeff}, and
then take the off-diagonal part, while performing the needed replica
matrix multiplications and inversions, according to the rules
described in \eqref{RNew}-\eqref{ProdInverse}. This leads to
\begin{eqnarray}\label{Rough24} 
\overline{ \langle u_\alpha(\ks) \rangle \langle u_\beta(-\ks)\rangle} \simeq \frac{\tilde \Delta_\mu({\bf k}) P^T_{\alpha \beta}({\bf k})}{ \tilde \mu({\bf k})^2 k^2} 
+ 
\left((D \tilde \mu({\bf k}) - 2 \tilde b({\bf k})) \tilde \Delta_\mu({\bf k}) + \tilde \mu({\bf k}) \tilde \Delta_b({\bf k})
 \right) \frac{ P^L_{\alpha \beta}({\bf k})}{2 (D-1) \tilde \mu({\bf k})^3 k^2}\;,
\end{eqnarray}
which, one can check is precisely equivalent to the above expression
\eqref{uuDisconnectedFourier}.  Our aim in subsequent sections is to
compute the universal thermal and disorder roughness exponents $\zeta$
and $\zeta'$, and related exponents, characterizing the power-law
rough critical phases for various types of disorder in the presence of
thermal fluctuations.

\section{SCSA of homogeneous membrane in the flat phase}
\label{PureMembrane}
\subsection{Background}

We now return to the model of a homogeneous polymerized membrane and
study its flat phase using the method of the Self-Consistent Screening
Approximation (SCSA), first introduced by Bray \cite{Bray} in the
context of the critical $O(N)$ model.  In the following sections we
will extend the methods developed here to treat the crumpling
transition of phantom membranes and to study disordered membranes. The
idea behind SCSA is that instead of performing a perturbative
expansion (in nonlinear couplings, $\epsilon$ or $1/d_c$) one can
study a particularly-truncated closed set of integral Dyson equations
satisfied by the correlation functions. These are built by elevating
the $1/d_c$-expansion into a set of self-consistent
equations. Although generally quite intractable, as we will show,
these integral equations can be solved analytically in closed form in
the long wavelength limit using the fact that theory is critical and
correlation obey simple isotropic scaling forms.

The first pioneering attempt at this problem was by Nelson and Peliti
(NP) \cite{NP} who introduced a simplified truncation of Dyson
equations involving a renormalized bending rigidity $\kappa_R(\ks)$.
Their self-consistent treatment in $D=2$ predicted $\kappa_R(\ks) \sim
k^{-1}$, i.e., $\eta=1$.  Perturbatively, their approximation amounts
to a partial resummation of a class of one-loop Feynman graphs for the
renormalized height correlator. Their study crucially neglected the
nontrivial renormalization of the Lam\'e coefficients, $\mu_R(\ks)$
and $\lambda_R(\ks)$, which arises because the in-plane stresses can
also be relaxed by 'soft' out-of-plane displacements. As a result,
curvature fluctuations soften these in-plane elastic constants,
thereby screening the phonon-mediated nonlinearities. This was first
appreciated by Toner (in an unpublished work) and explicitly
calculated by Aronowitz-Lubensky (AL) who showed through a detailed
one-loop RG calculation that at long wavelength the renormalized
Lam\'e coefficients acquire a non-trivial universal wavevector
dependence $\mu_R(\ks) \sim k^{\eta_u}$ and $\lambda_R(\ks) \sim
k^{\eta_u}$, and obtained the exponents $\eta_u$ and $\eta$ to
accuracy $O(\epsilon=4-D)$ in a dimensional expansion. This was
complemented by a $1/d_c$-expansion calculation by David, et
al. \cite{GDLP}, predicting $\eta = 2/d_c + O(1/d_c^2)$ in $D=2$.

Significant discrepancies between these methods and numerical
simulations, associated with the physical dimensionality $D=2$, $d=3$
being far away from the expansion points $D=4$ and $d_c=+\infty$,
respectively, motivated us to introduce a more accurate,
self-consistent method, the SCSA, which takes into account the physics
of screening neglected in NP.  Using the SCSA we derived and solved
analytically \cite{LRprl} {\it two independent} integral equations for
the correlator of the $d_c$-component out-of-plane fluctuations and
for the renormalized elastic interaction, which determine the
renormalized moduli $\kappa_R(\ks)$, $\mu_R(\ks)$, and
$\lambda_R(\ks)$.

We note three important properties of the SCSA applied to the membrane
problem. First, the SCSA is {\em exact} by construction for large
codimension $d_c=d-D$ to first order in $1/d_c$ and arbitrary
$D$. Second, in the opposite extreme limit of $d_c=0$ it gives
$\eta=(4-D)/2$ which is also the {\em exact} result since clearly
$\eta_u=0$ for $d=D$ and the two exponents are related according to
\eqref{exponentrelation}.  This second property is special to the
problem of a polymerized membrane, contrasting the $O(N)$ model where
the SCSA is not exact for $N=0$. Thus, the SCSA approximation as a
function of $d_c$ is tightly constrained by these two extremes.
Finally, we find that our SCSA results are also {\em exact} to first
order in $\epsilon=4-D$, for arbitrary $d_c$, a feature again special
to the membrane problem, arising from the Ward identities associated
with the rotational invariance in the embedding space. Hence, since
within a single analysis the SCSA reproduces all previously known
exact limits \cite{AL,GDLP}, we expect it to be considerably more
accurate for the physical membrane predictions. It includes the
physics of the lower- as well as of the upper-critical dimension, and
summarizes all the information contained in a one-loop calculation.
These properties are at the heart of its successful agreement with
numerics and experiments, as discussed in the Introduction.
%It is in fact also very similar to the improved self-consistent method
%of Kawasaki \cite{Kawasaki}
%for the critical dynamics of the binary fluid mixture, which is also
%exact to order $\epsilon$ also because of Ward identities associated with the
%Galilean invariance of the fluid. 
%It disagree however
%by a tiny amount to order $\epsilon^2$, and is overall remarkably accurate. 
%\cite{CommentKawasaki}. 
%The results of 
%two-loop calculations that are in progress and indicate
%indicate that the SCSA is not exact, with the deviations appearing at 
%two-loop order \cite{TwoLoop}. 
%Here the approximation gives predictions which compare well with 
%numerical simulations and is exact in various extreme limits which 
%makes it quantitatively trustworthy. 

\subsection{Effective model for the out-of-plane height fluctuations}

We now use SCSA to study thermal fluctuations of a homogeneous membrane
described by the coarse-grained Hamiltonian ${\cal F}[\h,u\as]$, given
by Eqs.(\ref{Fflatii}) and \eqref{st2} in terms of the height $\h(\x)$
and phonon $u\as(\x)$ fields. Because in-plane phonon modes $u\as(\x)$
appear only linearly and quadratically, it is convenient to integrate
them out exactly \cite{NP,RNpra} and obtain an effective Hamiltonian
${\cal F}_{eff}[\h] = {\cal F}[\h,0] + \delta {\cal F}[\h]$ for the
height fluctuations $\h(\x)$.  From Eqs.(\ref{Fflatii}) and
\eqref{st2} the coupling of the phonon field to the height field can
be written as
\be 
{\cal F}_{u-h} = -
\int d^D x \, u_\gamma A_{\alpha \beta
  \gamma}({\bf \partial}) \partial_\alpha {\vec h}
\cdot \partial_\beta {\vec h}\;, 
\ee
where $A_{\alpha \beta \gamma}({\bf k}) = \frac{\lambda}{2}
\delta_{\alpha \beta} i k_\gamma + \frac{\mu}{2} (i k_\alpha
\delta_{\beta \gamma} + i k_\beta \delta_{\alpha
  \gamma})$. Integrating over the in-plane phonon field $u_\gamma$
using the harmonic Hamiltonian
\bea {\cal F}_u[u_\alpha] \equiv {\cal F}[0,u_\alpha]
=\frac{1}{2}
\int_k u_\alpha(-{\bf k})k^2\left[\mu P^T_{\alpha\beta}({\bf k}) 
+ (2\mu+\lambda)P^L_{\alpha\beta}({\bf k})\right]u_{\beta}({\bf k}), \label{freephonon} 
\eea
with $P^T_{\alpha\beta}({\bf k}) = \delta_{\alpha\beta} - k_\alpha
k_\beta/k^2$ and $P^L_{\alpha\beta}({\bf k}) = k_\alpha k_\beta/k^2$ the
transverse and longitudinal projection operators, leads to
\bea \delta {\cal F} = - \frac{1}{2} \int_k A_{\alpha \beta
  \gamma}({\bf k}) A_{\alpha' \beta' \gamma'}(-{\bf k}) \langle
u_\gamma({\bf k}) u_{\gamma'}(-{\bf k}) \rangle_0 \times
( \partial_\alpha {\vec h} \cdot \partial_\beta {\vec h} )({\bf k})
( \partial_{\alpha'} {\vec h} \cdot \partial_{\beta' }{\vec h} )(-{\bf
  k}), 
\eea
where we defined $\int_k = \int \frac{d^D k}{(2 \pi)^D}$ and $\langle
.. \rangle_0$ denotes the average with respect to
\eqref{freephonon}. Hence we obtain the following height-only
effective Hamiltonian,
\begin{equation}\label{FflatEff}
{\cal F}_{eff}[\h] ={1\over2}\kappa\intd(\partial^2\h)^{2} + 
{1\over4 d_c}\intd \left[\mu \left(P^T\ad\pt\as\h\cdot\pt\bs\h\right) \left(P^T\bc\pt\cs\h\cdot\pt\ds\h\right)
+{\mu\lambda\over 2\mu+\lambda}\left(P^T\ab\pt\as\h\cdot\pt\bs\h\right)^2
\right], 
\end{equation}
where here and below the summation over the repeated indices is
implied and $P^T\ab$ is the transverse projection operator
$P^T\ab=\delta\ab-\pt\as\pt\bs/\pt^2$.  For later convenience in
implementing the SCSA we have extracted a factor $1/d_c$ from the
interaction by redefining the Lam\'e coefficients, $\mu\rightarrow
\mu/d_c$, $\lambda\rightarrow \lambda/d_c$. We note that for physical
membranes, with $d_c=1$ this rescaling leaves the elastic moduli
unchanged.  The effective Hamiltonian above describes the physics of a
polymerized membrane purely in terms of the height fluctuations field
$\h(\x)$. Note that there is an implicit short length scale cutoff $a$
and correspondingly an ultraviolet momentum cutoff $\Lambda \sim 2
\pi/a$ (e.g., in graphene, set by the lattice constant in real space
and the size of the first Brillouin zone in reciprocal space), that we
will always consider $a$ shorter than any other length scale in the
problem.

We note that for $D=2$, $P^T(\partial)_{\alpha \beta} =(\hat z
\times \partial)_\alpha (\hat z \times \partial)_\beta$ and ${\cal
  F}_{eff}[\h]$ simplifies considerably, reducing to
\begin{equation}\label{FflatEff2D}
{\cal F}_{eff}[\h] ={1\over2}\kappa\int d^2 x (\partial^2\h)^{2} + 
{K_0 \over8 d_c}\int d^2 x
\left(P^T\ab\pt\as\h\cdot\pt\bs\h\right)^2\;, 
\end{equation}
where $K_0= {4 \mu(\mu + \lambda) \over (2\mu+\lambda)}$ is the
Young's modulus, the only combination of elastic coefficients that
characterizes the strength of the out-of-plane non-linearities.
However for the general theoretical analysis, we will keep $D$
arbitrary below, working with ${\cal F}_{eff}[\h]$, Eq.\eqref{FflatEff}.

The resulting effective nonlocal interaction has an interesting
geometrical interpretation. It can be rewritten as the long-range
interaction between local Gaussian curvatures mediated by the in-plane
phonons, an interpretation that is physically appealing \cite{NP}. In
the absence of surface tension or external forces acting on the
membrane the effective Hamiltonian above has the form of the
$O(d_c)\times O(D)$ invariant $\phi^4$-theory, exactly at criticality
(``mass'' terms exactly zero). This massless property of the theory is
preserved by the renormalization (coarse-graining) because the
effective Hamiltonian describes the theory of the interacting
Goldstone modes $\pt\as\h$ coming from the spontaneously broken $O(d)$
rotational symmetry in the embedding space.  The ``criticality'' is
strictly imposed by this nonlinearly realized rotational invariance.

The renormalization of the two transverse tensor parts of the quartic
interaction in Eq.\eqref{FflatEff} determines the two independent
renormalized Lam\'e coefficients $\mu_R(\ks)$, $\lambda_R(\ks)$, that
characterize fluctuating membrane's anomalous flat phase.  However, as
we will see below, while the amplitudes are independently
renormalized, the rotational invariance imposes the same power-law
wavevector dependence in these renormalized elastic moduli.  It is
convenient to work in Fourier space where $\F_{eff}[\h(\ks)]$ is
expressed in terms of the Fourier transform of the height fields,
$\h(\ks)=\intd\;\h(\x)\;{\rm e}^{-i\ks\cdot\x}$, and is given by
\begin{equation}\label{Feff}\F_{eff}[\h(\ks)]={\kappa \over 2} \int_k\;k^4 \mid\h(\ks)\mid^2 +
{1\over4 d_c}\int_{k_1, k_2, k_3} R_{\alpha \beta, \gamma \delta}(\q)
k_{1\alpha} k_{2\beta} k_{3\gamma} k_{4\delta}\; \h(\ks_1)\cdot\h(\ks_2)\;
\h(\ks_3)\cdot\h(\ks_4)\;, \end{equation}
with $\q=\ks_1+\ks_2$ and $\ks_1+\ks_2+\ks_3+\ks_4={\bf 0}$.  The
four-point coupling is a fourth-order tensor
$R_{\alpha\beta,\gamma\delta}(\q)$ that is transverse to $\q$, and,
from \eqref{FflatEff}, reads
\begin{equation}\label{RtensorAB}R_{\alpha\beta,\gamma\delta}(\q)=
\mu A_{\alpha\beta, \gamma\delta}(\q)+
{\mu\lambda\over 2\mu+\lambda}B_{\alpha\beta, \gamma\delta}(\q)\;, \end{equation}
where,
\begin{eqnarray}\label{ABtensors} A_{\alpha\beta, \gamma\delta}(\q)&=&
{1\over2}\left(P^T\ac(\q) P^T\bd(\q)+P^T\ad(\q)
P^T\bc(\q)\right)\;, \\
B_{\alpha\beta,\gamma\delta}(\q)&=&
P^T\ab(\q)P^T\cd(\q)\;,
\end{eqnarray}
with the longitudinal part having been eliminated through the phonon
integration. Note that because a uniform strain tensor has $D(D+1)/2$
independent components, the integration over the ${\bf q}=0$ in-plane
phonons eliminates all corresponding non-linearities, hence the mode
${\bf q}=0$ is {\it excluded} from the wavevector summation in the
quartic nonlinearity \eqref{Feff} \cite{NP,Jerusalem}.

A product of two transverse operators, $P^T\ab(\q)P^T\cd(\q)$,
together with two inequivalent tensors obtained by permutations of
indices, form a complete basis for a space of transverse fourth rank
tensors. The membrane's effective quartic interaction tensors,
$A_{\alpha\beta, \gamma\delta}(\q)$ and
$B_{\alpha\beta,\gamma\delta}(\q)$ thus form a two-dimensional,
reducible representation of this two-dimensional vector space.
Because of the reducibility of the representation the two coupling
coefficients of the two transverse operators in Eq.\eqref{FflatEff} do
not renormalize independently, with each feeding into the
renormalization of the other. This occurs because the corresponding
tensors $A_{\alpha\beta, \gamma\delta}(\q)$ and $B_{\alpha\beta,
  \gamma\delta}(\q)$ are not mutually orthogonal. Indeed, using the
following notation for the product of two tensors, $(A \star
B)_{\alpha\beta, \gamma\delta}= A_{\alpha\beta, \rho \sigma} B_{\rho
  \sigma, \gamma\delta}$ we see that $A\star A=A$, $A\star B=B \star
A=B$ and $B \star B=(D-1) B$.  Instead of working in this
two-dimensional space of coupling constants it is more convenient to
transform to an independently renormalizable set, i.e., an irreducible
representation.  This can be accomplished by introducing two new
tensors $M_{\alpha\beta, \gamma\delta}(\q)$ and $N_{\alpha\beta,
  \gamma\delta}(\q)$ that are mutually orthogonal projectors under the
tensor multiplication (i.e., $M \star M=M$, $N \star N=N$ and $M \star
N = N \star M = 0$) and are a linear combination of $A_{\alpha\beta,
  \gamma\delta}(\q)$ and $B_{\alpha\beta, \gamma\delta}(\q)$,
\begin{eqnarray}\label{MNtensors}N_{\alpha\beta,\gamma\delta}(\q)
&=&{1\over D-1}P^T_{\alpha\beta}(\q)P^T_{\gamma\delta}(\q)\;, \\
M_{\alpha\beta,\gamma\delta}(\q)&=&{1\over2}\left(P^T_{\alpha\gamma}(\q) 
P^T_{\beta\delta}(\q)+P^T_{\alpha\delta}(\q)P^T_{\beta\gamma}(\q)\right)
- N_{\alpha\beta,\gamma\delta}(\q)\;. 
\end{eqnarray}
The corresponding new elastic coupling constants are also linearly
related to the old ones,
\begin{eqnarray}\label{newConstants}\mu&=& \mu\;,\\
b&=& {\mu(2\mu+D\lambda)\over 2\mu+\lambda}\;. 
\end{eqnarray}
The elastic constant $\mu$ is the usual shear modulus and $b$ the
D-dimensional generalization of Young's modulus, proportional to the
bulk modulus of a D-dimensional solid.  It is physically reasonable
that the bulk and shear moduli renormalize independently, as we find
here mathematically.

In terms of these new tensors and coupling constants the vertex 
$R_{\alpha\beta,\gamma\delta}(\q)$ of Eq. \eqref{RtensorAB} becomes
\begin{equation}\label{RtensorMN}R_{\alpha\beta,\gamma\delta}(\q)=
\mu M_{\alpha\beta,\gamma\delta}(\q)+ b N_{\alpha\beta,\gamma\delta}(\q)\;. \end{equation}
In $D=2$, since $P^T_{\alpha \beta}({\bf q}) = q^T_{\alpha}
q^T_{\beta}$, where $ q^T_{\alpha} = \epsilon_{\alpha \gamma}
q_\gamma$, one sees that the tensor $M_{\alpha \beta, \gamma
  \delta}({\bf q})$ identically vanishes.  This is consistent with the
above observation that a single elastic constant, the Young modulus
$K_0 = 2 b$, characterizes the elastic non-linearities of a
two-dimensional sheet.

\subsection{Derivation of the SCSA equations for a homogeneous membrane}

\begin{figure}[htbp]
        \centering
        \includegraphics[width=0.7\textwidth,scale=1]{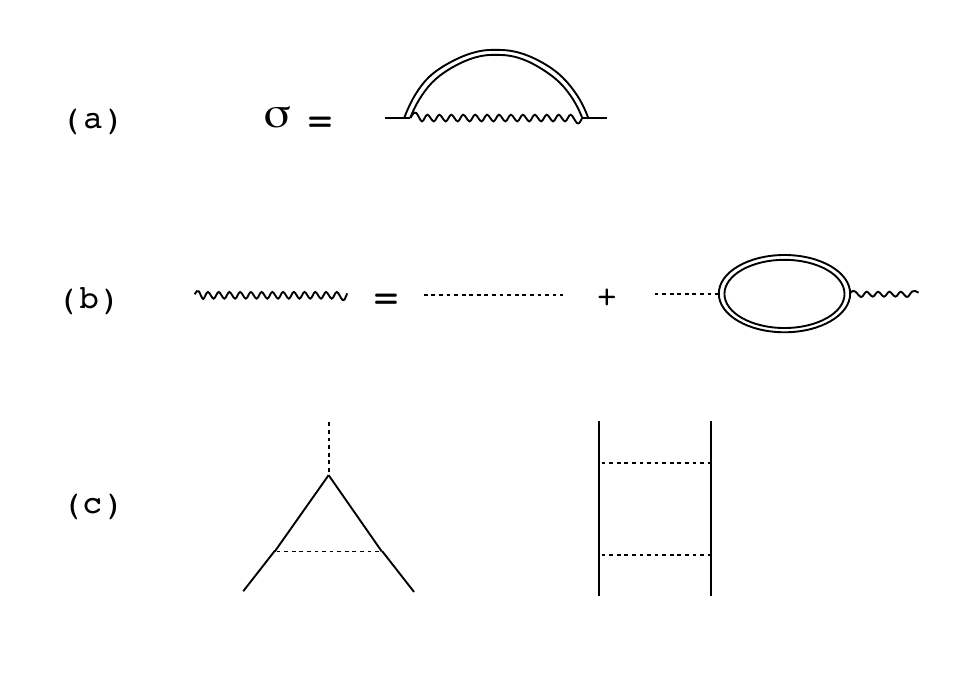}
        \caption{\label{FeynmanLoopsSCSAsr} Graphical representation
          of the SCSA: (a) self-energy correction, \eqref{SCSAeqn} to
          the out-plane-fluctuations propagator, that determines the
          renormalized bending rigidity, enhanced by fluctuations, (b)
          screened quartic elastic nonlinearity, \eqref{SCSAeqn{b}},
          that determines the renormalized in-plane elastic moduli,
          softened by fluctuations, (c) UV finite vertex and
          polarization diagram, neglected in SCSA.}
\end{figure}

The height fluctuations in the flat phase are described by the
Hamiltonian \eqref{Feff}. It is the sum of the quadratic bending
energy part and the quartic elastic nonlinearities, encoding
membrane's in-plane elasticity.  In the absence of non-linearities
(setting $\mu=\lambda=0$) the fluctuations of the height are Gaussian,
with the correlator in Fourier space given by
\bea
\langle h_i(\ks)h_j(\ks')\rangle_0 
= \delta_{ij} G_0({\bf k}) (2 \pi)^D \delta({\bf
  k}+{\bf k}')\ , \quad G_0({\bf k})= \frac{T}{\kappa k^4}.
\eea
This leads to the harmonic roughness exponent $\zeta=(2-D)/2$, that
implies the absence of flat phase order in $D=2$, consistent with
conventional wisdom based on the Hohenberg-Mermin-Wagner
theorems\cite{Hohenberg,MerminWagner,Coleman}.  To deal with the
quartic nonlinearity, schematically $R (\vec h \cdot \vec h)^2$, we
study, as usual, the perturbation theory in powers of the coupling
constant, here the tensor $R_{\alpha\beta,\gamma\delta}(\q)$, around
the quadratic theory. Each term in the expansion is represented by a
Feynman diagram (see Fig. \ref{FeynmanLoopsSCSAsr}), where the bare
propagator $G_0({\bf k})$ is represented by a solid line, and the
quartic nonlinearity $(\vec h \cdot \vec h)^2$ with interaction
vertex, $R_{\alpha\beta,\gamma\delta}(\q)$ by a dotted line joining
the two pairs of fields $\vec h \cdot \vec h$ on each side.

We aim to calculate the correlator of the height field $\h$
fluctuations
\begin{eqnarray}\label{hCorrelator}
\langle h_i(\ks) h_j(-\ks)\rangle = \delta_{i j} G(\ks)\; (2 \pi)^D \delta({\bf k}+{\bf k}') 
\quad , \quad G({\bf k}) = {T \over\kappa_R(\ks) k^4} = {T\over\kappa k^4+\sigma_c(\ks)}\;, 
\label{propren}
\end{eqnarray}
where $G({\bf k})$ is the propagator dressed by thermal fluctuations
in the presence of non-linearities. For convenience we have introduced
the self-energy $\sigma_c(\ks)$ which is the correction to the bare
propagator.

In addition to the self-energy there are also corrections to the
quartic non-linearities. To construct the SCSA equations, which will
determine both the propagator and the renormalized interaction vertex
$\tilde{R}_{\alpha\beta,\gamma\delta}(\q)$, it is useful to recall the
exact analysis in the limit of large $d_c$ (a
$1/d_c$-expansion). Since for each internal loop of unconstrained
$d_c$-component height fields, there is a factor $d_c$, the dominant
set of diagrams which correct the four point vertex
$R_{\alpha\beta,\gamma\delta}(\q)$ for $d_c=+\infty$ form a geometric
series of so-called polarization (or RPA) bubbles, which can be
resummed exactly. These are depicted in Fig. \ref{FeynmanLoopsSCSAsr}
and represent screening of the elastic interactions by out-of-plane
fluctuations.  This is equivalent to introducing a
Hubbard-Stratonovich field $\chi$ to decouple the quartic interaction
and formally integrate over ${\vec h}$ leading
\cite{ColemanLargeN,ZinnJustin} to a ${\rm Tr} \ln(G_0^{-1}+\chi)$
term in the action.

The self-energy is then determined exactly to first order in $1/d_c$
by a single sunset diagram involving the screened interaction
$\tilde{R}_{\alpha\beta,\gamma\delta}(\q)$, illustrated in
Fig. \ref{FeynmanLoopsSCSAsr}.  In the figure the double solid line
denotes the renormalized propagator, $G(\ks)$, the dotted line the
bare interaction vertex $R_{\alpha\beta,\gamma\delta}(\q)$ and the
wiggly line represents the ``screened'' interaction
$\tilde{R}_{\alpha\beta,\gamma\delta}(\q)$, dressed by the vacuum
polarization bubbles.  The resulting expression would obviously lead
to exponents and other physical quantities that diverge as
$d_c\rightarrow 0$ and therefore are not expected to be accurate at
the physical value of $d_c=1$. The SCSA corrects this deficiency by
self-consistently replacing all the bare $\h$ propagators by the
corresponding renormalized propagator, \eqref{propren}.
%%FIG\SCSAdiagrams{Graphical representation of the SCSA: 
%(a) Self-energy graph that
%determines renormalized bending rigidity $\kappa_R(\ks)$ and the corresponding
%anomalous and roughness exponents $\eta$ and $\zeta$, respectively. (b) Graphs
%leading to renormalization of the elastic 4-point interaction and the
%corresponding exponent $\eta_u$. (c) Examples of some of the higher order
%vertex and box diagrams that are UV finite. Finiteness of these graphs is the
%reason that the SCSA to first order in $\epsilon$ is exact to all orders in
%$1/d_c$.\eqref{SCSAdiagrams}.
The resulting closed set of two SCSA integral equations, corresponding
to Fig. \ref{FeynmanLoopsSCSAsr}, is given by
\begin{eqnarray}\label{SCSAeqn}\kappa_R(\ks) k^4 &=& \kappa k^4 + {2\over d_c}
k\as k\bs k\cs k\ds \int_q\tilde{R}_{\alpha \beta, \gamma \delta}(\q) G(\ks-\q)
\;, \\
\tilde{R}_{\alpha\beta,\gamma\delta}(\q)&=& R_{\alpha\beta,\gamma\delta}(\q)-
R_{\alpha\beta,\gamma_1\gamma_2}(\q)\Pi_{\gamma_1\gamma_2,\delta_1\delta_2}(\q)
\tilde{R}_{\delta_1\delta_2,\gamma\delta}(\q)\;, \label{SCSAeqn{b}}
\end{eqnarray}
where
\begin{equation}\label{Polarization}
\Pi_{\alpha \beta, \gamma \delta}(\q)
= \frac{1}{T}  \int_p\; p\as p\bs p\cs p\ds G(\pb) G(\q-\pb) 
\end{equation}
is the vacuum polarization bubble, that encodes screening of the
long-scale in-plane elasticy by short-scale out-of-plane
fluctuations. We note that the Dyson equation for the self-energy
contains an additional "tadpole" diagram contribution (which is
usually dominant, i.e., $O(1)$ in the $1/d_c$-expansion, and
corresponds to a shift in the critical point). Here this term is
strictly absent, since it involves zero momentum transfer for which
$R({\bf q}=0)=0$ (since the zero mode is excluded, as remarked
above). This is a reflection of the general principle that precludes
generation of a mass for a Goldstone mode, that in this case
corresponds to absence of surface tension for free boundary conditions
in the flat phase.

The resulting SCSA equations therefore amount to a partial resummation
of the $1/d_c$-expansion based on the result correct only to first
order in $1/d_c$. Although a priori this approximation is
uncontrolled, in retrospect it is expected to be quite accurate given
the properties of the SCSA described in the Introduction and in the
beginning of this section.

Let us now simplify the above equations using the symmetries of the
theory.  It can be verified that the dressed interaction has the same
tensorial form as the bare one
\begin{equation}\label{Rrenorm}\tilde{R}_{\alpha\beta,\gamma\delta}(\q)=\mu_R(\q) 
M_{\alpha\beta,\gamma\delta}(\q) + 
b_R(\q) N_{\alpha\beta,\gamma\delta}(\q)\;, \end{equation}
but with renormalized shear $\mu_R(\q)$ and bulk-like moduli
$b_R(\q)$.  To determine them we now study the polarization bubble.
From its definition the tensor $\Pi_{\alpha \beta, \gamma \delta}(q)$
is the sum of a symmetric component $\Pi_{\alpha \beta, \gamma
  \delta}^{sym}(\q)= T I(\q)S_{\alpha \beta, \gamma \delta}$
proportional to the fully symmetric tensor $S_{\alpha \beta, \gamma
  \delta}= \delta_{\alpha \beta} \delta_{\gamma \delta} +
\delta_{\alpha \gamma} \delta_{\beta \delta}+ \delta_{\alpha \delta}
\delta_{\beta \gamma}$ with 
\bea I({\bf q}) = \frac{1}{(D-1)(D+1) T^2}
\int_p\; (p_\alpha P_{\alpha \beta}^T({\bf q}) p_{\beta})^2 G(\pb)
G(\q-\pb),
\label{Iq} 
\eea
and a longitudinal component where some of tensor indices are carried
by $q\as$.  Clearly, from Eqs.\eqref{RtensorAB} and \eqref{ABtensors},
because of the transverse structure of the vertex $R_{\alpha \beta,
  \gamma \delta}(\q)$, only the symmetric component $\Pi_{\alpha
  \beta, \gamma \delta}^{sym}(\q)$ contributes.  Equation\;
\eqref{SCSAeqn{b}} can them be easily solved since in the subspace
spanned by $M_{\alpha\beta,\gamma\delta}(\q)$ and
$N_{\alpha\beta,\gamma\delta}(\q)$ we have
\begin{eqnarray}\label{NMproject}
N_{\alpha\beta,\gamma_1\gamma_2}(\q)\ S_{\gamma_1\gamma_2,\gamma_3\gamma_4}
N_{\gamma_3\gamma_4,\gamma\delta}(\q)
&=&(D+1)N_{\alpha\beta,\gamma\delta}(\q)\;,\\
M_{\alpha\beta,\gamma_1\gamma_2}(\q)\ S_{\gamma_1\gamma_2,\gamma_3\gamma_4}
M_{\gamma_3\gamma_4,\gamma\delta}(\q)
&=& 2M_{\alpha\beta,\gamma\delta}(\q)\;,\\
M_{\alpha\beta,\gamma_1\gamma_2}(\q)S_{\gamma_1\gamma_2,\gamma_3\gamma_4}
N_{\gamma_3\gamma_4,\gamma\delta}(\q)
&=&N_{\alpha\beta,\gamma_1\gamma_2}(\q)\ S_{\gamma_1\gamma_2,\gamma_3\gamma_4}
M_{\gamma_3\gamma_4,\gamma\delta}(\q)=0\;.
\end{eqnarray}

Using the above orthogonal projection relations inside Eq. \eqref{SCSAeqn{b}}, and
the form \eqref{Rrenorm} for $\tilde{R}_{\alpha\beta,\gamma\delta}(\q)$
we obtain the equations for the renormalized shear and bulk-like moduli,
\begin{eqnarray}\label{renormModuli} \mu_R(\q)&=&{\mu\over 1+2 T I(\q)\mu}
\;,\\
b_R(\q)&=&{b\over 1+ (D+1) T I(\q)b}\;. 
\end{eqnarray}
%
%The membrane model is therefore renormalizable at least to one-loop order.
Using Eqs. \eqref{Rrenorm} inside Eq. \eqref{SCSAeqn} we obtain,
\begin{equation}\label{SimpleSigma} \kappa_R(\ks)= \kappa + {2 T\over d_c}\int_q 
{b_R(\q) + (D-2) \mu_R(\q)\over D-1} 
 \frac{\left(\hat k\as P\ab^T(\q) \hat k\bs\right)^2}{\kappa_R(\ks-\q) |\ks-\q|^4}\;. \end{equation}
 We note again that for $D=2$ the renormalized bending rigidity is determined
 by the Young modulus $K_0({\bf q})= 2 b_R(\q)$ alone, consistent with \eqref{FflatEff2D}.

 The closed set of Eqs. \eqref{renormModuli}, \eqref{SimpleSigma},
 \eqref{Iq}, \eqref{propren} constitute the SCSA equations for an
 homogeneous membrane of internal dimension $D$, in embedding
 dimension $d$. They determine the out-of-plane propagator $G({\bf
   k})$, i.e. the wave-vector dependent renormalized bending rigidity
 $\kappa_R({\bf k})$ and the renormalized elastic moduli $\mu_R(\q)$
 and
\begin{equation}
b_R(\q)={\mu_R(\q)(2\mu_R(\q)+D\lambda_R(\q))\over
  2\mu_R(\q)+\lambda_R(\q)}.
\end{equation}

\subsection{Analysis of the SCSA equations for a homogeneous membrane}
\label{analysisSCSAhomo}
The SCSA equations above predict the full non-trivial momentum
dependence of the elastic moduli of the homogeneous fluctuating
membrane. There are two main regimes corresponding to large and small
$q$, respectively.

For sufficiently large $q \gg q_{\rm nl}$, where $q_{\rm nl}$ is
determined below, we expect that perturbation theory converges and
that $G({\bf k}) \approx G_0({\bf k})$. In this regime, from
Eq. \eqref{Iq}, we see that $I({\bf q}) \sim 1/(\kappa^2 q^{4-D})$ for
$D<4$, hence from Eqs. \eqref{renormModuli}, \eqref{SimpleSigma} the
thermal fluctuation corrections to the elastic moduli are small and
are given by
\bea
\label{q_nlT}
&&\frac{b_R(\q)- b}{b} \sim - \frac{b T}{\kappa^2 q^{4-D}},\\
&&\frac{\kappa_R(\kb) -\kappa}{\kappa} \sim \frac{b T}{d_c \kappa^2
  k^{4-D}}.
\eea 

This high $q$ regime extends down to $q_{\rm nl}$ defined by the above
dimensionless corrections being small compared to unity, which is given by 
\bea
q_{\rm nl} \sim\left (\frac{b T}{\kappa^2}\right)^{1/(4-D)}, 
\eea 
the analog of Ginzburg criterion in critical
phenomena\cite{ChaikinLubensky}. We note that $q_{\rm nl} \to 0$ for
small temperature and/or large $\kappa$, with the nontrivial
strong-fluctuation regime below $q_{\rm nl}$ squeezed out in this
stiff regime limit.  For $q \ll q_{\rm nl}$ we expect that the effect
of non-linearities become important and the system crosses over to the
so-called universal anomalous elasticity regime which we now study.

We now study the regime of small wavevector $q$.  Anticipating that
the solution will describe a critical fixed point, with power-law
correlations, we search a solution for the height-height correlation
function, equivalently the propagator $G(\ks)$, as a power of $k$,
\begin{equation}\label{Gscale}G(\ks)=\frac{T}{\kappa_R(\ks) k^4} \simeq T Z_\kappa^{-1} k^{-4+\eta}\;
\quad , \quad \kappa_R(\ks) \simeq Z_\kappa k^{-\eta}, 
\end{equation}
where $Z_\kappa$ is an amplitude, which in the present case will be
non-universal, and $\eta$ determines the universal roughness exponent
through Eq.\eqref{ZetaIdentity}. Substituting this ansatz into
Eq.\eqref{Iq} we find that the vacuum polarization integral diverges
as
\begin{equation}\label{Idiverge}
I(\q) \simeq \frac{1}{(D^2-1) Z_\kappa^2} \int_p
{(p_\alpha P_{\alpha \beta}^T({\bf q}) p_{\beta})^2
\over|\pb|^{4-\eta}|\pb+\q|^{4-\eta}}\; =  \frac{\Pi(\eta,D)}{(D^2-1)
Z_\kappa^{2}} q^{-\eta_u}\ ,
\end{equation}
with
\begin{equation}
\label{WardIdentity}
\eta_u=4-D-2 \eta
\end{equation}
the anomalous exponent of the in-plane phonons modes, and in the
second equality in \eqref{Idiverge} we have set the short scale cutoff
$a \to 0$ since the integral is ultraviolet convergent.  For future
use we have defined
\bea
\Pi(\eta,\eta',D) = \int_p {(p_\alpha P_{\alpha \beta}^T(\hat {\bf q}) p_{\beta})^2
  \over|\pb|^{4-\eta}|\pb+\hat \q|^{4-\eta'}}\;  \quad , \quad \Pi(\eta,D) = \Pi(\eta,\eta,D), \label{PiD} 
\eea 
and the integral is calculated in the Appendix B, with the amplitude
$\Pi(\eta,D)$ found to be
%
%
%previously defined, and 
%%
%\begin{equation}\label{Aintegral}A(D,\eta)\;q^{-\eta_u}={P^T\ab P^T\cd\over P^T_{\alpha_1\alpha_2}
%P^T_{\alpha_3\alpha_4} S_{\alpha_1\alpha_2\alpha_3\alpha_4}}
%\int_p{p\as p\bs p\cs p\ds\over|\pb|^{4-\eta}|\pb+\q|^{4-\eta}}\; \end{equation}
%%
%Using Feynman parameter integration
%\cite{Fintegration}
%the integral gives,
%
\begin{equation}\label{Aresult} \Pi(\eta,D) = (D^2-1) {\Gamma(2-\eta-D/2)\Gamma(D/2+\eta/2)
\Gamma(D/2+\eta/2)\over4(4\pi)^{D/2}\Gamma(2-\eta/2)\Gamma(2-\eta/2)
\Gamma(D+\eta)}\;. \end{equation}

Substituting Eq.\eqref{Idiverge} into Eqs.\eqref{renormModuli} and
assuming $\eta_u>0$, (i.e., that we are looking for a self-consistent
solution with $\eta<(4-D)/2$, to be checked a posteriori), we find for
$q \ll q_{\rm n l}$ that the $I({\bf q})$ terms in the denominators in
Eqs.\eqref{renormModuli} dominate, giving 
\bea && \mu_R({\bf q}) \simeq
\frac{1}{2 T I({\bf q})} \sim \frac{(D^2-1) Z_\kappa^2}{2 T
  \Pi(\eta,D)} q^{\eta_u}, 
\eea 
and similarly for $b_R({\bf q})$ leading to the small $q$ behavior of
the renormalized elastic constants 
\bea
\label{muR2}
&& \mu_R({\bf q}) \simeq Z_\mu q^{\eta_u} \quad , \quad Z_\mu =
\frac{(D^2-1) Z_\kappa^2}{2 T \Pi(\eta,D)}\ ,\\
&& b_R({\bf q}) \simeq Z_b q^{\eta_u} \quad , \quad Z_b = \frac{(D-1)
  Z_\kappa^2}{T \Pi(\eta,D)}\ .  
\eea

Using the SCSA equation \eqref{SimpleSigma} for $\kappa_R({\bf q})$ we obtain 
\bea
\kappa_R({\bf k}) = \kappa + \left(  {2 T\over d_c} {Z_b + (D-2) Z_\mu \over (D-1) Z_\kappa} \int_q 
\frac{\left(\hat k\as P\ab^T(\q) \hat k\bs\right)^2 |\q|^{\eta_u} }{|\hat \ks-\q|^{4-\eta}}\ \right) k^{-\eta}\;.
\label{kren1} 
\eea 
For future use we define
\bea\label{Sigma_etaetap}
\Sigma(\eta,\eta',D) = \int_q {(\hat k_\alpha P_{\alpha \beta}^T(\hat {\bf q}) \hat k_{\beta})^2
|\q|^{4-D - 2\eta} 
\over |\hat \ks+ \q|^{4-\eta'}}\;  \quad , \quad \Sigma(\eta,D) :=
\Sigma(\eta,\eta,D)\ ,
\eea 
and the integral is calculated in the Appendix B, with the amplitude
$\Sigma(\eta,D)$ found to be
\begin{eqnarray}
\Sigma(\eta,D)
%k^{-4+\eta}\int_q |\q|^{\eta_u} |\ks-\q|^{-(4-\eta)} 
%\left(k\as P^T\ab(q) k\bs\right)^2\;,\\
&=&{(D^2-1)\Gamma(\eta/2)\Gamma(D/2+\eta/2)\Gamma(2-\eta)
\over4(4\pi)^{D/2}\Gamma(2-\eta/2)\Gamma(D/2+\eta)\Gamma(D/2+2-\eta/2)} 
\; .
\label{SigmaResult}
\end{eqnarray}
We now note that the second term in \eqref{kren1} dominates at small
$k$ leading to $\kappa_R({\bf q}) \simeq Z_\kappa k^{-\eta}$
consistent with the ansatz.  Hence the solution is self-consistent.

Substituting $Z_b$ and $Z_\mu$ into \eqref{kren1} we find that,
crucially, the amplitude $Z_\kappa$ cancel out. The remaining equality
gives a transcendental equation, depending only $d_c$ and $D$,
determines the exponent $\eta(D,d_c)$, that is therefore universal,
\be \frac{D(D-1)}{d_c}\frac{\Sigma(\eta,D)}{\Pi(\eta,D)} = 1\ .
\label{eqSCSApure} 
\ee
More explicitly, this equation reads
\be
D(D-1) \frac{\Gamma (2-\eta ) \Gamma
   \left(2-\frac{\eta }{2}\right)
   \Gamma \left(\frac{\eta
   }{2}\right) \Gamma (D+\eta
   )}{\Gamma \left(-\frac{D}{2}-\eta
   +2\right) \Gamma
   \left(\frac{1}{2} (D-\eta
   +4)\right) \Gamma
   \left(\frac{D}{2}+\eta \right)
   \Gamma \left(\frac{D+\eta
   }{2}\right)} = d_c\ .
\label{scsa1} 
\ee
Analysis of the left hand side shows that there is a unique solution
continuously related to $\eta=0$ for either $d_c=+\infty$ or $D=4$,
which we call $\eta(D,d_c)$.

%
%
%and using the result
%together with the scaling ansatz Eq.\ref{Gscale} inside Eq. \ref{SimpleSigma} we find
%that for $\mu>0$, $b>0$ the unknown nonuniversal factor $Z$ as well 
%powers of the wavevector $k$
%cancel. We then obtain the self-consistent equation for the 
%exponent $\eta$ as a function of dimensionalities $d_c$ and $D$
%%
%\begin{equation}\label{SCSAeqnii}d_c=\left({D\over D+1}\right){B(D,\eta)\over A(D,\eta)}\;, \end{equation}
%%
%where
%%
%\begin{eqnarray}\label{Bresult} B(D,\eta)&=&
%k^{-4+\eta}\int_q |\q|^{\eta_u} |\ks-\q|^{-(4-\eta)} 
%\left(k\as P^T\ab(q) k\bs\right)^2\;,\\
%&=&{(D^2-1)\Gamma(\eta/2)\Gamma(D/2+\eta/2)\Gamma(2-\eta)
%\over4(4\pi)^{D/2}\Gamma(2-\eta/2)\Gamma(D/2+\eta)\Gamma(D/2+2-\eta/2)}
%\;.
%\end{eqnarray}
%%
%Upon using the definition of $A(D,\eta)$ and after some simplifications we
%obtain,
%%
%\begin{equation}\label{SCSAeqniii}d_c={2 \over \eta} 
%D(D-1) { { \Gamma[1+{1\over 2}\eta] \Gamma[2-\eta] \Gamma[\eta+D] \Gamma[2-{1\over 2}\eta] }
%\over {\Gamma[{1\over 2}D + {1\over 2}\eta] \Gamma[2-\eta-{1\over 2}D] 
%\Gamma[\eta+{1\over 2}D] \Gamma[{1\over 2}D+2-{1\over 2}\eta]}}\;. \end{equation}
%
For membranes of physical dimensionality $D=2$ and arbitrary $d_c$, we
can explicitly solve Eq. \eqref{scsa1} and we obtain for
$\eta(D=2,d_c)$
\begin{equation}\label{Eta}\eta(D=2,d_c)={4\over
    d_c+\sqrt{16-2d_c+d_c^2}}\; ,
\end{equation}
with the corresponding values for $\eta_u=2-2 \eta$ and
$\zeta=1-\frac{\eta}{2}$ obtained from the rotational identity,
Eq. \eqref{WardIdentity}, and from Eq. \eqref{ZetaIdentity},
respectively.  Thus for physical membranes $D=2$, $d=1$ we obtain
\bea
\label{EtaZetaPhysical}
 \eta={4\over 1+\sqrt{15}} &\approx&  0.821\;, \\
 \eta_u= \frac{2}{7} (9 - 2 \sqrt{15}) &\approx& 0.358\;, \\
 \zeta = \frac{1}{7} (8 - \sqrt{15})&\approx& 0.590\;.
\end{eqnarray}

\subsection{Discussion of the SCSA predictions for a homogeneous
  membrane}
\label{discussionSCSAhomo}

We can now examine various limiting approximations of our SCSA
result. For the membranes of large codimension we can expand the
solution of \eqref{scsa1} in powers of $1/d_c$ obtaining,
\begin{eqnarray}\label{dcExpand} \eta(D,d_c)&=& {8\over d_c}\left({D-1\over D+2}
\right){\Gamma[D]\over{\Gamma[{D \over 2}]^3 \Gamma[2-{D \over 2}]}}
+O\left({1 \over{d_c^2}}\right)\;,\\
\eta(D=2,d_c)&=& {2 \over d_c}+O\left({1 \over {d_c^2}}\right)\;. 
\end{eqnarray}
This result agrees with the findings of Refs.\;\onlinecite{AL,GDLP}.  As
discussed above, this limiting property was expected from the
construction of the SCSA, built on $1/d_c$-expansion.

Expansion of the solution of \eqref{scsa1} to first order in
$\epsilon=4-D$ gives
\begin{equation}\label{epsExpand}\eta(D=4-\epsilon,d_c)
={\epsilon\over2+d_c/12}\;, 
\end{equation}
which is also in agreement with the result of Ref.\ \onlinecite{AL},
exact to $O(\epsilon)$ for all $d_c$. This is not a general property
of SCSA. Here it can be traced to the fact that the vertex diagram (c)
in Fig. \ref{FeynmanLoopsSCSAsr} is {\it convergent}, due to the
structure of the theory. Because of the transverse projectors in
\eqref{MNtensors} one can always extract one power of external
momentum from each external $\h$ legs. As a result the only
counter-terms needed are for two-point functions. This special
property can be traced to a Ward identities based on the underlying
embedding-space rotational invariance. The results of the two-loop
calculation are in progress, and already indicate that the SCSA is not
exact, with the deviations appearing at the two-loop order.
\cite{TwoLoopLR}

We also observe from \eqref{scsa1} that the solution for $d_c=0$ is
$\eta(D,d_c=0)=\frac{4-D}{2}$, i.e., $\eta_u=0$, which is the exact
result for $d_c=0$, as discussed in previous section.

Thus, as advertised, the SCSA is indeed {\em exact} in three distinct
complementary limits. These strong constraints are at the heart of its
quantitative accuracy in the physical dimension.

\begin{figure}[htbp]
        \centering
        \includegraphics[width=0.7\textwidth,scale=1]{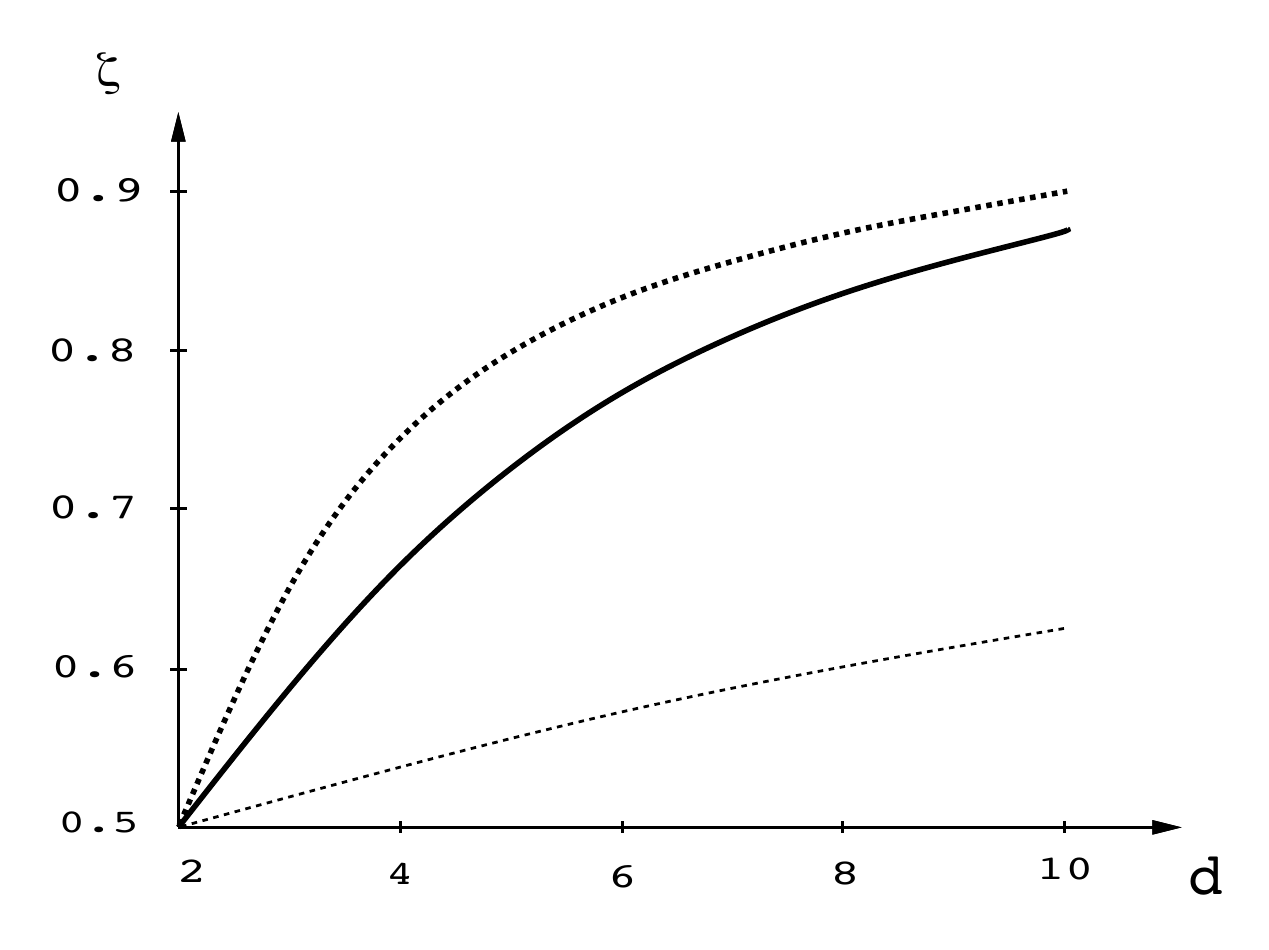}
        \caption{\label{zetaSCSAsr} A roughness $\zeta$ as a function
          of embedding dimension, $d$ for two-dimensional ($D=2$)
          elastic membranes. The solid curve is the SCSA result,
          Eq.\eqref{Eta}. The long-dashed-short-dashed curve is the
          $O(\epsilon)$ result, setting $\epsilon=2$.  The dashed
          curve corresponds to $\eta=2/d$ chosen (somewhat
          arbitrarily) in Ref.\onlinecite{AL,GDLP} as a possible
          interpolation to finite $d$ (asymptotic to the solid curve
          for $d\rightarrow\infty$).}
\end{figure}

%{\red recheck, add reference} 
%It is interesting to compare
%with simulations by Grest for $D=2$ membranes with
%self-avoidance in increasing co-dimensions, which gives a flat membrane
%in $d=3,4$ with $\zeta(d_c=1)=0.64 \pm 0.04$, $\zeta(d_c=2)=0.77 \pm
%0.04$, whereas we obtain $0.59$,$0.67$, respectively. From the
%simulations $d=5$ seems very close to a marginal case but the membrane
%is crumpled with $\nu=0.8 \pm0.06$. Note that we find $\nu=0.8$ at the
%crumpling transition (see below) for $d=5$, and $\zeta(d=5)=\nu=0.73$
%in the flat phase.

The present method also gives interesting predictions for the
lower-critical dimension $D_{lc}(d_c)$ for orientational order, i.e.,
order in $\nabla {\bf h}$. The fluctuations of the latter can be
calculated as 
\be \langle (\nabla h)^2 \rangle \sim T \int_q
\frac{q^2}{Z_\kappa q^{4-\eta}} \sim T L^{2-D-\eta}\ ,
\ee 
which is found to diverge with system size $L$ for $D < D_{lc}$
determined by the equation
\bea 
\label{Dlc}
2-\eta(D_{lc},d_c)=D_{lc}\ .
\eea 
From \eqref{scsa1} one easily finds that this equation is equivalent
to $d_c=D_{lc} (D_{lc}-1)/(2-D_{lc})$. Thus, we find the lower-critical
dimension as a function of $d_c$ to be
\bea 
D_{lc}(d_c)={1\over
  2}(1-d_c+\sqrt{d_c^2+ 6d_c+1})\;.
\eea 
In particular for $d_c=1$ we find $D_{lc}=\sqrt{2}$. We also observe
that $D_{lc}(d_c)$ increases from $D_{lc}=1$ for $d_c=0$, to
$D_{lc}=2$ when $d_c \rightarrow \infty$.  On the other hand if one
keeps the embedding dimension $d$ fixed, solving
$2-\eta(D_{lc},d-D_{lc})=D_{lc}$ gives $D_{lc}=3/2$ for $d=3$. Hence
in either case, the flat phase of a physical two-dimensional membrane,
that spontaneously breaks a continuous rotational symmetry, is thus
stable and is in fact stabilized by thermal fluctuations (through
renormalization of its elastic moduli), thus evading the
Hohenberg-Mermin-Wagner-Coleman theorems 
\cite{Hohenberg,MerminWagner,Coleman}, thereby exhibiting 
the aforementioned order-from-disorder phenomenon.

Similarly, one can ask how the anomalous thermal fluctuations affect
the in-plane translational order, i.e., in the $u_{\alpha}$ fields, by
calculating the exponent of the Debye-Waller factor
\be
\langle (u_\alpha)^2 \rangle \sim T \int_q \frac{1}{Z_b q^{2+\eta_u}} \sim T L^{2 \zeta_u}\;, \quad 
\zeta_u=\frac{2-D+\eta_u}{2}\;,
\ee
using \eqref{Rough2}.  Hence we find that the fluctuations in the
in-plane positions of the atoms diverge for $D<D_u$ where $D_u$ is
determined by the equation $2+\eta_u(D_u,d_c)=D_u$. By virtue of Ward
identity $\eta_u = 4 - D - 2\eta$, \eqref{exponentrelation}, this is
equivalent to $\eta(D_u)=3-D_u$, and from \eqref{scsa1} we obtain that
$D_u=2.24$ for $d_c=1$. Hence, for physical membranes the
quasi-long-range positional order of a flat two-dimensional crystal is
made unstable by out-of-plane fluctuations. It suggests that
dislocations in a crystalline membrane at finite temperature cost a
finite energy and if so, will always unbind at sufficient long scales
(which of course can be very large depending on the core energy
determined by the bond strength). Strictly speaking, it therefore
precludes the existence of a crystalline fluctuating membrane in the
thermodynamic dynamic. For the idealized case of permanent bonds the
length at which this occurs becomes infinite.

Another related mechanism for proliferation of dislocations was
analyzed by Nelson and Seung (NS) \cite{NS}, who showed that at $T=0$
above a scale $R_b^0 \sim 122 \frac{\kappa}{K_0 a}$ a dislocation will
buckle out of plane, lowering its energy to a finite value. At finite
temperature, our analysis shows that this estimate holds only for
temperature such that $R_b^0 \ll L_{\rm nl}$, where $L_{\rm nl} \sim
1/q_{\rm nl} \sim \kappa/\sqrt{K_0 T}$ is the length scale beyond
which anomalous elasticity sets in, as the system crosses over to the
non-trivial fixed point of the thermal membrane studied above. At
higher temperature, the renormalization of the bending rigidity and
the Young modulus by thermal fluctuations must be taken into account,
and one finds the thermally-modified estimate,
\bea
R_b \sim R_b^0 \left(\frac{R_b^0}{L_{n l}} \right)^{\delta}, 
\eea 
with $\delta=\eta-\eta_u/(1-\eta+\eta_u) \approx 0.862$.  We find
$R_b^0/L_{n l}= 122 \sqrt{T/(K_0 a^2)}$ (which is $\approx 4.3$ in
graphene at room temperature $T=1/40$ eV, using the standard
parameters $a=1.4 \AA$, $a^2 K_0=20 eV$, $\kappa=1 eV$).  This
demonstrates that the effective buckling length (beyond which
dislocations unbind by buckling) at finite temperature grows as almost
the square of the NS length (valid at $T=0$, at the Gaussian fixed
point). Hence temperature has actually a stabilizing affect on
in-plane order at intermediate scale. At longer scales, in principle,
dislocations unbind and lead to a hexatic fluctuating
membrane\cite{NP}, whose flat phase is expected to be unstable to
out-of-plane fluctuations. However, in the idealized case of a
tethered membrane, and even in graphene, whose covalent carbon bonds
are much more energetic than the relevant thermal energy, dislocation
core energies are very large and for finite size membrane the flat
phase remains stable with properties discussed here.

Let us now comment on the values of the amplitudes $Z_\kappa$, $Z_\mu$
and $Z_b$ in \eqref{Gscale}, \eqref{muR2}. Although each of them is
non-universal one can form two universal ratio from them, respectively
\bea
&& \frac{2 T Z_b}{Z_\kappa^2} =  \frac{2(D-1)}{\Pi(\eta,D)},\\
&& \frac{2 Z_b}{Z_\mu} = \frac{4}{D+1}.
\eea 
For the physical membrane, $D=2$, we find $2 T Z_b/Z_\kappa^2
\approx 11.23$ and $2 Z_b/Z_\mu = 4/3$.  Examining
further the elastic properties of a membrane from our SCSA solution
we find from Eqs. \eqref{renormModuli}, that in the long wavelength
limit $\frac{\lambda_R({\bf q})}{\mu_R({\bf q})} \simeq - {2 \over
  {D+2}}$. Hence the membrane is described by a negative universal
Poisson ratio,
\bea
\label{PoissonRatio} 
%&& \frac{\lambda_R({\bf q})}{\mu_R({\bf q})} \simeq 
% - {2 \over {D+2}} \;. \\
\sigma_R({\bf q}) &=& \frac{\lambda_R({\bf q})}{ 2 \mu_R({\bf q}) +
   (D-1) \lambda_R({\bf q})} = - \frac{1}{3}\;.
\eea
Although at first this result seems counterintuitive, it has a nice
physical interpretation. A regular solid in the absence of thermal
fluctuations is characterized by a positive Poisson ratio because when
it is stretched in one direction it must contract in the other to
minimize large density changes. However, as we have seen, a
two-dimensional tensionless membrane in the presence of thermal
fluctuations exhibits wild out-of-plane fluctuations. On average it
therefore occupies a smaller projected area with the shrinkage factor
determined by the scalar order parameter $t$,
Eq. \eqref{MFTtangent}. If we stretch such a membrane in one direction
we necessarily suppress its transverse fluctuations and therefore the
membrane stretches in all in-plane directions, corresponding to a
negative Poisson ratio.

We note that the SCSA equations admit two other, peculiar, fixed
points. Indeed from \eqref{renormModuli} the choice of a zero bulk
modulus $b=0$ leads to $b_R({\bf q})=0$, and is a solution of the SCSA
equations at the limit of mechanical stability of the elastic
manifold. Similarly, the choice $\mu=0$ in \eqref{renormModuli} with
$b \neq 0$, leads to another solution, with $\mu_R({\bf q})=0$. It
corresponds to anomalous elasticity of a nematic
elastomer\cite{WarnerBook}, studied extensively in Ref.\onlinecite{LRelastomer,LubenskyElastomer}. As is easily seen by inserting
into \eqref{SimpleSigma}, the equation for the associated exponent
$\eta$ is obtained by multiplying the left hand side of \eqref{scsa1}
by $(D+1)(D-2)/(D(D-1))$ for the first one ($b=0$) and $2/(D(D-1))$
for the second one ($\mu=0$).  Hence we find 
\bea
&& \eta(D,d_c;b=0) = \eta(D, \frac{D(D-1)}{(D-2)(D+1)} d_c)\ ,  \\
&& \eta(D,d_c;\mu=0) = \eta(D, \frac{D(D-1)}{2} d_c)\ , 
\eea
where $\eta(D,d_c)$ is the standard SCSA solution analyzed above.  For
the first, fully compressible $b=0$ fixed point in $D=3$ we find
$\eta=0.446$. Since $\eta \to 0$ as $D \to 2^+$, we conclude that
$D_{lc}=2$ for this fixed point, which corresponds to a membrane fine
tuned to the edge of its mechanical stability.  The second, $\mu=0$
fixed point gives, for $D=2$ and $d_c=1$, $\eta(2,1;\mu=0) = \eta(2,
1)= {4\over 1+\sqrt{15}} \approx 0.821$, the same value as for the
standard SCSA fixed point. It is in principle possible to realize a
vanishing shear modulus in a nematic elastomer
membrane.\cite{LRelastomer} However, it remains to be
further studied whether the neglected in-plane phonon nonlinearities
(that have been shown to be irrelevant at the Gaussian fixed point)
remain irrelevant at the present SCSA fixed point.\cite{LRelastomer}

Finally, one can ask about properties of a membrane in presence of a
finite, but small tension and compression, i.e., slightly
off-criticality.  Here we will not reanalyze this buckling transition
via the SCSA calculation, but will use the scaling relations derived
in Ref.\ \onlinecite{CrumplingBucklingGuitter,GDLP} to predict the
buckling critical exponents within our SCSA for the $\eta, \eta_u$
exponents.  There are various ways to study deviations from the
critical tensionless, rotational invariant membrane. One is to put the
membrane under stress, without breaking the embedding-space rotational
invariance, by adding to the free energy \eqref{Fflat} the term
\bea 
\label{Ftension} 
\delta {\cal F} = \tau \int d^D x ~
(\partial_\alpha u_\alpha + \frac{1}{2}\partial_\alpha \vec h
\cdot \partial_\alpha \vec h)\ , 
\eea 
which leads to the change $\kappa q^4 \to \kappa q^4 + \tau q^2$ in
the quadratic part of the free energy \eqref{FflatEff2D} for the
out-of-plane height modes.  For $\tau>0$ the membrane is stretched,
while for $\tau<0$ it tends to buckle.  Alternatively, one can use
constrained boundary conditions, imposing a projected area $(t L)^D$
for the membrane, where $t$ is different from the spontaneous
equilibrium value $t_{\rm sp}(T)<1$. As argued in \cite{GDLP} this
generates a tension, with $\tau_R \sim t-t_{\rm sp}(T)$, where
$\tau_R$ is the renormalized value of the tension. Finally, one can
set $\tau_R=0$, but introduce a linear term $\delta {\cal F} = - f
\int d^D x \partial_\alpha u_\alpha$, which explicitly breaks
rotational invariance (as a frame would do). 

It is shown in \cite{CrumplingBucklingGuitter,GDLP} that this leads to
a finite internal correlation length, which diverges as $\xi \sim
\tau_R^{-\nu}$, in the tensionless limit, and to $t-t_{\rm sp}(T) \sim
f^{1/\delta}$, where the two exponents $\nu$ and $\delta$ are given by
\bea 
\frac{1}{\nu} = D-2 + \eta\;, \quad \delta=(2-\eta)\nu\;.
\label{scalingdelta} 
\eea 
Combining these with our SCSA values for the $\eta$ exponent leads to
$\nu=\frac{1}{4}(1+\sqrt{15}) = 1.2182$, and $\delta= \frac{1}{2}
(\sqrt{15}-1)=1.4365$, for a physical membrane.  Note that the theory
of Ref.\ \onlinecite{CrumplingBucklingGuitter,GDLP} also predicts $t
\sim f^{1/\delta}$ at $T=T_c$, where $T_c$ is the crumpling transition
temperature. It would be interesting to extend the SCSA calculation to
derive (and confirm) these results from first principles, and also to
obtain a better description of the buckled state at $\tau_R<0$.

\section{The crumpling transition} \label{sec:crumpling} 
\subsection{Derivation of the SCSA equations}

In this Section we analyze the crumpling transition using the SCSA. We
focus at the critical point, and thus tune all the parameters to sit
exactly at criticality.  We will thus determine the only exponent at
criticality, the exponent $\eta$ (see below).  Calculation of the
other independent exponent (e.g. $\nu$) requires an independent
calculation. Because the crumpling transition occurs at nonzero
temperature, for convenience in this section we work in units in which
$T=1$. We start from the isotropic theory for a polymerized phantom
membrane tuned at its critical crumpling point
\cite{KKN,PKN,KantorNelson,ARP}.  The Hamiltonian is given by
\begin{equation}
\label{hamEFF2}
{\cal F}[\vec{r}] =\intd\left[{\kappa\over2}(\partial^2
\vec{r})^{2} 
+ \frac{\mu_0}{4 d} (\partial_{\alpha}\vec{r}\cdot\partial_{\beta}\vec{r})^{2} +\frac{\lambda_0}{8 d} 
(\partial_{\alpha}\vec{r}\cdot\partial_{\alpha}\vec{r})^{2}\right]\;,
\end{equation}
where, as compared to \eqref{hamEFF}, we have denoted
$u=\frac{\mu_0}{4 d}$ and $\tilde v=\frac{\lambda_0}{8 d}$,
%$$
%F= \int d^Dx  {\kappa \over 2} (\nabla^2{\bf r})^2
%+ {\mu_0 \over 4} ({\partial _{\alpha} {\bf r}}.{\partial _{\beta} {\bf r}} )^2
%+ {\lambda_0 \over 8} ({\partial _{\alpha} {\bf r}}.{\partial _{\alpha} {\bf r}} )^2
%\eqno(3.1)
%$$
where we have used subscripts to distinguish the coefficients of the
anharmonic terms from the true Lam\'e coefficients in the flat
phase. Equation \eqref{hamEFF2} can be rewritten in terms of Fourier
components, in a form analogous to \eqref{Feff}
\bea
{\cal F}[\vec{r}]={\kappa \over 2} \int_{k}  k^4 \mid {\vec r}(\kb) \mid^2 +
{1 \over {4 d} } \int_{k_1,k_2,k_3} R_{\alpha \beta, \gamma \delta}(\q)~
k_{1 \alpha} k_{2 \beta} k_{3 \gamma} k_{4 \delta}~
{\vec r}(\kb_1)\cdot{\vec r}(\kb_2 )~
{\vec r}(\kb_3)\cdot{\vec r}(\kb_4 ),
\label{(3.2)}
\eea
with $\q=\ks_1+\ks_2$ and $\ks_1+\ks_2+\ks_3+\ks_4={\bf 0}$, and we
use $\int_k$ to denote $\int d^Dk/(2 \pi)^D$.  Note that we work
exactly at the crumpling transition point where the renormalized
quadratic term has been tuned to and kept at zero. The bare four-point
fourth order tensorial interaction has the form: 
\be R_{\alpha \beta,
  \gamma \delta}
=\frac{\mu_0}{2}(\delta_{\alpha\gamma}\delta_{\beta\delta}+\delta_{\alpha\delta}\delta_{\beta\gamma})
+ \frac{\lambda_0}{2} \delta_{\alpha\beta}\delta_{\gamma\delta}.
\label{(3.3)}
\ee
This parameterization is insufficient to express the renormalized
$\tilde{R}_{\alpha \beta, \gamma \delta} ({\bf q})$ (i.e., after
dressing by the vacuum polarization bubbles), since the momentum $\q$
can carry indices and tensors, for instance, $q_{\alpha}q_{\beta}
\delta_{\gamma\delta}$ can appear. The most general parameterization
that we will need, in terms of the irreducible representations under
tensor multiplication, which is defined as $R''=R' * R$, i.e.,
$R''_{\alpha\beta,\gamma\delta}
=R'_{\alpha\beta,\alpha'\beta'}R_{\alpha'\beta',\gamma\delta}$,
is :
\be
R= \sum_{i=1}^5 w_i W_i\;,
\label{(3.4)}
\ee
in terms of five "elastic constants" $w_i$ and five "projectors"
$W_i$, $i=1,\ldots,5$, defined as
\bea
&& (W_3)_{\alpha\beta,\gamma\delta}(\q)
={1 \over {D-1}} P^T_{\alpha \beta} P^T_{\gamma \delta}\;,\;\;\;
(W_5)_{\alpha\beta,\gamma\delta}(\q)=P^L_{\alpha \beta}P^L_{\gamma\delta}\;,\\
&& 
(W_4)_{\alpha\beta,\gamma\delta}(\q) = (W_{4a})_{\alpha\beta,\gamma\delta}(q) 
+ (W_{4b})_{\alpha\beta,\gamma\delta}(\q)\;,\\
&& (W_{4a})_{\alpha\beta,\gamma\delta}(\q) = {1 \over {\sqrt{D-1}}} P^T_{\alpha \beta} P^L_{\gamma \delta}\;,
\quad 
(W_{4b})_{\alpha\beta,\gamma\delta}(\q) ={1 \over {\sqrt{D-1}}} P^L_{\alpha \beta} P^T_{\gamma \delta}\;,\\
&& (W_2)_{\alpha\beta,\gamma\delta}(\q)=
{1 \over 2}(P^T_{\alpha \gamma} P^L_{\beta \delta } 
+ P^T_{\alpha \delta} P^L_{\beta \gamma} + P^L_{\alpha \gamma}
P^T_{\beta \delta } + P^L_{\alpha \delta} P^T_{\beta \gamma})\;, \\
&& 
W_1(\q)={1\over 2}(\delta_{\alpha\gamma}\delta_{\beta\delta}
+\delta_{\alpha\delta}\delta_{\beta\gamma})
- W_3(\q) - W_5(\q) - W_2(\q)\;,
\label{(3.6)}
\eea 
where $P^T_{\alpha \beta}=\delta_{\alpha \beta} -
q_{\alpha}q_{\beta}/q^2$ and $P^L_{\alpha
  \beta}=q_{\alpha}q_{\beta}/q^2$ are the standard transverse and
longitudinal projection operators associated to $\q$. The first two
projectors $W_1, W_2$ are mutually orthogonal and orthogonal to the
other three. Note that while $R$, being symmetric, can be expressed in
terms of the symmetric tensors $W_i$, $i=1,..5$, we will need at some
intermediate stages of the calculations some products (such as $\Pi*R$
see below), which are not symmetric. Hence we introduced $W_4^a$ and
$W_4^b$, which together with $W_i$, $i=1,2,3$ and $W_5$ make the
representation complete under tensor multiplication. The rules for the
tensor multiplication $T''=T'*T$ of the more general tensors $T =
\sum_{i=1}^3 w_i W_i + w_{4a} W_{4a} + w_{4b} W_{4b} + w_{5} W_{5}$
and $T' = \sum_{i=1}^3 w'_i W_i + w'_{4a} W_{4a} + w'_{4b} W_{4b} +
w'_{5} W_{5}$ are then
\bea
w_1''=w'_1 w_1\;,\;\;\; w_2'' = w'_2 w_2\;,\;\;\;
\begin{pmatrix}
w_3''&w_{4a}''\\ w_{4b}''&w_5'' 
\end{pmatrix}
=
\begin{pmatrix}
w'_3&w_{4a}' \\
w_{4b}' &w'_5 
\end{pmatrix}
\begin{pmatrix}
w_3&w_{4a}\\
w_{4b}&w_5
\end{pmatrix}\;,
\label{3.5}
\eea 
with $T'' = \sum_{i=1}^3 w''_i W_i + w''_{4a} W_{4a} + w''_{4b}
W_{4b} + w''_{5} W_{5}$.  For the tensor $R_{\alpha \beta,\gamma \delta}$, which
belongs to the space of tensors which are symmetric under $\alpha
\leftrightarrow \beta$, under $\gamma \leftrightarrow \delta$ and
under $(\alpha,\beta) \leftrightarrow (\gamma,\delta)$ (let us call
this symmetry ${\cal S}$) the five couplings in \eqref{(3.4)} are
sufficient, and their bare values (so that \eqref{(3.4)} reproduces
\eqref{(3.3)}) are given by 
\bea w_1= w_2 = \mu_0,\;\;
w_3= \frac{1}{2}(D-1)\lambda_0+ \mu_0,\;\;
w_4=\frac{1}{2} \sqrt{D-1}\lambda_0,\;\;
w_5=\frac{1}{2} \lambda_0+ \mu_0\;.
\label{(3.7)}
\eea
Note that the two eigenvalues of the matrix formed by the $w_i$,
$i=3,4,5$, are then $\mu_0$, and $\mu_0 + \frac{1}{2} D \lambda$.

%old result (wrong)
%$$
%w_1= w_2 = 2 \mu_0~~~w_3=(D-1)\lambda_0+ 2 \mu_0~~~w_4=\sqrt{D-1} \lambda_0~~~w_5=\lambda_0+2\mu_0
%\label{(3.7)}
%$$

For the crumpling transition the SCSA equations now take the form:
\bea
\Gc(\kb)^{-1} - \kappa k^4 =: \sigma(\kb)&=& {2 \over d} \int_q k_{\alpha}(k_{\beta}-q_{\beta})(k_{\gamma}-q_{\gamma})k_{\delta}
\tilde{R}_{\alpha \beta, \gamma \delta}(\q) \Gc(\kb-\q)\;,
\label{(3.8a)} \\
\tilde{R}(\q) &=& R(\q) - R(\q) \Pi(\q) \tilde{R}(\q)\;,
\label{(3.8b)}
\eea
where $\Gc(\kb)$ is the propagator of the $\vec r$ field, i.e., $\langle
r_i(\kb) r_j(\kb') \rangle = \Gc(\kb) (2 \pi)^D \delta(\kb + \kb')
\delta_{ij}$, and in the second line tensor-product notation is
implied.  The complete Dyson equation for the self-energy contains an
additional UV divergent "tadpole" diagram contribution, which scales
as $k^2$. The integral in \eqref{(3.8a)} also contains a component
that scales as $k^2$ at small $k$. Both contributions have been
subtracted by the shift in a critical point. This is a standard
procedure when dealing with a critical theory, where a parameter
(distance to crumpling transition parameters) must be tuned.

The vacuum polarization tensor is also a symmetric tensor (with the
symmetry ${\cal S}$ defined above)
\be
\Pi_{\alpha \beta, \gamma \delta}(\q)
= {1 \over 4} \int_p \left(p_{\alpha} (q_\beta-p_{\beta}) + p_{\beta} (q_\alpha-p_{\alpha})\right)
\left(p_{\gamma} (q_\delta-p_{\delta}) + p_{\delta} (q_\gamma-p_{\gamma})\right)
\Gc(\pb) \Gc(\q-\pb)\;.
\label{(3.9)}
\ee
Hence it can be written as
\be
\Pi(\q) = \sum_{i=1}^5 \pi_i(\q) W_i(\q)\;,
\label{(3.10)}
\ee
where the $\pi_i(\q)$ are ``polarization bubble'' integrals in the
$W_i$ basis.  The renormalized interaction $\tilde{R}(\q)$ also
exhibits ${\cal S}$ symmetry, and can thus be written as
\bea \label{renR} 
\tilde R(\q) = \sum_{i=1}^5 \tilde w_i(\q) W_i(\q)\;,
\eea
where the renormalized couplings $\tilde w_i(\q)$ read
\be
\tilde{w}_1(\q)={ w_1 \over {1 + w_1 \pi_1(\q)}}\;,\;\;\;
\tilde{w}_2(\q)={ w_2 \over {1 + w_2 \pi_2(\q)}}\;,
\ee
\bea
\begin{pmatrix} \tilde{w}_3(\q) & \tilde{w}_4(\q) 
\\ \tilde{w}_4(\q) & \tilde{w}_5(\q) \end{pmatrix} 
=
\begin{pmatrix}  w_3 & w_4 \\ w_4 & w_5 \end{pmatrix} 
\left( 
\begin{pmatrix}  1 & 0 \\ 0 & 1  \end{pmatrix}   +
\begin{pmatrix} \pi_3(\q) & \pi_4(\q) \\ \pi_4(\q) & \pi_5(\q) \end{pmatrix}
\begin{pmatrix} w_3&w_4 \\ w_4&w_5 \end{pmatrix}\right)^{-1}\;\;.
\label{(3.11)}
\eea
Note that in this section we denote with $\tilde w_i(\q)$ the renormalized
couplings. Substituting the form \eqref{renR} in \eqref{(3.8a)} we
obtain
\bea
\sigma(\kb)= {2 \over d} \sum_{i=1,5} \int_q \tilde{w}_i(\q) \Gc(\kb-\q)
k_{\alpha}(k_\beta-q_{\beta}) (W_i)_{\alpha\beta,\gamma\delta}(\q) k_{\gamma} (k_\delta-q_{\delta})\;,
\label{(3.12)}
\eea
which allows us to obtain $\kappa_R({\bf q})$ (after the subtraction
of the $k^2$ term).  The above equations form a closed set of SCSA
equations for the five renormalized elastic coupling constants,
together with the renormalized bending rigidity.

\subsection{Analysis and results}

We now analyze the SCSA equations for the crumpling transition,
following closely the analysis in the previous section for the flat
phase.  To solve these equations at the critical point, we use the
long wavelength form of the critical propagator,
$\Gc(\kb)=Z^{-1}_\kappa/k^{4-\eta}$. The $\pi_i(\q)$ are integrals similar to
the ones studied for the flat phase, and are calculated in the
Appendix \ref{app:crumpling}. They diverge for small $q$ as:
\bea
\pi_i(\q) \simeq Z_\kappa^{-2} a_i(\eta,D) q^{-(4-D-2\eta)}\;.
\label{(3.13)}
\eea
For the amplitudes $a_i(\eta,D)$ we find 
\begin{eqnarray}
&&a_1= 2A\;,\;\; a_2=A {2(2-\eta) \over{D+ \eta - 2}}\;,\;\;
a_3=A (D+1)\;,\;\; a_4= A \sqrt{D-1} (D+ 2\eta - 3)\;,\\
&&a_5={A \over {D-2+\eta}}(-22+31D-10D^2+D^3+43\eta-32D\eta+5 D^2\eta-24\eta^2+8D\eta^2+4 \eta^3)\;,\nonumber
\end{eqnarray}
with 
\begin{equation}
A = \frac{\Pi(\eta,D)}{D^2-1}\;,
\label{(3.14)}
%={ \Gamma(\eta/2+D/2)^2 \Gamma(2-\eta-D/2) \over { 4 (4\pi)^{{D \over{2}}}  \Gamma(2-\eta/2)^2 \Gamma(\eta+D) }} 
\end{equation}
where the ``polarization bubble'' integral, $\Pi(\eta,D)$ is defined in
\eqref{PiD},\eqref{Aresult} and derived in Appendix \ref{FinalResults}.

To compute the self-energy we define the amplitudes $b_i(\eta,D)$
through:
\bea
 \int_q q^{4-D- 2\eta} |\kb - \q|^{-(4-\eta)}
k_{\alpha}(k_\beta-q_{\beta}) (W_i)_{\alpha\beta,\gamma\delta}(\q) k_{\gamma} (k_\delta-q_{\delta})
= b_i(\eta,D) k^{4-\eta}\;.
\label{(3.15)}
\eea
%where $Q_i$ are normalisation factors $Q_1=\frac{1}{2} (D-2)(D+1)$, $Q_2=D-1$, $Q_3=Q_5=1$, $Q_4=1/2$. 
The explicit calculation in the Appendix \ref{app:crumpling}
gives
\begin{eqnarray}
&&b_1=B(D-2)(D+1)\;,\;\;
b_2=-B{(D-1)(D^2-4+2 \eta) \over {D-2+\eta}}\;,\;\;
b_3=B(D+1)\;,\\
&&b_4=2B\sqrt{D-1}(2 \eta-3)\;,\;\;
b_5={B \over {D-2+\eta}}
(-22+15D-2D^2+43\eta-16D\eta-24\eta^2+4D\eta^2+4 \eta^3)\;,\nonumber
\end{eqnarray}
where
%=
%{ \Gamma(\eta/2) \Gamma(2-\eta) \Gamma(D/2+\eta/2)
%\over { 4 (4\pi)^{D \over{2}}  \Gamma(2-\eta/2) \Gamma(\eta+D/2) \Gamma(D/2 - \eta/2 +2 )}}
\begin{equation}
B = \frac{\Sigma(\eta,D)}{D^2-1}\;,
\label{(3.16)}
\end{equation}
with $\Sigma(\eta,D)$ defined in
\eqref{Sigma_etaetap},\eqref{SigmaResult} and derived in Appendix
\ref{FinalResults}.

In the limit $\q \to 0$ we find
\bea
&& \tilde w_1(\q) \simeq \frac{1}{\pi_1(\q)}\;,\quad \tilde w_2(\q) \simeq \frac{1}{\pi_2(\q)}\;,\\
&& \begin{pmatrix} \tilde{w}_3(\q) & \tilde{w}_4(\q)\;,\\
\tilde{w}_4(\q) & \tilde{w}_5(\q) \end{pmatrix} 
\simeq \begin{pmatrix} \pi_3(\q) & \pi_4(\q) \\ \pi_4(\q) &
  \pi_5(\q) \end{pmatrix}^{-1}\;.
\label{inverse1} 
\eea 
Substituting Eqs.\eqref{(3.11)},\eqref{(3.14)},\eqref{(3.16)}
into \eqref{(3.12)} we see that factors of $Z_\kappa$ cancel and we find the
self-consistent equation:
\begin{equation}
\frac{d}{2} = \sum_{i=1,2} {b_i(\eta,D) \over a_i(\eta,D) } + { b_3(\eta,D)a_5(\eta,D)-b_4(\eta,D)a_4(\eta,D)
+b_5(\eta,D)a_3(\eta,D) \over {a_3(\eta,D) a_5(\eta,D) - a_4(\eta,D)^2 }}\;.
\label{(3.17)}
\end{equation}

Putting everything together, after considerable simplifications, the
equation determining the crumpling transition exponent
$\eta=\eta_{cr}(D,d)$ is found to be:
\bea
d={ { D(D+1)(D-4+\eta)(D-4+2\eta)(2D-3+2 \eta)\Gamma[{1\over 2}\eta] \Gamma[2-\eta] \Gamma[\eta+D] \Gamma[2-{1\over 2}\eta] }
\over {2(2-\eta)(5-D-2\eta)(D+\eta-1)\Gamma[{1\over 2}D + {1\over 2}\eta] \Gamma[2-\eta-{1\over 2}D] 
\Gamma[\eta+{1\over 2}D] \Gamma[{1\over 2}D+2-{1\over 2}\eta]}}\;,
\label{(3.18)}
\eea
which in $D=2$ reduces to finding the root of a cubic equation
\bea \label{cr2} 
d= \frac{24 (\eta -1)^2 (2 \eta +1)}{(\eta -4) \eta  (2 \eta -3)}\;.
\eea
For $d=3$ and $D=2$ we find 
\bea
\eta_{cr}(2,3) =0.5352..
\eea
At the crumpling transition the radius of gyration of the membrane scales as
\bea
\langle (\vec r(L) - \vec r(0))^2 \rangle = R_G^2 \quad , \quad R_G\sim L^{\nu_G} \quad , \quad 
\nu_G=\frac{4-D-\eta_{cr}}{2}\;, 
\eea 
which leads for $d=3$, $D=2$ to the radius of gyration exponent and the Haussdorf dimension, respectively,
\bea
\nu_G=0.7324\;,\quad d_H=D/\nu_G = 2.7308.
\eea 

To compare our prediction with the result of the $1/d$-expansion 
we expand it to lowest order in $1/d$, 
\bea
&& \eta_{cr}(D,d) \simeq \frac{C(D)}{d} + O(1/d^2)\;,\;\;\text{where} \\
&& C(D) = \frac{(D-4)^2 (2 D-3) \Gamma (D+2)}{2 (5-D) (D-1)
   \Gamma \left(2-\frac{D}{2}\right) \Gamma
   \left(\frac{D}{2}+2\right) \Gamma
   \left(\frac{D}{2}\right)^2}\;,\nonumber
\eea
which can be shown to coincide with that of Ref.\;\onlinecite{AL}. 
We can also expand our SCSA prediction in $\epsilon = 4-D$, finding
\bea 
\eta_{cr}(D,d) \simeq \frac{25}{3 d} (4-D)^3 + O((4-D)^3),
\eea
%x
though it cannot be checked against a previous $\epsilon$-expansion as
$\eta_{cr}$ was neglected in Ref.\;\onlinecite{PKN}, consistent with
its vanishing to $O(\epsilon)$.

It is also interesting to determine the embedding dimension $d_u(D)$
above which self-avoidance is ${\it irrelevant}$ for the membrane
${\it at}$ the crumpling transition. The self-avoiding membrane
interaction, that we have so far neglected, is proportional to $\int
d^D x d^D x' \delta^{(d)}(\vec r(x)-\vec r(x'))$.  Hence by power
counting it scales as $L^{2D - d \nu_G}$ with membrane's internal
dimension $L$. It is thus expected to be irrelevant when
\be
d > 2 D/\nu_G = 4D/(4-D-\eta_{cr}(D,d))  \quad \Leftrightarrow \quad
\text{self-avoidance {\em irrelevant}}.
\ee
Solving this equation, using \eqref{cr2}, for $D=2$ amounts to solving
the following equation for $\eta$:
\bea
d_u(2) = \frac{8}{2-\eta} =  \frac{24 (\eta -1)^2 (2 \eta +1)}{(\eta -4) \eta  (2 \eta -3)}\;.
\eea 
There are four roots with only physical one given by $\eta=0.3956$.
%and $\eta=1.3672$, but can
%discard the second one as unphysical, and 
Hence we find
\be
d_u(2) = 4.986\;.
\ee
This shows that for a physical membrane, $d=3$, self-avoidance is {\em
  relevant} at the crumpling transition.  Indeed the scaling dimension
of the self-avoiding interaction of a physical membrane is $2 D - d
\nu_G = 4 - 3 \nu_G = 1.803$, namely, strongly relevant at long scales.

One can also obtain the lower-critical dimension $D_{lc,cr}(d)$ for
the crumpling transition, defined by $2-\eta(D_{lc,cr},d)=D_{lc,cr}$,
when the correction to the critical temperature is driven to negative
infinity. From \eqref{(3.18)} we find the equation
$d=D_{lc,cr}/(2-D_{lc,cr})$, or, equivalently,
$D_{lc,cr}(d)=2d/(1+d)$. Since $d=D/(2-D)$ is equivalent to
$d_c=D(D-1)/(2-D)$ we find that the lower-critical dimensions of both
the crumpling transition and the flat phase (see \eqref{Dlc}) fixed points
are identical. Although originating from an approximation on different
theories, and resulting from a priori very different calculations, we
find $D_{lc}(d_c+D_{lc}(d_c))=D_{lc}(d_c)$, for arbitrary $d_c$,
which shows that the SCSA is quite consistent.

%{\red to assess the validity of this text: It is interesting to note that the %above discussion shows explicitly, and exactly to first order 
%in $1/d$,
%that the field theory of the crumpling
%transition is not renormalizable in the usual sense and that the analysis
%of Packzuski, Kardar and Nelson in \cite{PKN} may be incorrect. For large $d$ %we should recover
%the result of \cite{PKN} where a fixed point was found to order $\epsilon$
%for $d>219$. This is not the case however, because within the SCSA new %four-points couplings are
%generated by renormalization, which were not considered in \cite{PKN}.}
%To first order in $1/d$, one can keep only the polarization loops
%(which should give the contribution proportional to $d$ in the RG flow
%equations (5a-5b) of \cite{PKN}) and one already needs five coupling
%constants relevant below $D=4$. 

To first order in $\epsilon$ from the above result, \eqref{(3.14)},
neglecting $\eta = O(\epsilon^3)$, we find
\be
a_1=a_2=2 A\;,\;\;a_3=5A\;,\;\;a_4=\sqrt{3}A\;,\;\;a_5=3A\;,
\label{epsA} 
\ee
where
\begin{equation}
A = \frac{1}{192 \pi ^2 \epsilon}\;.
\end{equation}
%** old result 
%\be
%A={1 \over {3 \epsilon}}~~~a_1=a_2=A~~a_3=5A~~a_4=\sqrt{3}A~~a_5=3A
%\label{epsA} 
%\ee
Using expression \eqref{(3.13)} for $\pi_i(\q)\sim a_i$ and 
\eqref{inverse1} we see that to this order, near $D=4$,
\bea
&& \tilde w_1(\q)\simeq \tilde w_2(\q) 
\simeq\frac{Z_\kappa^2 q^{\epsilon-2\eta}}{2A}\\
&& \begin{pmatrix} \tilde{w}_3(\q) & \tilde{w}_4(\q) \\ \tilde{w}_4(\q) & \tilde{w}_5(\q) \end{pmatrix} 
\simeq\frac{Z_\kappa^2 q^{\epsilon-2\eta}}{A}
\left(
\begin{array}{cc}
 \frac{1}{4} & -\frac{1}{4 \sqrt{3}} \\
 -\frac{1}{4 \sqrt{3}} & \frac{5}{12} \\
\end{array}
\right)\;.
\eea
Remarkably, comparing with \eqref{(3.7)}, or equivalently computing
$\tilde R(\q)$ in \eqref{renR} we see that (to this lowest order in
$\epsilon$) the renormalized vertex recovers the tensor structure in
\eqref{(3.3)}, parameterizable by only two renormalized Lam\'e moduli
(as in the bare theory with local elasticity), with 
\begin{equation}
\tilde \mu_0(\q) = 96\pi^2\epsilon Z_\kappa^2 q^{\epsilon -
  2\eta}=-3\tilde \lambda_0(\q).
\end{equation}
These renormalized moduli agree {\em exactly} with those derived from
the $d\rightarrow\infty$ limit of the fixed-point couplings at the
crumpling transition critical point, studied to lowest-order in
$\epsilon$-expansion in Ref.\;\onlinecite{PKN} (after a correction to
their constant $K_4$, taking it to be $S_4/(2\pi)^4 = 1/(8\pi^2)$
rather than $1/(2\pi^2)$). However, in contrast to this
$\epsilon$-expansion (valid near $D=4$), that found a
fluctuation-driven first-order transition when extended to a physical
membrane ($D=2$ and $d = 3 < 219$), our SCSA analysis predicts a
continuous crumpling transition, that survives in a physical membrane,
as in other applications of SCSA.\cite{SCtransitionLR}

Based on the above ratio of the renormalized long wavelength moduli
$\tilde\lambda_0({\bf q})/\tilde\mu_0({\bf q}) \simeq - 1/3$, (found
to hold to leading order in $\epsilon$ in the limit of large $d$) we
predict a negative universal Poisson ratio at the crumpling transition
of a phantom membrane near the upper critical dimension
\bea
\label{PoissonRatioCrumple} 
\tilde\sigma_c({\bf q}) &=& \frac{\tilde\lambda_0({\bf q})}
{2\tilde\mu_0({\bf q}) + \tilde\lambda_0({\bf q})} 
|_{D\simeq 4,d \to +\infty} = - \frac{1}{5}\;.
\eea

\section{Flat phase of an elastic membrane with quenched
  disorder}
\label{MembraneDisorder}

We now turn our study to a flat phase of a heterogeneous elastic
membrane. Motivated by the quantitative success of the SCSA when
applied to a disorder-free homogeneous membrane, here we extend this
calculation method to a treatment of a membrane with quenched
disorder. This will complement earlier studies of Ref.\cite{RNpra,ML}
using the renormalization group, controlled by an $\epsilon$-expansion.

\subsection{Effective flat-phase model of a heterogeneous elastic
  membrane}

We recall the model of a disordered membrane introduced in Section \ref{sec:disM}. It
is defined by free energy Eq.\eqref{FDisorder}
\begin{eqnarray}\label{DFree}
{\F}[\h,u\as]&=&\intd\left[{1\over2}
\;\kappa(\pt^2\h)^{2} + 
\mu (u\ab)^2 + {1\over2}\lambda(u\aas)^2 -\kappa\pt^2\h\cdot\f(\x)-2\mu u\ab c\ab(\x) - 
\lambda u\aas c\bbs(\x)\right]\; ,
\end{eqnarray}
with random zero-mean Gaussian stress and curvature disorder, 
fully characterized by the correlators Eq.\eqref{DisorderCorr}
\begin{eqnarray}\label{DCorr}
\overline{c\ab(\x) c\cd(\x^\prime)}&=&
\hat \Delta_{\lambda}(\x-\x^\prime) \delta\ab \delta\cd
+ \hat \Delta_{\mu}(\x-\x^\prime)\left (\delta\ac\delta\bd +
\delta\ad\delta\bc\right)\;,\\
\overline{f_i(\x) f_j(\x^\prime)}&=&\delta_{i j}
\hat \Delta_{\kappa}(\x-\x^\prime)\; .
\end{eqnarray}
As for the homogeneous membrane in Sec.\ref{PureMembrane} it is useful
to integrate out the in-plane phonon degrees of freedom $u\as(\x)$,
that can be done exactly as they appear only linearly and
quadratically in the free energy above. We thereby obtain the
extension of the effective free energy in Eq.\eqref{FflatEff} to
disordered membranes

\begin{eqnarray}\label{FDflatEff}
{\F}[\h]&=&\intd\left\{{1\over2}\kappa(\partial^2\h)^{2} + 
{1\over4}
\left[\mu \left(P^T\ad\pt\as\h\cdot\pt\bs\h\right) \left(P^T\bc\pt\cs\h\cdot\pt\ds\h\right)
+{\mu\lambda\over 2\mu+\lambda}\left(P^T\ab\pt\as\h\cdot\pt\bs\h\right)^2
\right]\;,\right.
\\
&&\left.-\kappa\pt^2\h\cdot\f(\x)
- \left[\mu \left(P^T\ad\pt\as\h\cdot\pt\bs\h\right) \left(P^T\bc c\cd(\x)\right)
+{\mu\lambda\over 2\mu+\lambda}\left(P^T\ab\pt\as\h\cdot\pt\bs\h\right)
\left(P^T\cd c\cd(\x) \right)
\right] \right\}\;.
\end{eqnarray}

A system with quenched disorder is characterized by a distribution
function of physical observables. Averages of an observable over
possible realizations of disorder will be very close to its typical
value if the corresponding distribution function is narrowly
distributed. It turns out that the distribution function for the free
energy density has a vanishing variance in the thermodynamic limit,
while the partition function ${\cal Z}$ is widely
distributed. Therefore in a quenched disordered system, like the
membrane considered here, the average and also sample representative,
typical free energy is given by
\begin{equation}\label{aveF}
F=%\overline{F_d[\f(\x),c\ab(\x)]}=
 -T\overline{\ln {\cal Z}[\f(\x),c\ab(\x)]}\;,
\end{equation}
rather than by a logarithm of the disorder-average of the
partition function $\ln\overline{{\cal Z}}$, as it would be for an
annealed form of disorder. In above equations the overbar denotes the
average over disorder realizations. The difficulty in computing the
average of $\ln {\cal Z}$ can be handled in the standard way using the
replica `trick' \cite{EdwardsAnderson} (although there are alternative
methods \cite{NRpra}). Using the identity
\begin{equation}\label{identity}
\ln {\cal Z}=\lim_{n\rightarrow 0}{{\cal Z}^n-1\over n}\;,
\end{equation}
and assuming the validity of the interchange of the thermodynamic and
$n\rightarrow 0$ limits, we reduce a difficult average over a
logarithm into an average of an $n$-replicated theory represented by
${\cal Z}^n$. In the replicated theory described by $n$ %annealed
height fields $\h_a(\x)$, $(a=1,2,\ldots,n)$, the effective %annealed
averages over disorder fields $\f(\x)$ and $c\ab(\x)$ (which couple
linearly) can now be easily performed. The effective replicated free
energy takes the form of a critical (massless) $O(d_c) \times O(N)$
symmetric theory, with a nonlocal, tensor quartic interaction,
generalizing Eq.\eqref{Feff} to a disordered membrane\cite{RNpra,LRprl}
%As we have seen in the study of homogeneous membranes in Sec.\ref{PureMembrane}
%the SCSA is easiest to implement when the free energy has the form of 
%an $O(d_c)$ model, (with a possible tensor product extension). 
%
\be \label{FDeff}
\F^r_{eff}[\h\as(\kb)]={1\over 2} 
\int_k\;\kappa_{ab}(\kb)  k^4\h^a\as(\kb)\cdot\h^b\bs(-\kb) +{1\over4 d_c}\int_{k_1, k_2, k_3} R^{ab}_{\alpha \beta, \gamma \delta}(\q)
k_{1\alpha} k_{2\beta} k_{3\gamma} k_{4\delta}\; \h_a(\kb_1)\cdot\h_a(\kb_2)\ 
\h_b(\kb_3)\cdot\h_b(\kb_4)\;,
\ee
with $\q=\kb_1+\kb_2$ and $\kb_1+\kb_2+\kb_3+\kb_4={\bf 0}$, and sum
over repeated indices is implied throughout. As in the pure case we
have rescaled uniformly by a factor $1/d_c$ all quartic terms (which
is immaterial for $d_c=1$) in order to obtain a well defined large
$d_c$ limit. Let us now describe the different terms in
\eqref{FDeff}. For convenience we define 
\bea \Delta_\kappa(\q) =
\kappa^2 \hat \Delta_\kappa(\q) \quad , \quad \Delta_\mu(\q)= 4\mu^2
\hat \Delta_\mu(\q) \quad , \quad \Delta_\lambda(\q)= 4\mu^2 \hat
\Delta_\lambda(\q)\ .
\eea
%For notational convenience and in anticipation of 
%fixed points of order $1/d_c$ we have rescaled the $\h_a(\x)$ fields 
%and the coupling constants
%%
%\begin{eqnarray}\label{Defs}
%\h_a(\kb)\h_b(-\kb)/\T&\rightarrow& \h_a(\kb)\h_b(-\kb)\;,\\
%\Delta_\kappa(\q)\kappa^2/\T&\rightarrow& \Delta_\kappa(\q)\;,\\
%\Delta_\mu(\q)4\mu^2d_c&\rightarrow& \Delta_\mu(\q)\;,\\
%\Delta_\lambda(\q)4\mu^2d_c
%&\rightarrow&\Delta_\lambda(\q)\;,\\
%\mu\T d_c&\rightarrow& \mu\;,\\
%\lambda\T d_c&\rightarrow& \lambda\;.
%\end{eqnarray}
%%
The replica matrix $\kappa_{ab}(\kb)$ is the bending rigidity analog
in the replicated theory and has the form,
\begin{equation}\label{Rigidity}
\kappa_{ab}(\kb)=\kappa\delta_{ab}-\frac{\Delta_\kappa(\kb)}{T} J_{ab}\;,
\end{equation}
where $J_{ab}$ is an $n\times n$ matrix with all entries equal to
$1$. Here we will only be interested in the replica symmetric
solutions, with all correlators invariant under permutation of replica
indices. In this subspace all the replica matrices can be represented
as a two component supervector, with the first and second component
being the coefficients of $\delta_{ab}$ and $J_{ab}$, respectively.
In this notation
\begin{equation}\label{RNew}
\kappa_{ab}(\kb)=\left(\kappa\;,\; - \frac{\Delta_\kappa(\kb)}{T} \right)\;.
\end{equation}
Using the definition in terms of the original matrices in replica
space, the replica matrix product $u_{ab} u_{bc}$ and the replica
matrix inversion in the new vector space is easily defined and in the
$n\rightarrow 0$ limit is given by,
\begin{eqnarray}\label{ProdInverse}
u_{ad} v_{db} = (u_c\;,u\;)(v_c\;,\;v)&=&(u_c v_c\;,\; u v_c + u_c v )\;,\\
(u^{-1})_{ab} = (u_c\;,\;u)^{-1}&=&(1/u_c\;,\;-u/u_c^2)\;.
\end{eqnarray}

The bare four-point coupling tensor
$R^{ab}_{\alpha\beta,\gamma\delta}(\q)$ is an $n$-replica
generalization of the quartic interaction of the pure membrane,
Eq.\eqref{RtensorAB} to the disordered case,
\begin{equation}\label{Rtensor}
R^{ab}_{\alpha\beta,\gamma\delta}(\q)=
{1\over2}\mu_{ab}(\q)\left(P^T\ac(\q)P^T\bd(\q)+P^T\ad(\q)P^T\bc(\q)\right)
+\rho_{ab}(\q)P^T\ab(\q)P^T\cd(\q)\;,
\end{equation}
where in our notation the replicated elastic moduli tensors are 
\begin{eqnarray}\label{PPcoeff}
\mu_{ab}(\q)&=&\left(\mu\;,\; -\frac{1}{T} \Delta_\mu(\q)  \right)\;,\\
\rho_{ab}(\q)&=&\left({\mu\lambda\over 2\mu+\lambda}\;,\; 
-{\lambda(4\mu+(D+1)\lambda)\over T(2\mu+\lambda)^2}\Delta_\mu(\q)
-\frac{1}{2T} \left({2\mu+D\lambda\over2\mu+\lambda}\right)^2
\Delta_\lambda(\q)\right)\;.
\end{eqnarray}

\subsection{The SCSA equations for the heterogeneous membrane}

We next present the SCSA equations for the flat phase of an
  elastic membrane with quenched internal disorder using the effective
  replicated free energy $\F^r_{eff}[\h\as(\kb)]$, (\ref{FDeff}),
  derived above.  As for the disorder-free case it is more convenient
to work in the orthogonal representation of tensors
$M_{\alpha\beta,\gamma\delta}(\q)$ and
$N_{\alpha\beta,\gamma\delta}(\q)$ defined in Eq.\eqref{MNtensors}. In
terms of these the quartic interaction becomes,
\begin{equation}\label{RtensorMNhetero}
R^{ab}_{\alpha\beta,\gamma\delta}(\q)=
\mu_{ab}M_{\alpha\beta,\gamma\delta}(\q)
+b_{ab}N_{\alpha\beta,\gamma\delta}(\q)\;,
\end{equation}
where,
\begin{eqnarray}\label{MNcoeff}
\mu_{ab}(\q)&=&\left(\mu\;,\; - \frac{1}{T} \Delta_\mu(\q)\right)\;,\\
b_{ab}(\q)&=&\left(b\;,\; - \frac{1}{T} \Delta_b(\q)\right)\;,\\
&=&\left({\mu(2\mu+D\lambda)\over 2\mu+\lambda}\;,\;
-\frac{1}{2 T} \left({2\mu+D\lambda\over2\mu+\lambda}\right)^2
\left[2\Delta_\mu(\q)+(D-1)\Delta_\lambda(\q)\right]\right)\;,
\end{eqnarray}
which defines $\Delta_b(\q):= \frac{1}{2} \left({2\mu+D\lambda\over2\mu+\lambda}\right)^2
(2\Delta_\mu(\q)+(D-1)\Delta_\lambda(\q))$.

As described in Sec. \ref{sec:disM} the roughness of the membrane in
the flat phase is determined by two correlation functions of the
height-field, $\h(\kb)$,
\begin{eqnarray}
\label{hh}
\overline{\langle h^i(\kb) h^j(\kb')\rangle}-\overline{\langle
h^i(\kb)\rangle\langle h^j(\kb')\rangle}&=&G_\kappa(\kb)
\delta_{ij}(2\pi)^D\delta^{(D)}(\kb+\kb')\;,\\
\label{hh_d}
\overline{\langle h^i(\kb)\rangle\langle h^j(\kb')\rangle}&=&G_\Delta(\kb)
\delta_{ij}(2\pi)^D\delta^{(D)}(\kb+\kb')\;,
\end{eqnarray}
where the first connected correlation function describes thermal
roughness and the thermally disconnected piece corresponds to zero
temperature roughness due to disorder. It is easy to show that in the
replicated theory in our notation the two components of the replicated
correlator
\begin{eqnarray}\label{Rcorr}
\lim_{n\rightarrow 0}
\langle h^i_a(\kb) h^j_b(\kb')\rangle&=&
G_{ab}(\kb)\delta_{ij}(2\pi)^D\delta^{(D)}(\kb+\kb')\;,\\
&=&\left(G_{\kappa}(\kb) \;,\;G_{\Delta}(\kb) \right)
\delta_{ij}(2\pi)^D\delta^{(D)}(\kb+\kb')\;,
\end{eqnarray}
correspond to the correlators defined in Eqs.\eqref{hh},\eqref{hh_d}. For
these correlators one can define the renormalized bending rigidity
$\tilde \kappa(\kb)$ and renormalized curvature disorder $\tilde
\Delta_\kappa(\kb)$ as follows
\bea
G_\kappa(\kb)= \frac{T}{\tilde \kappa(\kb) k^4} \quad , \quad G_\Delta(\kb) = \frac{\tilde \Delta_\kappa(\kb)}{\tilde \kappa(\kb)^2 k^4}. \label{renormG} 
\eea 
For notational convenience, in this Section and the next two, we will
denote the renormalized parameters (obtained from SCSA) with a tilde,
instead of the subscript $R$ as in the previous sections, but these
denote the same quantities, e.g., $\tilde \kappa(\q)=\kappa_R(\q)$.

To study the physics of the flat phase of a disordered membrane we set
up two coupled tensor integral equations for the renormalized
propagator $G_{ab}(\kb)$ and the renormalized four point interaction
$\tilde{R}^{a b}_{\alpha \beta, \gamma \delta}(q)$, generalizing the
SCSA method to the replicated theory. The $1/d_c$-expansion for
the quartic vertex and two-point height correlator, when made
self-consistent, gives 
\begin{eqnarray}
\label{SCSAeqnG}
 T (G^{-1})_{ab}(\kb)&=& \kappa_{a b}(\kb) k^4 +
{2 \over d_c} k_{\alpha}k_{\beta}k_{\gamma}k_{\delta}
\int_q \tilde{R}^{a b}_{\alpha \beta, \gamma \delta}(\q) G_{a b}(\kb-\q)\;,\\
\tilde{R}^{a b}_{\alpha \beta, \gamma \delta}(\q)&=& 
R^{a b}_{\alpha \beta, \gamma \delta}(\q) - 
R^{a c}_{\alpha \beta, \gamma_1 \gamma_2}(\q) 
\hat \Pi^{c d}_{\gamma_1 \gamma_2, \gamma_3 \gamma_4}(\q)
\tilde{R}^{d b}_{\gamma_3 \gamma_4, \gamma \delta}(\q)\;, \label{SCSAeqnR2} 
\end{eqnarray}
where we have defined the replicated polarization bubble matrix
\bea
\label{Polarization_ab}
\hat \Pi^{a b}_{\alpha\beta,\gamma\delta}(\q)
&=&\frac{1}{T} \int_p p_{\alpha}p_{\beta}p_{\gamma}p_{\delta}G_{a b}(\pb) 
G_{a b}(\q-\pb)\;,
\eea
which in our compact two component notation reads 
\bea 
\left(\Pi_1^{\alpha\beta,\gamma\delta}(\q)\;,
\;\Pi_2^{\alpha\beta,\gamma\delta}(\q)\right)
&=& \frac{1}{T}\int_p p_{\alpha}p_{\beta}p_{\gamma}p_{\delta}
\big(G_\kappa(\pb)G_\kappa(\q-\pb)+G_\kappa(\pb)G_\Delta(\q-\pb)\\
&&+G_\kappa(\q-\pb)G_\Delta(\pb)\;,\;G_\Delta(\pb)G_\Delta(\q-\pb)\big)
\;,
\eeq
%
%The above is the vacuum polarization bubble in Fig. 
Note that the sum over repeated indices is implied except for the
equation \eqref{Polarization_ab} for $\hat \Pi^{a
  b}_{\alpha\beta,\gamma\delta}(\q)$, where indices $a,b$ are not
summed.

As for the homogeneous membrane, $\hat \Pi^{a b}_{\alpha\beta,\gamma\delta}(\q)$
has a following tensor structure (see Appendix A, for more details)
\begin{equation}\label{TensorP}
\hat \Pi^{a b}_{\alpha\beta,\gamma\delta}(\q)=
\hat \Pi_{a b}^{sym}(\q) S_{\alpha\beta,\gamma\delta}+
\hat \Pi_{a b}^{long1}(\q)\left(\delta\ab q\cs q\ds+
\delta\cd q\as q\bs\right)+
\hat \Pi_{a b}^{long2}(\q) q\as q\bs q\cs q\ds\;,
\end{equation}
where
\begin{equation}\label{Ssymm}
S_{\alpha \beta, \gamma \delta}=
\delta_{\alpha \beta} \delta_{\gamma \delta} +
\delta_{\alpha \gamma} \delta_{\beta \delta}+
\delta_{\alpha \delta} \delta_{\beta \gamma}
\end{equation}
is a fully symmetric tensor previously defined. Because the quartic
interaction is a transverse projector in indices $\alpha,\beta,\gamma,\delta$
upon tensor multiplication in Eq.\eqref{SCSAeqnR2} only the symmetric part of 
$\hat \Pi^{a b}_{\alpha\beta,\gamma\delta}(\q)$ survives
\begin{eqnarray}\label{TensorPsym}
\hat \Pi_{ab}^{sym}(\q)
&=&\big(\Pi_1^{sym}(\q)\;,\;\Pi_2^{sym}(\q)\big)\;,\\
&=&\big(T\Pi_{\kappa\kappa}(\q)+2\Pi_{\kappa\Delta}(\q)\;,\;
\frac{1}{T}\Pi_{\Delta\Delta}(\q)\big)\;,
\end{eqnarray}
where we have defined the integrals
\begin{equation}\label{Pij}
\Pi_{nm}(\q)={1\over (D^2-1) T^{a_{n,m}}}\int_p\left(p\as P^T\ab(\q) p\bs\right)^2
G_n(\pb) G_m(\q-\pb)\;,
\end{equation}
with $n,m=\kappa,\Delta$ and the $a_{n,m}$ are chosen as
$a_{\kappa,\kappa}=2$, $a_{\Delta,\kappa}=a_{\kappa,\Delta}=1$ and
$a_{\Delta,\Delta}=0$, so that the $\Pi_{nm}(\q)$ have well defined
$T=0$ limits.

Using the above results together with the orthogonality property of tensors 
$M_{\alpha\beta,\gamma\delta}(\q)$ and $N_{\alpha\beta,\gamma\delta}(\q)$
from which the quartic vertex is composed, we find again that
the renormalized interaction can be written as
\begin{equation}\label{RtensorMNhetero2}
\tilde R^{ab}_{\alpha\beta,\gamma\delta}(\q)=
\tilde \mu_{ab}(\q) M_{\alpha\beta,\gamma\delta}(\q)
+\tilde b_{ab}(\q) N_{\alpha\beta,\gamma\delta}(\q)\;,
\end{equation}
where
\begin{eqnarray}\label{MNcoeff2}
\tilde \mu_{ab}(\q)&=&\left(\tilde \mu(\q) \;,\; - \frac{1}{T} \tilde \Delta_\mu(\q)\right)\;,\\
\tilde b_{ab}(\q)&=&\left(\tilde b(\q)\;,\; - \frac{1}{T} \tilde \Delta_b(\q)\right)\;,
\end{eqnarray}
and $\tilde \mu(\q)$ and $\tilde b(\q)$ are the renormalized shear and
bulk moduli (denoted by $\mu_R(\q)$ and $b_R(\q)$ in the previous
sections) and $\tilde \Delta_\mu(\q)$ and $\tilde \Delta_b(\q)$ are
the associated renormalized disorder variances. To determine these
renormalized couplings we note that the tensor equation
(\ref{SCSAeqnR2}) is then equivalent to two independent replica matrix
equations for $\tilde{\mu}_{ab}(\q)$ and $\tilde{b}_{ab}(\q)$

%$\mathbb{1}$ problem for Leo's latex
\begin{eqnarray}\label{bmu}
\tilde{\mu}_{ab}(\q)&=& 
\bigg(\hat{1}  + 2 \, \hat \mu(\q) \cdot
\hat\Pi^{sym}(\q)\bigg)_{ac}^{-1}\mu_{cb}\;,\\
\tilde{b}_{ab}(\q)&=&
\bigg(\hat{1}  + (D+1) \, \hat b(\q) \cdot  
\hat \Pi^{sym}(\q)\bigg)_{ac}^{-1}b_{cb}\;,
\end{eqnarray}
%
%{\red in the final iteration replace the tensors above back to mathbb
%  macro of Pierre's version, because somehow they does not work for
%  equations above with my latex}
%BAD
%
%\begin{eqnarray}\label{bmu}
%\tilde{\mu}_{ab}(\q)&=& 
%\bigg(\mathbb{1}  + 2 \, \hat \mu(\q) \cdot \hat %\Pi^{sym}(\q)\bigg)_{ac}^{-1}\mu_{cb}
%\;,\\
%\tilde{b}_{ab}(\q)&=&
%\bigg(\mathbb{1}  +(D+1) \, \hat b(\q) \cdot  \hat %\Pi^{sym}(\q)\bigg)_{ac}^{-1}b_{cb}
%\;,
%\end{eqnarray}
%
%BAD 
where the hat denotes replica matrices, and replica matrix
multiplication and inversion are implied.

We now use the multiplication and inversion rules that hold within our
replica symmetric subspace, Eq.\eqref{ProdInverse}, to further reduce
the above two matrix equations to four scalar equations for
$\tilde{\mu}(\q)$, $\tilde{b}(\q)$, $\tilde{\Delta}_\mu(\q)$ and
$\tilde{\Delta}_b(\q)$ (whose bare values are defined in
Eq.\eqref{MNcoeff}).  Beginning with $\tilde{\mu}_{ab}(\q)$ we obtain
\begin{eqnarray}\label{ReduceMU}
\tilde{\mu}_{ab}(\q)&=&\left(\tilde{\mu}(\q)\;,\;
- \frac{1}{T} \tilde{\Delta}_\mu(\q)\right)\;,\\
&=&\bigg(1+2\mu\Pi_1^{sym}(\q)\;,\;2\mu\Pi_2^{sym}(\q)
-\frac{2}{T} \Delta_\mu(\q)\Pi_1^{sym}(\q)\bigg)^{-1}\
\cdot \bigg(\mu\;,\;- \frac{1}{T} \Delta_\mu(\q)\bigg)\;,\\
&=&\left({1\over1+2\mu\Pi_1^{sym}(\q)}\;,\;-{2\mu\Pi_2^{sym}(\q)
- \frac{2}{T} \Delta_\mu(\q)\Pi_1^{sym}(\q)\over\left(1+2\mu\Pi_1^{sym}(\q)\right)^2}
\right)\ \cdot \bigg(\mu\;,\;- \frac{1}{T} \Delta_\mu(\q)\bigg)\;,\\
&=&\left({\mu\over1+2\mu\Pi_1^{sym}(\q)}\;,\;
-{\frac{1}{T}  \Delta_\mu
\over1+2\mu\Pi_1^{sym}(\q)}
-{2\mu^2\Pi_2^{sym}(\q)-2\mu \frac{1}{T}  \Delta_\mu(\q)\Pi_1^{sym}(\q)
\over\left(1+2\mu\Pi_1^{sym}(\q)\right)^2}
\right)\;,\\
&=&\left({\mu\over1+2\mu
\left[T\Pi_{\kappa\kappa}(\q)+2\Pi_{\kappa\Delta}(\q)\right]}\;,\;
- \frac{1}{T} {\Delta_\mu+2\mu^2 T\Pi_{\Delta\Delta}(\q)\over
\left(1+2\mu\left[
T\Pi_{\kappa\kappa}(\q)+2\Pi_{\kappa\Delta}(\q)\right]\right)^2}
\right)\;.
\end{eqnarray}
Identical manipulations give,
\begin{eqnarray}\label{ReduceB}
\tilde{b}_{ab}(\q)&=&\left(\tilde{b}(\q)\;,
\;- \frac{1}{T} \tilde{\Delta}_b(\q)\right)\;,\\
&=&\left({b\over1+(D+1)b
\left[ T \Pi_{\kappa\kappa}(\q)+2\Pi_{\kappa\Delta}(\q)\right]}\;,\;
- \frac{1}{T} { \Delta_b+(D+1)b^2\Pi_{\Delta\Delta}(\q)
\over\big(1+(D+1)b\left[
T \Pi_{\kappa\kappa}(\q)+2\Pi_{\kappa\Delta}(\q)\right]\big)^2}
\right)\;.
\end{eqnarray}

Having determined the renormalized four-point coupling constants we
turn to Eq.\eqref{SCSAeqnG} to determine the two components of the
propagator, $G_\kappa(\kb)$ and $G_\Delta(\kb)$, which will in turn
determine the roughness exponents $\zeta$ and $\zeta'$, respectively.
Using the index permutation symmetry of the tensor $k\as k\bs
  k\cs k\ds$ and the definition of $\tilde{R}^{a b}_{\alpha \beta,
    \gamma \delta}(\q)$ in terms of $\tilde{\mu}_{ab}(\q)$ and
  $\tilde{b}_{ab}(\q)$ inside the second term on the right-hand side
  of Eq.\eqref{SCSAeqnG}, we find,
\begin{equation}\label{Simple}
k\as k\bs k\cs k\ds \tilde{R}^{a b}_{\alpha \beta, \gamma \delta}
(\q)=\left(k\as P^T\ab(\q) k\bs\right)^2 \left[{D-2\over D-1}\
\tilde{\mu}_{ab}(\q)+{1\over D-1}\ \tilde{b}_{ab}(\q)\right]\;.
\end{equation}
Substituting this result into Eq.\eqref{SCSAeqnG}, and 
inverting $G_{ab}(\kb)$ on the right-hand side we obtain two
self-consistent scalar equations for $G_\kappa(\kb)$ and
$G_\Delta(\kb)$. These, together with their definition in
\eqref{renormG} give
\begin{eqnarray}
\label{GGeqn}
 \tilde \kappa(\kb)&=&\kappa + 
{2\over d_c}\int_q (\hat k\as P\ab^T(\q) \hat k\bs)^2
\left\{\left[{D-2\over D-1}\
\tilde{\mu}(\q)+{1\over D-1}\ \tilde{b}(\q)\right]
\left[\frac{T}{\tilde \kappa(\kb-\q) (\kb-\q)^4} + \frac{\tilde \Delta_\kappa(\kb-\q)}{\tilde \kappa(\kb-\q)^2 (\kb-\q)^4} \right]\nonumber \right.\\
&&\left.\;\;\;\;\;\;\;\;\;\;\;\;-  \left[{D-2\over D-1}\
\tilde{\Delta}_\mu(\q)+{1\over D-1}\
\tilde{\Delta}_b(\q)\right] \frac1{\tilde \kappa(\kb-\q) (\kb-\q)^4} \right\}\;,\\
\tilde \Delta_\kappa(\kb) 
&=&\Delta_{\kappa}(\kb) + {2\over d_c}\int_q (\hat k\as P\ab^T(\q) \hat k\bs)^2
\left[{D-2\over D-1}\
\tilde{\Delta}_\mu(\q)+{1\over D-1}\
\tilde{\Delta}_b(\q)\right] \frac{\tilde \Delta_\kappa(\kb-\q)}{\tilde \kappa(\kb-\q)^2 (\kb-\q)^4} \;. \label{GGeqn2}
\end{eqnarray}
Recalling the second set of SCSA equations (obtained above)
\bea 
\label{couplings2} 
&& \tilde{\mu}(\q) = {\mu\over1+2\mu \left[ T
    \Pi_{\kappa\kappa}(\q)+2\Pi_{\kappa\Delta}(\q)\right]}\ ,  \\
\label{couplings2b} 
&& \tilde{b}(\q) = {b\over1+(D+1)b
\left[ T \Pi_{\kappa\kappa}(\q)+2\Pi_{\kappa\Delta}(\q)\right]}\ , \\
\label{couplings2c} 
&& \tilde{\Delta}_\mu(\q) ={ \Delta_\mu(\q)+2\mu^2  \Pi_{\Delta\Delta}(\q)\over
\left(1+2\mu\left[T
\Pi_{\kappa\kappa}(\q)+2\Pi_{\kappa\Delta}(\q)\right]\right)^2}\ , \\
\label{couplings2d} 
&& \tilde{\Delta}_b(\q) =
{ \Delta_b(\q)+(D+1)b^2  \Pi_{\Delta\Delta}(\q)
\over\big(1+(D+1)b\left[
T \Pi_{\kappa\kappa}(\q)+2\Pi_{\kappa\Delta}(\q)\right]\big)^2}\ ,
\eea 
and the three polarization bubbles,
\bea \label{Pijexplicit}
&& \Pi_{\kappa \kappa}(\q)={1\over (D^2-1)}\int_p\left(p\as P^T\ab(\q) p\bs\right)^2
\frac{1}{\tilde \kappa(\pb) p^4 \tilde \kappa(\q-\pb) |\q - \pb|^4 }\ , \\
&& \Pi_{\kappa \Delta}(\q)={1\over (D^2-1)}\int_p\left(p\as P^T\ab(\q) p\bs\right)^2
\frac{\tilde \Delta_\kappa(\pb)}{\tilde \kappa(\pb)^2 p^4 \tilde
  \kappa(\q-\pb) |\q - \pb|^4 }\ , \\
&& \Pi_{\Delta \Delta}(\q)={1\over   (D^2-1)}\int_p\left(p\as P^T\ab(\q) p\bs\right)^2
\frac{\tilde \Delta_\kappa(\pb) 
\tilde \Delta_\kappa(\q-\pb)}{\tilde \kappa(\pb)^2 p^4 \tilde
\kappa(\q-\pb)^2 |\q - \pb|^4 }\ , 
\eea
equations \eqref{GGeqn}, \eqref{GGeqn2},
\eqref{couplings2}-\eqref{couplings2d}, \eqref{Pijexplicit} form a
closed set of SCSA equations for a heterogeneous membrane.  They
reduce to the homogeneous SCSA equations, Eqs. \eqref{renormModuli},
\eqref{SimpleSigma}, \eqref{Iq}, \eqref{propren} for
$\Delta_\kappa(\kb)=\Delta_\mu(\kb)=\Delta_b(\kb)=0$.

One important remark which can be made immediately from these
equations is that curvature disorder, i.e., a non-zero
$\Delta_\kappa(\kb)$, generates a non-zero $\tilde
\Delta_\kappa(\kb)$, hence a non-zero $\Pi_{\Delta\Delta}(\q)$, which
in turn generates a non-zero stress disorder, $\tilde{\Delta}_\mu(\q)
>0$ and $\tilde{\Delta}_b(\q) >0$, even if stress disorder is absent
in the bare model (i.e., even if we take ${\Delta}_\mu(\q) =
\Delta_b(\q)=0$). Hence the three types of disorders must be
considered simultaneously. Note, however, that the converse is not
true. Namely, curvature disorder is {\em not} generated by stress-only
disorder, protected by the up-down, $\h\rightarrow -\h$
symmetry. Thus, the case of stress-only disorder,
$\Delta_\kappa(\q)=0$ is special and requires a separate discussion,
as we do below.

\section{Analysis of the SCSA equations for the heterogeneous
    membrane with short-range disorder} 
\label{sec:analysis} 

We begin by studying the problem of short-range disorder, 
characterized by three bare (curvature and two stress) disorder
variances, that are $\q$ independent at long scales,
\begin{eqnarray}\label{PowerLaw0}
\Delta_\kappa(\q)=\Delta_\kappa\;,\quad\quad \Delta_\mu(\q)=\Delta_\mu \;,
\quad\quad \Delta_\lambda(\q)=\Delta_\lambda \;.
\end{eqnarray}

In principle the SCSA integral equations can be solved numerically, to
obtain the full $q$ dependence of the height correlators
$G_\kappa(\kb),\ G_\Delta(\kb)$ and phonon correlators, on all the
microscopic parameters, $\mu, \lambda, \kappa, \Delta_\kappa(\q),
\Delta_\mu(\q), \Delta_\lambda(\q), T$ and $\kb$.  Again, as in the
homogeneous case, in the large and small $q$ regimes we can explore
the asymptotics analytically. 

\subsection{Perturbative regime of short length, large $q\gg q_{\rm nl}$ scales}

Because of the complexity of the SCSA equations
(\ref{couplings2})-\eqref{couplings2d} there are number of crossover
momentum scales, to which we will generically refer as $q_{\rm
  nl}$. For sufficiently large $q \gg q_{\rm nl}$, where $q_{\rm nl}$
is determined below, we expect that perturbation theory converges and
that the correlators are approximately equal to the bare, Gaussian
fixed point propagators 
\bea
&& G_\kappa(\kb) \approx G^0_\kappa(\kb)=\frac{T}{\kappa k^4}\;, \\
&& G_\Delta(\kb) \approx G^0_\Delta(\kb) =
\frac{\Delta_\kappa(k)}{\kappa^2 k^4}\;, 
\eea
for short-range disorder, $\Delta_\kappa(k) = \Delta_\kappa$,
characterized by the roughness exponents $\zeta=\zeta'=(4-D)/2$.

In this regime from the denominator of Eqs.\eqref{couplings2b} and
from \eqref{Pijexplicit}, we see that the correction to the bulk
modulus at $T=0$ (coming from the screening by frozen out-of-plane
undulations is given by
\bea
&& \frac{\tilde b(\q)- b}{b} \sim - \frac{b \Delta_\kappa}{\kappa^3 q^{4-D}}\;.\\
\eea
The perturbation theory for a heterogeneous membrane at $T=0$ remains
convergent when such correction to the above Gaussian fixed point
description remain small. For the case of physical interest
$D<4$, this regime extends down to $q_{\rm nl,\Delta}$, given by
\beq
q_{\rm nl,\Delta} \sim \left (\frac{b \Delta_\kappa}{\kappa^3}\right)^{1/(4-D)},
\eeq
to be contrasted with the thermal nonlinear crossover wavevector 
in Eq.\eqref{q_nlT}, which we can now denote $q_{\rm nl,T}$
to differentiate it from $q_{\rm nl,\Delta}$.
We note that $q_{\rm nl,\Delta}$ reassuringly vanishes in the
large $\kappa$, small $\Delta_\kappa$ limit, corresponding to a stiff
and/or homogeneous membrane. The nonlinear ($T=0$, disorder-driven) crossover
wavevector scale for the shear modulus is of the same form but with
the bulk modulus, $b$ replaced by $\mu$.

Finally, comparing the two terms in the numerators of Eqs.\eqref{couplings2c}, 
\eqref{couplings2d} and \eqref{Pijexplicit} one finds that 
the corrections to the bare stress disorder $\Delta_{b}$ due to the curvature disorder remains small
as long as
\bea
\frac{b^2 \Delta_\kappa^2}{\kappa^4 q^{4-D}} &\ll& \Delta_b,
\eea
hence there is a distinct length scale, $q_\Delta^{-1}$, above which they cannot be neglected.
The characteristic wavevector is 
\beq
q_{\rm \Delta} \sim \left (\frac{b^2 \Delta_\kappa^2}{\kappa^4 \Delta_b}\right)^{1/(4-D)},
\eeq
where we recall that $2 b = K_0$ is the Young modulus
for a physical membrane $D=2$.
There is a similar condition and wavevector associated to $\Delta_\mu$.

\subsection{Non-perturbative regime of long-length, small $q\ll q_{\rm
    nl}$ scales}
\label{sec:4b} 

We now study the non-perturbative regime of wavevectors smaller than
$q_{\rm nl}$, where the system crosses over to a nontrivial,
fluctuations and nonlinearity-controlled fixed point. At these long
scales the solutions to the SCSA equations are universal, independent
of the microscopic parameters and can be obtained analytically.
Anticipating, as in the case of a homogeneous membrane, that the
solution describes a critical fixed point, we search for a
height-height correlator, the propagator $G(\ks)$, that, for small $k$
is a power-law in $k$
\bea \label{Gscale2}
&& G_\kappa(\ks)=\frac{T}{\tilde \kappa(\ks) k^4} \simeq T Z_\kappa^{-1} k^{-4+\eta}\;
\quad , \quad \tilde \kappa(\ks) \simeq Z_\kappa k^{-\eta}, \\
&& G_{\Delta}(\ks) =  \frac{\tilde \Delta_\kappa(\kb)}{\tilde \kappa(\kb)^2 k^4}  \simeq Z_\Delta^{-1}k^{-4+\eta'}\;
\quad , \quad \tilde \Delta_\kappa(\kb) \simeq Z_{\Delta_\kappa} k^{\eta' - 2 \eta}, \label{Gscale3}
\eea
where $Z_\kappa$ and $Z_\Delta$ are thermal and disorder non-universal
amplitudes, with $Z_\Delta^{-1} = Z_{\Delta_\kappa}/Z_\kappa^2$. The
universal exponents $\eta$ and $\eta'$ also determine (via
\eqref{zetazetaprime}) the scaling of the roughness of the membrane
due to temperature and disorder, respectively, (see also
Eqs. \eqref{Roughii} and \eqref{Rcorr}), with exponents
\begin{eqnarray}\label{etazeta2}
\zeta=2-D/2-\eta/2 \quad , \quad \zeta'=2-D/2-\eta'/2\;.
\end{eqnarray}
We note that $G_{\kappa}$ vanishes at $T=0$, while $G_{\Delta}$
remains non-zero, vanishing for a homogeneous membrane,
$\Delta_\kappa=\Delta_\mu=\Delta_\lambda=0$.

%In the next section we will analyse the SCSA equations in many physical 
%limits. We will find that these equations admit a surprising rich set of
%solutions that describe scaling properties of 
%disordered membranes at finite temperature. The long wavelength
%generic solutions that we will find correspond to globally stable 
%infrared fixed points of the renormalization group flow 
%characterizing the flat wrinkled phases of a membrane. 
%As we will see, solutions corresponding to fixed points with some relevant 
%directions can also be found by an appropriate choice of 
%microscopic (bare) coupling constants.

Using the above Ansatz \eqref{Gscale2}, \eqref{Gscale3}, in the long
wavelength limit we can now compute the three polarization
``bubbles'', defined in Eq.\eqref{Pij}. The details of the calculation
are presented in Appendix B, with the result \eqref{PiD}
\begin{eqnarray}\label{Pscale}
\Pi_{\kappa\kappa}(\q)&=&\frac{1}{D^2-1} Z_\kappa^{-2} q^{2\eta-4+D} \, \Pi(\eta, D)
%\equiv Z_\kappa^{-2} q^{2\eta-4+D}\Pi(\eta)\;,
, \quad \text{for} \quad \eta < \frac{4-D}{2} ,
\\
\Pi_{\kappa\Delta}(\q)&=& \frac{1}{D^2-1} Z_\kappa^{-1}Z_\Delta^{-1} 
q^{\eta+\eta'-4+D}\, \Pi(\eta, \eta',D)\;,\quad \text{for} \quad \eta +\eta' < 4-D ,\nn  \\
\Pi_{\Delta\Delta}(\q)&=& \frac{1}{D^2-1} Z_\Delta^{-2} q^{2\eta'-4+D}\, \Pi(\eta',D),
\quad \text{for} \quad \eta' < \frac{4-D}{2} ,\nn
\end{eqnarray}
where, from \eqref{Pieteeta2} 
\bea
 \Pi(\eta,\eta',D) &= &  
(D^2-1){\g(2-\frac{\eta+\eta'}{2}-\frac{D}{2})
\g(\frac{D}{2}+\frac{\eta}{2})\g(\frac{D}{2}+\frac{\eta'}{2})
  \over4\fpi\g(2- \frac{\eta}{2})\g(2- \frac{\eta'}{2})\g(D+\frac{\eta+\eta'}{2} )} 
  \quad , \quad \Pi(\eta,D)=\Pi(\eta,\eta,D) . \label{Pieteeta2} 
\eea 

We note that the result given for each integral $\Pi_{ij}(\q)$ in
\eqref{Pscale} holds only when the corresponding integral is {\it
  divergent} at small $q$, as indicated there by exponent
inequalities. When the above exponent inequality does not hold, the
corresponding integral is convergent and equal to a finite number at
$q=0$, which depends on the full $q$-dependence of $G_\kappa(\q)$ and
$G_\Delta(\q)$ and cannot be expressed simply through the small $q$
behaviors of the propagators. The condition for all three
$\Pi_{ij}(\q)$ in \eqref{Pscale} to diverge as $\q \to 0$ is thus
given by 
\bea
\eta , \eta' < \frac{4-D}{2},
\label{cond1} 
\eea 
and will be found to hold in most (but not all) cases, that we will
check a posteriori.

We now insert results in \eqref{Pscale} into equations
\eqref{couplings2}-\eqref{couplings2d} for the renormalized elastic
moduli and disorder variances. We observe that different solutions
emerge depending on which of the polarization bubbles dominates in the
denominator of the renormalized coupling constants in
\eqref{couplings2}-\eqref{couplings2d}. From Eqs.\eqref{Pscale} we see
that there are three cases which are possible {\it a priori} (assuming
all bare couplings non-zero):

\begin{enumerate}

\item $\eta < \eta'$, then $T \Pi_{\kappa\kappa}(\q) \gg \Pi_{\kappa\Delta}(\q)$ as $\q \to 0$,
corresponding to a {\it temperature dominated fixed point.}

\item either $T=0$, or $\eta' < \eta$ for any $T$, then, in both cases 
$T \Pi_{\kappa\kappa}(\q) \ll \Pi_{\kappa\Delta}(\q)$ as $\q \to 0$,
corresponding to a {\it disorder-dominated fixed point.}

\item $\eta'=\eta$, then $T \Pi_{\kappa\kappa}(\q) \sim
  \Pi_{\kappa\Delta}(\q)$ as $\q \to 0$, {\it a marginal fixed point},
  meaning that {\it both disorder and temperature play a role.} Its
  $T=0$ limit is also called marginal when $\eta'=\eta$, as an
  infinitesimal temperature would play a role (in fact, as discussed
  below, it is even marginally relevant in the present case, meaning
  that any infinitesimal temperature eventually flows to the thermal
  fixed point). %{\blue do we know this marginal relevance of T for sure?}

\end{enumerate} 
We will now examine all three cases to determine which actually occur
as self-consistent solutions of above SCSA equations.

%One defines the exponents
%\bea
%\eta_u=4-D-2\eta \quad , \quad \eta'_u=4-D-2\eta'
%\eea 
%The condition \eqref{cond1} is equivalent to $\eta_u>0$ and $\eta'_u>0$
%which, as we discuss below, means that the thermal and disorder fluctuations 
%{\it reduce}, i.e. screen, the in-plane elastic moduli. Note also that for SR disorder
%the factors $\Delta_\mu(\q)=\Delta_\mu$ and 
%$\Delta_b(\q)=\Delta_b$ in the numerators of Eqs. \eqref{couplings2} are irrelevant in comparison with
%the diverging $\Pi_{\Delta\Delta}(\q)\sim q^{-\eta_u'}$ and therefore can be
%dropped at long wavelengths. 

%It is important to note that 
%In the flat phase there is a strict
%physical bound $2-D<\eta, \eta'< (4-D)/2$, which for real (2D)
%membranes reduces to $0<\eta, \eta'<1$. The lower bound arises simply from
%the definition of the flat phase having $\zeta,\zeta'<1$, while the upper
%bound is the requirement that fluctuations {\it reduce} the in-plane 
%elastic moduli, giving $\eta_u=4-D-2\eta>0$ and $\eta'_u=4-D-2\eta'>0$.
%Given these bounds it is important to observe from Eqs.\eqref{Pscale}
%that that at nontrivial fixed points $\Pi_{ij}(\q)$ all diverge 
%at long length scales.

\subsubsection{Disorder-dominant, $T=0$, short-range correlated disordered fixed point}

We start with the case 2,\\ 

(i) $\eta' < \eta$ for arbitrary $T$ \\

(ii) $T=0$ with a weaker assumption on $\eta,\eta'$
(which turns out to be $\eta' < 2 \eta$, as seen below). \\

The two cases can be studied simultaneously since in both one
can neglect $T \Pi\kk(\q)$ compared to $\Pi\kd(\q)$ in the denominator
of the renormalized coupling constants. We assume that
\bea
\label{cond3} 
\eta + \eta' < 4 - D,
%\quad {\blue \eta' < \eta} 
\eea 
so that $\Pi\kd(\q)$ and $\Pi\dd(\q)$ diverge and thus dominate at small $q$,
checking this assumption a posteriori.  From
Eqs. \eqref{couplings2}-\eqref{couplings2d} and \eqref{Pscale} 
we find at small $q$
\begin{eqnarray}\label{MLcoupling}
\left(\tilde{\mu}(\q)\;,
\;\tilde{\Delta}_\mu(\q)\right)&\simeq&
\left({1\over 4\Pi\kd(\q)}\;,\;{\Pi\dd(\q)\over
    8\Pi\kd(\q)^2}\right)\; ,\\
&\simeq& (D^2-1) \left({Z\kap Z\Del\over 4 \Pi(\eta,\eta',D)}q^{4-D-\eta-\eta'}\;,\;
{Z\kap^2\Pi(\eta',D)\over 8
\Pi(\eta,\eta',D)^2}q^{4-D-2\eta}\right)\;,\\
\left(\tilde{b}(\q)\;,
\;\tilde{\Delta}_b(\q)\right)&\simeq& \frac{2}{D+1} \left(\tilde{\mu}(\q)\;,
\;\tilde{\Delta}_\mu(\q)\right) . 
%\\
%\left(\tilde{b}(\q)\;,
%\;\tilde{\Delta}_b(\q)\right)&\simeq&
%\left({1\over2(D+1)\Pi\kd(\q)}\;,\;
%{\Pi\dd(\q)\over4(D+1)\Pi\kd(\q)^2}\right)\;\\
%&\simeq& (D^2-1)  \left({Z\kap Z\Del\over2(D+1)\Pi(\eta,\eta',D)}q^{4-D-\eta-\eta'}\;,\;
%{Z\kap^2\Pi(\eta',D)\over4(D+1)
%\Pi(\eta,\eta',D)^2}q^{4-D-2\eta}\right)\;,
\end{eqnarray}
Inserting these expressions into the equations \eqref{GGeqn},
\eqref{GGeqn2} and utilizing the integral 
\bea
&&\Sigma(\eta,\eta',D) := \int_q {(\hat k_\alpha P_{\alpha \beta}^T(\hat {\bf q}) \hat k_{\beta})^2
|\q|^{4-D - 2\eta} 
\over |\hat \ks+ \q|^{4-\eta'}}\; ,
\eea
defined in Eq.\eqref{Sigma_etaetap} and calculated in the Appendix B, 
\bea
&&\Sigma(\eta,\eta',D) = \frac{\left(D^2-1\right) \Gamma
   (2-\eta) \Gamma
   \left(\frac{D}{2}+\frac{\eta'}{2}\right
   ) \Gamma
   \left(\eta-\frac{{\eta}'}{2}\right
   )}{4 (4 \pi)^{D/2} \Gamma
   \left(2-\frac{{\eta}'}{2}\right
   ) \Gamma
   \left(\frac{D}{2}+\eta\right) \Gamma
   \left(\frac{D}{2}-
   \eta+\frac{{\eta}'}{2}+2 \right)} \quad , \quad \Sigma(\eta,D):= \Sigma(\eta,\eta',D) ,\label{Sigdef} 
\eea 
we obtain the self-consistency equations
\begin{eqnarray}\label{MLggEqns}
Z\kap k^{-\eta}&=&\kappa +{D (D-1) \over
4d_c}\left[{2Z\kap\over\Pi(\eta,\eta',D)}
\Sigma({\eta+\eta'\over2},\eta',D)k^{-\eta}-
{Z\kap\Pi(\eta',D)\over\Pi(\eta,\eta',D)^2}\Sigma(\eta,D)
k^{-\eta}\right]\;,\;\;\;\;\;\;\\
{Z\kap^2\over Z\Del}k^{-2\eta+\eta'}&=&\Delta\kap +{D(D-1)\over
4d_c}{Z\kap^2\Pi(\eta',D)\over
Z\Del\Pi(\eta,\eta',D)^2}\Sigma(\eta,\eta',D)
k^{-2\eta+\eta'}\;.\;\;\;\;\;\; \label{MLggEqns2}
\end{eqnarray}
In the first line, we have dropped the first correction term to
$\kappa$ in \eqref{GGeqn} (either subdominant, or absent in the
present case). We now assume $\eta' < 2 \eta$ at this
disorder-controlled fixed point, which is implied by $\eta'<\eta$ for
case (i) and added as an assumption for case (ii). Then we can safely
neglect the bare value $\Delta\kap$ in Eq.\eqref{MLggEqns2}.
Canceling out the remaining factors of powers of $k$ and nonuniversal
amplitudes, we obtain two equations implicitly determining $\eta$ and
$\eta'$ as a function of $D,d_c$,
\begin{eqnarray}\label{MLetaEqns}
1&=&{D(D-1)\over4d_c}\left[
{2\Sigma({\eta+\eta'\over2},\eta',D)\over\Pi(\eta,\eta',D)}
-{\Pi(\eta',D)\Sigma(\eta,D)\over\Pi(\eta,\eta',D)^2}\right]\;,\\
1&=&{D(D-1)\over4d_c}{\Pi(\eta',D)\Sigma(\eta,\eta',D)
\over\Pi(\eta,\eta',D)^2}\;.
\end{eqnarray}
These equations can be solved using the definitions of the functions
$\Pi(\eta,\eta',D)$ and $\Sigma(\eta,\eta',D)$ in
\eqref{Pieteeta2} and \eqref{Sigdef}. As is easily
checked numerically, or via a series expansion in small $\eta,\eta'$,
the only solution continuously related to 
$\eta=\eta'=0$ at $d_c=+\infty$ is a {\it marginal} disorder dominated
fixed point, i.e., with 
\be
\eta=\eta'\;.
\ee
Hence, for short-range disorder, there are no disorder dominated fixed
point with $\eta'<\eta$, case (i) above, and the only solution we find
within SCSA is a $T=0$ fixed point, i.e., the case (ii) above (with,
$\eta '=\eta < 2 \eta$ consistent with our assumptions).  As discussed
below, this SCSA solution corresponds to the $T=0$ fixed point which
was obtained in Ref. \cite{ML} using renormalization group methods,
controlled in an expansion in $\epsilon=4-D$.  As for the homogeneous
membrane, the present method is expected to be more accurate in the
physical dimension.

For $\eta=\eta'$, we find that both equations reduce to the same,
simpler equation,
\begin{equation}\label{MLfinalEqn}
1={D (D-1) \over4d_c} \frac{\Sigma(\eta,D)}{\Pi(\eta,D)}\;.
\end{equation}
Comparing with \eqref{eqSCSApure} we see that this is exactly the same
equation as the one for the homogeneous membrane with $d_c$ replaced
by $4 d_c$, hence we find the remarkable result at the $T=0$
disorder-dominated fixed point \cite{comment4dc}
\be
\eta'_{\rm dis}(D,d_c) = \eta_{\rm dis}(D,d_c) = \eta_{\rm pure}(D,4d_c), \label{magic} 
\ee
which holds within the SCSA for any $D, d_c$. Here and below we add
the subscript ${\rm pure}$ to denote all exponents of the thermal
fixed point of the homogeneous, i.e. pure, flat phase studied in
Section \eqref{PureMembrane}. It also implies for the thermal and
disorder roughness exponents
\be
\zeta'_{\rm dis}(D,d_c) = \zeta_{\rm dis}(D,d_c) = \zeta_{\rm pure}(D,4 d_c)\;.
\ee
Note that at $T=0$, strictly only the disorder-driven roughness is 
non-zero, i.e., the minimum energy configuration $h_{\min}(x)$ is
rough with exponent $\zeta'$
\be
\overline{ (h_{\min}(x)-h_{\min}(0))^2 } \sim |x|^{2 \zeta'}\ .
\ee
However, the fact that we find $\zeta=\zeta'$ means that, for
infinitesimal temperature $T>0$, the thermal fluctuations and the
disorder fluctuations scale with the same exponent. This is the sense
in which the temperature is marginal at this fixed point. This
marginality of temperature, in an RG sense, with thermal crossover
exponent $\phi_T=0$, was also observed in \cite{ML}, consistent with
the SCSA result above. We defer the discussion of whether the
temperature is marginally relevant, or marginally irrelevant to the
next section.

Let us now give more details on this $T=0$ fixed point. For a $D=2$ membrane
the expression simplifies to
\be
\eta_{\rm dis}(D=2,d_c)= \frac{1}{d_c + \sqrt{1 - \frac{1}{2} d_c + d_c^2}},
\ee
and in the physical dimensions $D=2$, $d_c=1$ we find
\bea
&& \eta'_{\rm dis} = \eta_{\rm dis}=\frac{2}{2+\sqrt{6}} \approx 0.449, \\
&& \zeta'_{\rm dis} = \zeta_{\rm dis} = 2-\sqrt{\frac{3}{2}} \approx 0.775.
\eea 
We thus predict that a heterogeneous membrane, characterized by
short-range quenched disorder, exhibits a wrinkled ground state
($T=0$) that is qualitatively rougher than a homogeneous membrane at
finite temperature.

As we discussed for a homogeneous membrane, the SCSA is exact to leading
order in the $1/d_c$-expansion for any $D$, and also to leading order
in the $\epsilon=4-D$ expansion for any $d_c$. The same applies here
for a disordered membrane. Expanding our result around $D=4$ we find
\bea
\eta_{\rm dis}(D=4-\epsilon,d_c) = \frac{\epsilon}{2 + d_c/3} 
\eea 
for any $d_c$, in agreement with the $O(\epsilon)$ result of Ref.\cite{ML}. 
Expanding now our solution for large codimension,
in powers of $1/d_c$, we obtain
\begin{eqnarray}\label{dcExpand_dis} \eta_{\rm dis} (D,d_c)&=& {2\over d_c}\left({D-1\over D+2}
\right){\Gamma[D]\over{\Gamma[{D \over 2}]^3 \Gamma[2-{D \over 2}]}}
+O\left({1 \over{d_c^2}}\right)\;,\\
\eta_{\rm dis}(D=2,d_c)&=& {1 \over  2 d_c}+O\left({1 \over
    {d_c^2}}\right)\ ,
\; 
\end{eqnarray}
an {\em exact} result, which, to our knowledge, is new.  Finally the result
for $d_c=0$, $\eta_{\rm dis}=(4-D)/2$ is also exact.  It is the limiting
case for our starting assumption \eqref{cond3}, which is thus verified
for all $d_c>0$.

Similar to our discussion for a homogeneous membrane we compute 
the lower critical dimension $D_{lc}$ of the $T=0$ disorder fixed
point.  Since $\eta=\eta'$, it reduces to a single equation $2 -
\eta_{dis}(D_{lc},d_c) = D_{lc}$ and using \eqref{magic} we find
\begin{eqnarray}
D_{lc}(d_c) &=& \frac{1}{2} (1- 4 d_c + \sqrt{16 d_c^2 + 24 d_c +1})\ ,\\ 
D_{lc}(1) &=& \frac{1}{2} (-3+\sqrt{41}) \approx 1.70\ ,
\end{eqnarray}
while we recall that $D_{lc}=1.41$ for a homogeneous membrane. 

We can now study the renormalized elastic moduli and random stress
variances at this $T=0$ fixed point. From \eqref{MLcoupling} and
$\eta'=\eta$, we see that they share the same screening exponent
$\eta_u=\eta'_u$, where $\eta'_u=4-D - 2 \eta'$, giving
\bea
&& \tilde{\mu}(\q) \simeq Z_\mu q^{\eta_u} \quad , \quad \tilde{b}(\q) \simeq Z_b q^{\eta_u} \quad , \quad  \tilde{\Delta}_\mu(\q) \simeq Z_{\Delta_\mu} q^{\eta_u} \quad , \quad 
\tilde{\Delta}_b(\q) \simeq Z_{\Delta_b} q^{\eta_u}, 
\eea
where here
\bea
&& \eta_u =  4-D-2 \eta \approx_{\substack{D=2 \\ d_c=1}} \, 1.101\ .
\eea
The roughness of the in-plane-phonon deformations at this $T=0$ fixed
point is given by
\bea
&& \overline{ (u_\alpha(\x)-u_\alpha(0)) (u_\alpha(\x)-u_\alpha(0))} \sim x^{2 \zeta'_u}
\quad , \quad \zeta'_u = \frac{2-D+\eta'_u}{2} \approx_{\substack{D=2
    \\ d_c=1}} \, 0.55\ .
\eea
since here $\eta=\eta'$. 

Above, we have introduced the amplitudes $Z_\mu$, $Z_b$,
$Z_{\Delta_\mu}$ and $Z_{\Delta_b}$, which, together with $Z_\kappa$
and $Z_{\Delta_\kappa}$, defined in \eqref{Gscale2},
\be
\tilde \kappa(\ks) \simeq Z_\kappa k^{-\eta} \quad , \quad 
\tilde \Delta_\kappa(\kb) \simeq Z_{\Delta_\kappa} k^{\eta' - 2 \eta}
= Z_{\Delta_\kappa} k^{- \eta}\;,
\ee
form a set of six a priori non-universal amplitudes for the disordered
membrane. The fixed point, however, is characterized by four universal
amplitude ratios. From \eqref{MLcoupling} we find
\bea
&& \frac{2 Z_b Z_{\Delta_\kappa}}{Z_\kappa^3} =
\frac{4 Z_{\Delta_b}}{Z_\kappa^2} = \frac{D-1}{\Pi(\eta,D)}
\approx_{\substack{D=2 \\ d_c=1}} \,  13.706\ , \\
&& \frac{2 Z_b}{Z_\mu} =  \frac{2 Z_{\Delta_b}}{Z_{\Delta_\mu}} =
\frac{4}{D+1}\ .  
\eea
Following the analysis and discussion in Sec.\ref{analysisSCSAhomo} we
see that the Poisson ratio is also $\tilde\sigma(\q)=-1/3$ for the disordered
membrane at this fixed point.

We can now address the question of the anomalous strain response to a
stress for a heterogeneous membrane. This is an extension to the
disordered case of the theory of the so-called buckling transition
mentioned in the pure case around Eq.\eqref{Ftension}. Here we will
simply give a ``back of the envelope'' derivation (that can be made
precise by the RG matching procedure) of the anomalous exponent
$1/\delta$, that relates the strain and the stress. Denoting
$\varepsilon\equiv \partial u$ the strain and $\sigma$ the applied
stress, one has, schematically $\varepsilon\sim
\frac{1}{\mu_R(q_\sigma)} \sigma \sim \frac{\sigma}{\mu}
q_{\sigma}^{-\eta_u}$, where $q_\sigma$ is the cutoff wave-vector
induced by the applied stress $\sigma$. The latter is determined by
balancing curvature energy against the stress energy of the height
fields (the last term in \eqref{Ftension} with $\sigma$ playing the
role of $\tau$), $\kappa_R(q_\sigma) q_\sigma^4 \sim \sigma
q_\sigma^2$.  This gives $q_\sigma \sim \sigma^{1/(2-\eta)}$. Hence we
find a {\em universal anomalous} nonlinear strain-stress relation for
arbitrary small stress $\sigma$ (cut off only by $\sigma_{NL}\sim
1/L^{2-\eta}$ due to membrane's finite extent, $L$), $\varepsilon \sim
\sigma^{1/\delta}$, with $\delta^{-1} = 1- \frac{\eta_u}{2-\eta}$.
Using $\eta_u=4-D-2 \eta$, this reproduces \eqref{scalingdelta}.

For a homogeneous (disorder-free) membrane in physical dimension
$D=2$, $d=1$ we find a {\em universal} nonlinear strain-stress
relation with $\eta=\eta_{\rm pure}(D=2,d_c=1)=4/(1+ \sqrt{15})$,
controlled by the thermal fixed point, with
\bea
\varepsilon\sim \sigma^{\frac{\eta}{2-\eta}} \sim \sigma^{0.6961}\;.
\eea 
For a randomly heterogeneous membrane at the $T=0$ fixed point we
replace the exponent $\eta=\eta_{\rm pure}(D=2,d_c=1)$ by
$\eta=\eta'=\eta_{\rm dis}(D=2,d_c=1) = \eta_{\rm pure}(D=2,4)=2/(2 +
\sqrt{6})$, leading to a universal anomalous strain-stress relation
for the disordered membrane
\bea
\varepsilon\sim \sigma^{\frac{\eta_{\rm dis}}{2-\eta_{\rm dis}}} \sim \sigma^{0.2899}\;,
\eea 
corresponding to an exponent $\delta=3.4495$.

Finally let us note that we have assumed generic disorder with
$\Delta_\kappa>0$. The case of disorder $\Delta_\kappa=0$, symmetric
under $\h\to -\h$ is more delicate at $T=0$, and, as was discussed in
\cite{RNpra} leads to runaway RG flows near $D=4$.  One possibility is
that the resulting flow to strong stress-only disorder leads to a
spontaneous breaking of the $\h \to - \h$ symmetry (or more generally
of the $O(d_c)$ symmetry). Such a scenario may further lead to a
crumpled spin-glass like order in the normals, as was explored in
\cite{RLjphys}. Here, within the flat phase, and from our SCSA
equations, we see a strong tendency towards such symmetry breaking,
since even a very small $\Delta_\kappa>0$, at very large scale,
eventually leads to the $T=0$ fixed point studied above.

\subsubsection{Search for a $T>0$ marginal fixed point} 
\label{sec:search} 

We now explore case 3, that is $\eta'=\eta$ and $T>0$, i.e., we search for a
marginal glass fixed point. We assume again that
$\eta=\eta'<\frac{4-D}{2}$, so that all $\Pi_{ij}(\q)$ integrals diverge
and dominate at small $q$.

From Eqs. \eqref{couplings2} and \eqref{Pscale} we find at small $q$
\begin{eqnarray}\label{MLcoupling2}
\left(\tilde{\mu}(\q)\;,
\;\tilde{\Delta}_\mu(\q)\right)&\simeq&
\left({1\over 2 ( T \Pi\kk(\q) + 2\Pi\kd(\q))}\;,\;{\Pi\dd(\q)\over 2 ( T \Pi\kk(\q) + 2\Pi\kd(\q))^2}\right)\;,\\
&\simeq& \frac{ (D^2-1)q^{4-D-2\eta}}{2 \Pi(\eta,D)} \left(\frac{1}{T Z\kap^{-2} + 2 Z\Del^{-1} Z\kap^{-1} } \;,\;
{Z\Del^{-2} \over (T Z\kap^{-2} + 2 Z\Del^{-1} Z\kap^{-1})^2 }\right)\;,\\
\left(\tilde{b}(\q)\;,
\;\tilde{\Delta}_b(\q)\right)&\simeq& \frac{2}{D+1} \left(\tilde{\mu}(\q)\;,
\;\tilde{\Delta}_\mu(\q)\right).
\end{eqnarray}

Substituting these expressions into Eqs.\eqref{GGeqn},
\eqref{GGeqn2}, we obtain the self-consistency equations
\begin{eqnarray}\label{MLggEqnsT}
Z\kap k^{-\eta}&=&\kappa +{D (D-1) \over
d_c} \left[ {T Z_\kappa^{-1} + Z_{\Delta}^{-1}  \over T Z\kap^{-2} + 2 Z\Del^{-1} Z\kap^{-1}}
 - {Z\Del^{-2} Z_\kappa^{-1}\over (T Z\kap^{-2} + 2 Z\Del^{-1} Z\kap^{-1})^2 }
 \right]   \frac{\Sigma(\eta,D)}{\Pi(\eta,D)} k^{-\eta}
\;,\;\;\;\;\;\;\\
{Z\kap^2\over Z\Del}k^{-\eta}&=&\Delta\kap +{D(D-1)\over
d_c} {Z\Del^{-3} \over (T Z\kap^{-2} + 2 Z\Del^{-1} Z\kap^{-1})^2 }
\frac{\Sigma(\eta,D)}{\Pi(\eta,D)}
k^{-\eta}\;.\;\;\;\;\;\; \label{MLggEqns2T}
\end{eqnarray}
Identifying the leading terms, we now obtain the equations
\bea
&& 1 = {D(D-1)\over
d_c} \frac{\Sigma(\eta,D)}{\Pi(\eta,D)}
 \left[ {a_T + a_\Delta \over a_T + 2 a_\Delta}
 - {a_\Delta^2 \over (a_T + 2 a_\Delta)^2 }
 \right]\ , \\
 && 1= {D(D-1)\over
d_c} \frac{\Sigma(\eta,D)}{\Pi(\eta,D)}
{a_\Delta^2  \over (a_T + 2 a_\Delta)^2 }\ ,
\eea 
where we have defined the thermal and disorder (positive) amplitudes
\bea
a_T = T Z\kap^{-2} \quad , \quad a_\Delta = Z\Del^{-1} Z\kap^{-1}.
\eea 
Clearly for these two equation to have a solution requires
\bea
{2 a_\Delta^2  \over (a_T + 2 a_\Delta)^2 } = {a_T + a_\Delta \over
  a_T + 2 a_\Delta}\rightarrow a_T(a_T + 3 a_\Delta)=0\ .
\eea 
Hence the only physical case of positive amplitudes when a solution
exists is
\be
a_T = 0 \quad \rightarrow \quad T=0\ ,
\ee
which is the zero-temperature solution studied in the previous section. 

This analysis shows that within the SCSA, there is no
finite-temperature $T > 0$ marginal case solution $\eta=\eta'$. The
only solution with $\eta=\eta'$ is the $T=0$ fixed point. This
excludes, within SCSA, a scenario with a line of fixed points with
continuously varying $T$. This is also an indication that (at least
within SCSA), the temperature is {\it marginally relevant} at the
$T=0$ fixed point. Indeed, since we also know that the thermal fixed
point is stable to disorder (as analyzed in the next
section)\cite{RNpra}, if instead temperature was irrelevant (i.e., the
flow of temperature was towards $T=0$), there would be a genuine glass
phase, and an additional critical fixed point separating high and low
$T$ phases. Since we have not found a solution to the SCSA equations
corresponding to such a fixed point, this possibility is unlikely. Our
conclusion is thus in agreement with the RG analysis of Ref.\cite{ML},
which also concluded that temperature is a marginally relevant
perturbation at the $T=0$ fixed point.  Since the RG flow away from
$T=0$ is marginal, this also implies an exponentially large (in $1/T$)
crossover length scale at low temperature. This crossover can, in
principle also be studied using the SCSA, but we do not pursue it
here.

\subsubsection{Thermal fixed point with short-range disorder}

Next we consider the case 1, $\eta < \eta'$ and $T>0$, where at long
wavelengths $T \Pi_{\kappa\kappa}(\q) \gg \Pi_{\kappa\Delta}(\q)$,
i.e., thermal fluctuations-induced screening dominates over the
disorder one.  From \eqref{couplings2} and \eqref{Pscale} we see that
for such a fixed point we have two cases to consider in the small $q$
limit (in all cases we are assuming $0 < \eta < \frac{4-D}{2}$, that
we will verify a posteriori):

\begin{itemize}

\item $0< \eta' < \frac{4-D}{2}$, then $\Pi\dd(\q)$ diverges at small
  $q$, and one can neglect the bare random stress variances
  $\Delta_\mu$ and $\Delta_b$ in the numerators of
  \eqref{couplings2c},\eqref{couplings2d}. In this case we find
\begin{eqnarray}\label{ALcoupling}
\left(\tilde{\mu}(\q)\;,
\;\tilde{\Delta}_\mu(\q)\right)&\simeq&
\left({1\over 2 T \Pi\kk(\q)}\;,\;{\Pi\dd(\q)\over 2 T^2
    \Pi\kk(\q)^2}\right)\; ,\\
&\simeq& (D^2-1) \left({Z\kap^2\over 2 T \Pi(\eta,D)}q^{4-D-2\eta}\;,\;
{Z\kap^4\Pi(\eta',D)\over 2 T^2
Z\Del^2\Pi(\eta,D)^2}q^{4-D+2\eta'-4\eta}\right)\;,\\
\left(\tilde{b}(\q)\;,
\;\tilde{\Delta}_b(\q)\right)&\simeq& \frac{2}{D+1} \left(\tilde{\mu}(\q)\;,
\;\tilde{\Delta}_\mu(\q)\right)\ .
%\left(\tilde{b}(\q)\;,
%\;\tilde{\Delta}_b(\q)\right)&\simeq& (D^2-1)
%\left({1\over T (D+1)\Pi\kk(\q)}\;,\;
%{\Pi\dd(\q)\over T^2 (D+1)\Pi\kk(\q)^2}\right)\;\\
%&=&\left({Z\kap^2\over T (D+1)\Pi(\eta,D)}q^{4-D-2\eta}\;,\;
%{Z\kap^4\Pi(\eta',D)\over T^2 (D+1)
%Z\Del^2\Pi(\eta,D)^2}q^{4-D+2\eta'-4\eta}\right)\;,
\end{eqnarray}

\item $\eta' > \frac{4-D}{2}$, then $\Pi\dd(\q=0)$ is finite and simply
adds to the bare values $\Delta_\mu$ and $\Delta_b$ in the
numerators of \eqref{couplings2}. To simplify notation we define
dressed dimensionless stress variances,
\be
d_\mu= \frac{\Delta_\mu+ 2 \mu^2 \Pi\dd(\q=0)}{2 \mu^2} \quad , \quad 
d_b = \frac{\Delta_b+ (D+1) b^2 \Pi\dd(\q=0)}{(D+1) b^2}\ .
\ee
In terms of these we find
\begin{eqnarray}\label{ALcoupling2}
\left(\tilde{\mu}(\q)\;,
\;\tilde{\Delta}_\mu(\q)\right)&\simeq&
\left({1\over 2 T \Pi\kk(\q)}\;,\;{d_\mu \over 2 T^2
    \Pi\kk(\q)^2}\right)\; ,\\
&\simeq& (D^2-1) \left({Z\kap^2\over 2 T \Pi(\eta,D)}q^{4-D-2\eta}\;,\;
{(D^2-1)  Z\kap^4 d_\mu \over 2 T^2
\Pi(\eta,D)^2}q^{2( 4-D-2 \eta)}\right)\;,\\
\left(\tilde{b}(\q)\;,
\;\tilde{\Delta}_b(\q)\right)&\simeq& \frac{2}{D+1} \left(\tilde{\mu}(\q)\;,
\; \frac{d_b}{d_\mu} \tilde{\Delta}_\mu(\q)\right)\;.
%\left(\tilde{b}(\q)\;,
%\;\tilde{\Delta}_b(\q)\right)&\simeq& (D^2-1)
%\left({1\over T (D+1)\Pi\kk(\q)}\;,\;
%{\Pi\dd(\q)\over T^2 (D+1)\Pi\kk(\q)^2}\right)\;\\
%&=&\left({Z\kap^2\over T (D+1)\Pi(\eta,D)}q^{4-D-2\eta}\;,\;
%{Z\kap^4\Pi(\eta',D)\over T^2 (D+1)
%Z\Del^2\Pi(\eta,D)^2}q^{4-D+2\eta'-4\eta}\right)\;,
\end{eqnarray}

\end{itemize} 
Inserting these expressions into Eqs.\eqref{GGeqn},
\eqref{GGeqn2} we find the following self-consistency equations,
\begin{eqnarray}\label{ALggEqns}
Z\kap k^{-\eta}&=&\kappa +{D(D-1) \over
d_c} {Z\kap \Sigma(\eta,D)\over\Pi(\eta,D)} k^{-\eta} + O(k^{\eta'-2 \eta}) 
- O(k^{\min(2 \eta',4-D)-3 \eta});, \\
{Z\kap^2\over Z\Del}k^{\eta'-2\eta}&=&\Delta\kap +
% {D(D-1)\over d_c}{Z\kap^4\Pi(\eta',D)\over T^2 Z\Del^3\Pi(\eta,D)^2}\Sigma(2\eta-\eta',\eta',D)
O(k^{\min(2 \eta',4-D) + \eta'-4\eta})\;,\;\;\;\;\;\; \nn
\end{eqnarray}
where the estimates are valid in both cases. In the first line of
\eqref{ALggEqns} we see that the dominant contribution to the
correction to the bending rigidity comes from purely thermal
fluctuations. We have indicated two other contributions (one positive,
one negative) in the order where they appear in \eqref{GGeqn}: they
originate from disorder that is subdominant for this case with
$\eta'>\eta$ and $4-D-2 \eta>0$.  In the second line of
\eqref{ALggEqns}, the second term on the right-hand side is always
subdominant compared to the two other terms if $\eta'<(4-D)/2$, and
also if $\eta'>(4-D)/2$, since then $4-D+\eta'-3 \eta> 3 (
\frac{4-D}{2} - \eta) >0$ from our assumption that $\eta<(4-D)/2$.
%\item $ \eta' > \frac{4-D}{2}$
%\begin{eqnarray}\label{ALggEqns4}
%Z\kap k^{-\eta}&=&\kappa +{D(D-1) \over
%d_c} {Z\kap \Sigma(\eta,D)\over\Pi(\eta,D)} k^{-\eta} + O(k^{\eta'-2 \eta}) 
%-  O(k^{4-D-3 \eta})  \\
%{Z\kap^2\over Z\Del}k^{\eta'-2\eta}&=&\Delta\kap +
%%{D(D-1)\over d_c}{Z\kap^4\Pi(\eta',D)\over T^2 Z\Del^3\Pi(\eta,D)^2}\Sigma(2\eta-\eta',\eta',D)
%O(k^{4-D + \eta'-4\eta}) \;,\;\;\;\;\;\;
%\end{eqnarray}
Balancing the dominant terms on each side of the first line, 
canceling the amplitude $Z_\kappa$, we obtain an implicit equation
\begin{equation}\label{ALeta}
\frac{D(D-1)\Sigma(\eta,D)}{d_c\Pi(\eta,D)}=1\;,
%{D\Sigma(\eta,D) \over d_c(D+1)\Pi(\eta,D)}=1\;.
\end{equation}
which determines $\eta(D,d_c)$. We note that it is identical to
Eq.\eqref{eqSCSApure} for a homogeneous membrane, and thus obtain
$\eta(D,d_c)=\eta_{\rm pure}(D,d_c)$. The second line gives
\begin{equation}\label{ALetaeta}
\eta'=2\eta\;,
\end{equation}
as well as $Z_{\Delta_\kappa}=1$. Thus we find that
$\tilde\Delta_{\kappa}(\q) \simeq \Delta_\kappa$, i.e., the bare
curvature disorder remains essentially un-renormalized, with only
subdominant corrections at long wavelength. These corrections are
$\sim k^{2 (\eta'-\eta)} \sim k^{2\eta}$ at small $k$ for $\eta' <
(4-D)/2$, i.e., $\eta<(4-D)/4$ and $\sim k^{4-D+\eta'-4 \eta}=k^{4-D-2
  \eta}$ if $\eta>(4-D)/4$. We note that above is valid for generic
disorder, such that $\Delta_\kappa >0$, and the symmetric case of
$\Delta_\kappa=0$ requires a special discussion given in the next subsection.

Clearly this solution of the SCSA equations corresponds to the
non-trivial thermal fixed point of the homogeneous membrane, studied
in Sec.\ref{analysisSCSAhomo}, with $\eta = \eta_{\rm pure}$ and
thermal roughness exponent $\zeta =(4-D- \eta)/2 = \zeta_{\rm pure}$,
weakly perturbed by random curvature and stress disorders. We recall
that for a physical membrane, $D=2,d_c=1$ we have $\eta \approx 0.821$
and $\zeta \approx 0.590$. The disorder is irrelevant at the thermal
fixed point. Irrelevance of disorder at the thermal fixed point can
also be obtained from a Harris criterion type of arguments, as discussed
in \cite{RNpra}.  

Note that a full detailed crossover from the nontrivial thermal fixed
point can in principle be studied using the SCSA. As can be seen from
\eqref{ALggEqns}, weak disorder corrections at the thermal fixed point
scale like $k^\phi$ (with the positive $\phi$ corresponding to irrelevant
disorder), with the crossover exponent given by
$\phi=\min(\phi_1,\phi_2)$, where
\bea
&& \phi_1=\eta'-\eta = \eta,\;\quad\quad \phi_1 \approx 0.821\;,\;\;\;{\rm for} \, \, \mbox{$D=2, d_c=1$}\;, \\
&& \phi_2= 4-D-2 \eta,\quad\quad \phi_2 \approx 0.358
\;,\;\;\;{\rm for} \, \, \mbox{$D=2, d_c=1$}\;. 
\eea 
While at this fixed point the asymptotic roughness is controlled by
thermal fluctuations and is the same as for a
homogeneous membrane, there is also a subdominant
%stress-disorder 
disorder contribution to the membrane roughness, as
described in \eqref{RoughFourier}. It is
controlled by another non-trivial exponent
\be 
\zeta' = \frac{4-D - \eta'}{2} =
\zeta - \eta/2 =_{D=2,d_c=1} \frac{1}{7} (9 - 2 \sqrt{15}) \approx 0.179\;,
\ee 
i.e., a smaller, but positive disorder roughness exponent for physical
membranes. Hence, quite remarkably, the effect of disorder can still
in principle be observed in the scaling of disorder-driven height
fluctuations corresponding to the rough wrinkled background
(disorder-controlled, disconnected component of height-height
correlator). Although challenging, such observation through
averaging of thermal fluctuations is in principle possible.

\subsubsection{Stress-only disorder ($\Delta_\kappa=0$) analysis}

We finish this section on short-range disorder by analyzing a special
case of $\Delta_\kappa=0$, i.e., stress-only disorder.  It is clear
that $\Delta_\kappa=0$ is an invariant subspace protected by
reflection symmetry, $\h \to - \h$, that, in the absence of spontaneous
breaking of this symmetry implies $\tilde \Delta_\kappa(\q)=0$. 
Thus we see from \eqref{Pijexplicit} that $\Pi_{\Delta
  \Delta}(\q)=\Pi_{\kappa\Delta}(\q)=0$, giving
\begin{eqnarray}\label{ALcoupling_stress}
\left(\tilde{\mu}(\q)\;,
\;\tilde{\Delta}_\mu(\q)\right)&\simeq&
\left({1\over 2 T \Pi\kk(\q)}\;,\;{\Delta_\mu \over 4 \mu^2 T^2 \Pi\kk(\q)^2}\right)\;,\\
&\simeq& (D^2-1) \left({Z\kap^2\over 2 T \Pi(\eta,D)}q^{4-D-2\eta}\;,\;
{(D^2-1) Z\kap^4 \Delta_\mu\over 4 \mu^2 T^2
\Pi(\eta,D)^2}q^{2(4-D-2\eta)}\right)\;,\\
\left(\tilde{b}(\q)\;,
\;\tilde{\Delta}_b(\q)\right)&\simeq&
\left({1\over (D+1)T\Pi\kk(\q)}\;,\;
{\Delta_b \over (D+1)^2 b^2 T^2\Pi\kk(\q)^2}\right)\;,\\
&=& (D^2-1)\left({Z\kap^2\over (D+1)T\Pi(\eta,D)}q^{4-D-2\eta}\;,\;
{(D^2-1) Z\kap^4 \Delta_b \over (D+1)^2 b^2 T^2
\Pi(\eta,D)^2}q^{2(4-D-2\eta)}\right)\;.
\end{eqnarray}
Using these expressions inside the equation for the renormalized
bending modulus, \eqref{GGeqn} (Eq.\eqref{GGeqn2} identically
vanishes) gives
\bea
\label{ALggEqns3}
Z\kap k^{-\eta}&=&\kappa +{D(D-1) \over
d_c} {Z\kap \Sigma(\eta,D)\over\Pi(\eta,D)} k^{-\eta} 
+ O(k^{4-D-3 \eta})\;.
\eea
The second correction term on the right-hand side vanishes identically
and the counting in $k$ of the last term has changed. Hence we again
find that the thermal fixed point of the homogeneous membrane is
stable to stress-only disorder, consistent with the result first found
in Ref.\cite{RNpra}. From (\ref{ALggEqns3}) we see that the crossover
exponent is again given by $\phi=\eta_u$, where $\eta_u=4-D-2 \eta$,
$\eta_u=0.358$ for the physical membrane.

% we find a temperature
%dominated behavior with several intermediate regimes and associated
%crossovers in the behavior of the renormalized coupling
%constants. These screened coupling constants are plotted in Fig.  for
%a generic set of microscopic parameters.  At very short wavelengths
%$1$ in denominator dominates both types of ``screening bubbles'', and
%the membrane is described by the bare (microscopic) coupling
%constants $\mu,\Delta_\mu,b,\Delta_b$. At longer intermediate
%wavelengths $1/q^*_{\mu1}\sim(\mu/Z\kap Z\Del)^{1/(\eta+\eta'+D-4)}$
%and $1/q^*_{b1}\sim(b/Z\kap Z\Del)^{1/(\eta+\eta'+D-4)}$ the
%renormalized coupling constants crossover to an intermediate regime
%controlled by disorder dominated fixed point. And finally, at
%$1/q^*_{\mu2}=1/q^*_{b2}\sim(Z\Del/Z\kap)^{1/(\eta-\eta')}$ there is a
%crossover to the infrared stable temperature dominated fixed point at
%which the renormalized coupling constants are given by

\section{Analysis of the SCSA equations for the heterogeneous membrane in long-range
  disorder} \label{sec:LRdis}

\subsection{Realization of long-range disorder}

In previous sections, our analysis and discussion of membrane's
quenched internal disorder was limited to local random heterogeneity,
that is {\em short-range} correlated in space. However, there are a
number of physical realizations that also motivate examination of
disorder that is {\em long-range} power-law correlated.

In soft-matter realization of polymerized phospholipid membranes,
long-range disorder has been previously addressed by Nelson and
Radzihovsky\cite{NRpra} in a context of experiments by D. Bensimon,
et. al\cite{Bensimon}, that observed a wrinkling transition of
vesicles partially polymerized with uv irradiation. They argued that
correlated disorder, characterized by a power-law exponent $z_\mu=2$
can arise from randomly distributed disclinations, with (in contrast
to a fluid membrane) partial polymerization preventing screening of
the long-range strains by immobile dislocations.  Also it was found
that randomly spaced and oriented lines of impurities (or vacancies or
interstitials) can give rise to long-range correlations in the strain
and curvature with $z_\mu=z\kap=1$.

Another possible mechanism for a generation of power-law correlated
disorder is through a frozen-in lipid tilt order. The latter is
well-known to exist in fluid membranes in the
$\beta'$-phase\cite{Jerusalem}. The associated in-plane vector
projection is an xy-order parameter, $\vec S$, that in the ordered
phase exhibits quasi-long-range correlations, characterized by a
temperature-dependent exponent $\eta$, with $\langle S\as(\x)
S\bs(0)\rangle \sim x^{-\eta}$. One would expect that upon
polymerization, these correlations will be frozen in and will couple
to elastic strains as quenched-in stress $c\ab(\x) \sim S\as(\x)
S\bs(\x)$. It will thus lead to quenched stress-disorder,
characterized by a power-law correlations, with a continuously tunable
exponent.

In solid state context, such as graphene and similar atom-thin elastic
sheets, long-range disorder can arise from a random distribution of
Coulomb impurities. It was also recently pointed out that adatoms,
which bind on one side of the membrane, or more generally in a
non-symmetric way with respect to the sheet, produce most naturally
random curvature disorder. The adatoms have been shown to interact via
power-law correlations\cite{AdatomsInteractions}, mediated by the
electrons (akin to RKKY interaction of spins), hence their spatial
distribution may be long-range correlated.
%We also expect a dislocations and grain boundaries
%\cite{DislocationsInGraphene,GrainboundariesGraphene}.  
The question of whether ripples, if they have an intrinsic origin,
will also produce long-range disorder is also interesting to explore,
especially since they can be controlled \cite{ControledRipples}.

Below we will study the effects of power-law correlated disorder and
will show that it has dramatic effects on the flat phase of an elastic
membrane, leading to a rich phase diagram.

\subsection{Long-range quenched disorder}
  
In this section we thus study the solutions to the SCSA equations for
short-range and long-range curvature and stress disorder. In Fourier
space both types of disorder can be characterized by disorder
variances that are power-laws in the wavevector
\begin{eqnarray}\label{PowerLaw}
\Delta_\kappa(\q) \simeq \Delta_\kappa q^{-z_\kappa}\; \quad , \quad 
\Delta_\mu(\q) \simeq  \Delta_\mu q^{-z_\mu}\;,\quad, \quad 
\Delta_\lambda(\q) \simeq  \Delta_\lambda q^{-z_\lambda}\;,
\end{eqnarray}
where the case of short-range disorder is described by the case 
$z_\kappa=z_\mu=z_\lambda=0$. We will specialize to the case 
$z_\mu=z_\lambda$, although it is
trivial to extend our analysis beyond this constraint. 
Our goal is to compute the roughness exponents 
$\zeta(z_\kappa, z_\mu),\ \zeta'(z_\kappa, z_\mu)$
in terms of these two independent range exponents.
In the following, instead of $\Delta_\lambda$, for convenience we will use 
\bea
&& \Delta_b(\q) \simeq  \Delta_b q^{-z_\mu} \quad , \quad  \Delta_b = \frac{b^2}{2 \mu^2} 
\left[2\Delta_\mu +(D-1)\Delta_\lambda \right] \quad , \quad  b={\mu(2\mu+D\lambda)\over 2\mu+\lambda}\;.
\eea
%b_{ab}(\q)&=&\left(b\;,\; - \frac{1}{T} \Delta_b(\q)\right)\;,\\
%&=&\left({\mu(2\mu+D\lambda)\over 2\mu+\lambda}\;,\;
%-\frac{1}{2 T} \left({2\mu+D\lambda\over2\mu+\lambda}\right)^2
%\left[2\Delta_\mu(\q)+(D-1)\Delta_\lambda(\q)\right]\right)\;.

\begin{figure}[htbp]
        \centering
        \includegraphics[width=0.7\textwidth,scale=1]{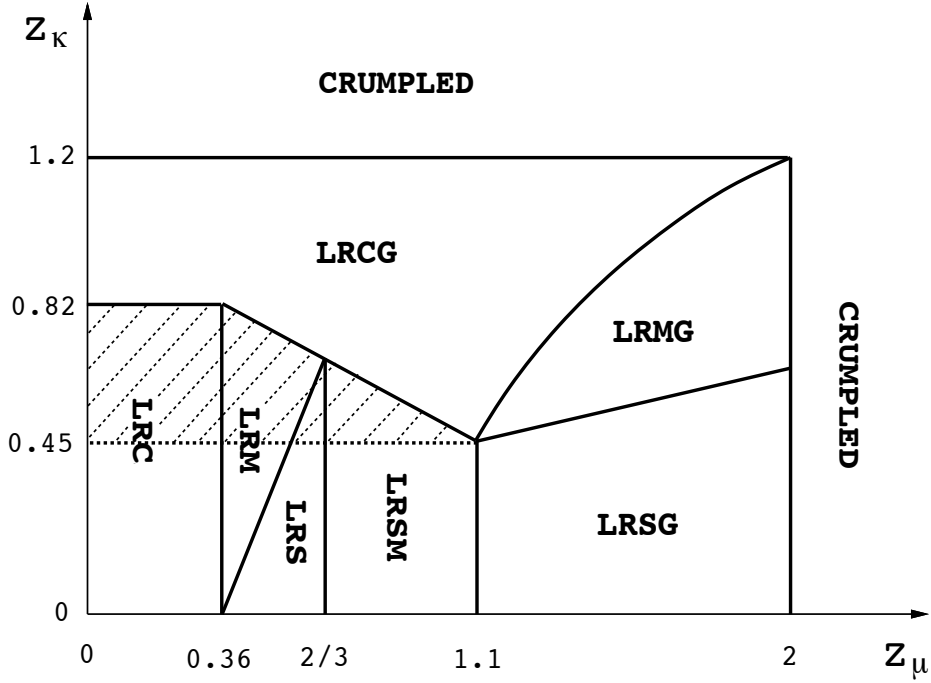}
        \caption{\label{Fig1_lr} Domain of stability (phase diagram)
          of the flat phases as a function of disorder range exponents
          $z_\mu,z_\kappa$.  (1) Disorder-dominated phases
          ($\zeta'>\zeta$): long-range stress glass (LRSG), long-range
          curvature glass (LRCG), long-range mixed glass (LRMG).  (2)
          Temperature-dominated phases ($\zeta'<\zeta$): LR curvature
          (LRC), LR mixed (LRM) and LR stress (LRS). (3) LRSM:
          marginal phase with $\zeta=\zeta'$.  The shaded area
          corresponds to a region of thermal phase transitions between
          several stable phases (LRCG and others). The region where
          the membrane crumples is also indicated.}
\end{figure}

In this section we begin with the simplest situation of stress-only
disorder, and build up in complexity, considering next the most
general situation of long-range disorder in both curvature and
stress. The richness of the problem leads us to consider many regimes
and identify several new phases, as shown in Fig. \ref{Fig1_lr}.  In
some regimes the long-range disorder is irrelevant, reducing to the
results of short-range (SR) disorder of the previous section. Upon
increasing the range exponents $z\kap,\ z_\mu$ we find that the
long-range disorder becomes relevant. In the phase diagram of Fig.
\ref{Fig1_lr} there are several cases. For $z\kap,\ z_\mu$ not too
large, there are temperature-dominated phases $\zeta >\zeta'$ (the
roughness of the membrane being then dynamic to leading order).  For
larger values of $z\kap,\ z_\mu$ there are disorder-dominated phases
$\zeta' >\zeta$, which are stable at any temperature (qualitatively
distinct phases from the case of SR disorder) and hence true glass
phases, where the roughness is frozen to leading order. We will refer
to these frozen ground state phases as ``flat-glasses'', by analogy
with spin-glass phases of magnetic spin systems. In between, there is
an intermediate range of stability for {\it marginal phases}, where
temperature and disorder play equal role $\zeta=\zeta'$. Each of these
three types of phases again subdivides into phases where either stress
disorder is more important, or curvature disorder is more important,
or, finally, mixed phases where both play a role. Finally upon
increasing $z\kap,\ z_\mu$ further, one enters the regime, where the
controlling roughness exponent $\zeta'$ exceeds unity, signaling an
instability to a crumpled-glass phase\cite{RLjphys}, presumably via a
disorder activated crumpling transition.

\subsection{Stress-only disorder}
\label{sec:stressonly}

Let us begin with the simpler case of stress disorder only,
$\Delta_\kappa(\q)=0$, which respects the up-down symmetry of the
membrane.  For $\Delta_\kappa(\q)=0$, if we assume the absence of
spontaneous breaking of the $h \to - h$ symmetry (or $O(d_c)$ symmetry
for $d_c>1$), we have, from \eqref{GGeqn}-\eqref{Pijexplicit}, $\tilde
\Delta_\kappa(\q)=G\Del(\kb)=\Pi\kd(\q)=\Pi\dd(\q)=0$ and our problem
is reduced to finding a single exponents $\eta(z_\mu)$ as a function
of the stress range exponent $z_\mu$.

In this subspace \eqref{couplings2}-\eqref{couplings2d} give
equations similar to Eq.\eqref{ALcoupling_stress} except that in the disorder
components in the numerator $\Delta_\mu(\q)$ and $\Delta_b(\q)$
dominate the vanishing $\Pi\dd(\q)$. Using \eqref{Pscale} this leads
to
\begin{eqnarray}\label{DeltaKapVanish}
\left(\tilde{\mu}(\q)\;,
\;\tilde{\Delta}_\mu(\q)\right)&=&
\left({1\over 2 T \Pi\kk(\q)}\;,\;{\Delta_\mu(\q)\over
4\mu^2 T^2 \Pi\kk(\q)^2}\right)\;,\\
&=& (D^2-1) \left({Z\kap^2\over 2 T \Pi(\eta,D)}q^{4-D-2\eta}\;,\;
{(D^2-1)\Delta_\mu Z\kap^4\over4\mu^2 T^2 \Pi(\eta,D)^2}q^{8-2D-4\eta-z_\mu}\right)\;,\nonumber  \\
\left(\tilde{b}(\q)\;,\;\tilde{\Delta}_b(\q)\right)&=&
\left({1\over (D+1) T \Pi\kk(\q)}\;,\;
{\Delta_b(\q)\over(D+1)^2 b^2 T^2\Pi\kk(\q)^2}\right)\;, \nonumber \\
&=& (D^2-1) \left({Z\kap^2\over(D+1) T \Pi(\eta,D)}q^{4-D-2\eta}\;,\;
{(D^2-1) \Delta_b Z\kap^4\over(D+1)^2 b^2 T^2
\Pi(\eta,D)^2}q^{8-2D-4\eta-z_\mu}\right)\;. \nonumber 
\end{eqnarray}
Using these expressions inside the equation for the renormalized
bending modulus, \eqref{GGeqn} (Eq.\eqref{GGeqn2} identically
vanishes) and dropping bare term that is
irrelevant at any fixed point with $\eta>0$, we obtain 
% {\red Please Leo can you recheck carefully all the amplitudes there,
%   they were not correct}
%
\begin{equation}\label{dkvEqn}
Z_\kappa k^{-\eta}=Z_\kappa {D(D-1) \over d_c} \frac{\Sigma(\eta,D)}{
\Pi(\eta,D)} k^{-\eta}
-{Z_\kappa A\kap} {2 \over d_c} 
{ \Sigma(2 \eta+z_\mu/2+D/2-2,\eta,D) \over {\Pi(\eta,D)^2}} 
k^{4-D-3\eta-z_\mu}\;,
\end{equation}
where we defined a universal dimensionless amplitude $A\kap$,
\begin{equation}\label{Akappa}
A\kap={(D-1) Z\kap^2\Delta\over b^2 T^2}
\end{equation}
in terms of an "effective disorder" bare variance defined as
\begin{equation}\label{effDelta}
\Delta= \Delta_b + \frac{(D-2) (D+1)^2 b^2}{4 \mu^2} \Delta_\mu = (D-1) \frac{b^2}{4 \mu^2} 
((2+D)(D-1) \Delta_\mu + 2 \Delta_\lambda)   \;,
\end{equation}
which is an explicitly positive quantity for $D>1$.\\

There are now two cases to consider. 

\begin{itemize}
\item
{\it Irrelevant disorder: homogeneous thermal fixed point.}
If $z_\mu<4-D-2\eta$ 
the disorder term is irrelevant and can be dropped at long wavelengths.
For this solution we find that balancing the remaining two
terms lead to the equation for $\eta$ 
\begin{equation}\label{ALagain}
\frac{D(D-1)}{d_c} \frac{\Sigma(\eta,D)}{\Pi(\eta,D)}=1\;,
\end{equation}
which is identical to Eq. \eqref{eqSCSApure} and determines the
thermal fixed point of the homogeneous membrane $\eta=\eta_{\rm
  pure}(D,d_c)$ studied in detail in Section \ref{analysisSCSAhomo}.

\item {\it Relevant disorder: disordered fixed point.}
If the stress disorder is longer range than the critical range, with
$z_\mu > z_\mu^{c1}=4-D-2\eta_{\rm pure}=\eta_u^{\rm pure}$, then 
the disorder term (last term) cannot be neglected in  Eq.\eqref{dkvEqn}.
Since it has a negative amplitude (it decreases the bending rigidity), 
the only choice is to balance all three terms by choosing
\begin{equation}\label{etaBalance}
\eta=2-\frac{D}{2}- \frac{z_\mu}{2} < \eta_{\rm pure} 
\end{equation}
to balance the powers of $k$. The result is an expression for 
a $z_\mu$-dependent universal amplitude $A\kap(z_\mu)$, 
which must be positive in
the range of stability of this new long-range fixed point,
\begin{equation}\label{univA}
A\kap(z_\mu)= \frac{D(D-1)}{2} \Pi(\eta(z_\mu),D) \left(1 - \frac{d_c \Pi(\eta(z_\mu),D)}{D(D-1) \Sigma(\eta(z_\mu),D)} \right)
\quad , \quad \eta(z_\mu)=2-\frac{D}{2}- \frac{z_\mu}{2}  < \eta_{\rm pure}\;.
\end{equation}
%%
%\begin{equation}\label{univA}
%A\kap(z_\mu)=\half D(D+1)\Pi(\eta(z_\mu))-
%\half d_c (D+1)\Pi(\eta(z_\mu))^2/\Sigma(\eta(z_\mu))\;.
%\end{equation}
%%

In the case of the physical membrane, $D=2,d_c=1$, we thus find within
SCSA that the LR disorder is relevant if $z_\mu > z_\mu^{c1} =
\eta_u^{\rm pure} = \frac{2}{7} \left(9-2 \sqrt{15}\right) \approx
0.3583$.  Denoting $z_\mu = z_\mu^{c1} + \delta$ we find that the
amplitude increases from zero as
\bea
A_\kappa(z_\mu) = 0.52874 \, \delta - 2.80214 \, \delta^2 +
O(\delta^3) \eea for small $\delta$. We have checked numerically that
it is positive for all values in $z_\mu^{c1}< z_\mu < 2$. Hence it is an
admissible (i.e., physical) solution of the SCSA equations.

Plugging these results back into \eqref{Gscale2} one obtains 
\bea \label{outofplane}
G_\kappa({\bf k}) \simeq \sqrt{ \frac{D-1}{A_\kappa(z_\mu)} \frac{\Delta}{b^2}} 
~ k^{- \frac{2+D+z_\mu}{2}} \quad , \quad \zeta = \frac{4-D}{4} + \frac{z_\mu}{4} > \zeta_{\rm pure}\;.
\eea

Since the effective bending rigidity is reduced by disorder, we find
that in this phase the roughness exponent of the out-of-plane
fluctuations is larger than the one of the pure homogeneous membrane.
The amplitude is non-universal, depending on the bare disorder (note
that the temperature dependence has dropped out since $Z_\kappa \sim
T$ at the fixed point). On the other hand one can check that the
renormalized elastic and disorder couplings scale as $\tilde \mu({\bf
  q}) \sim \tilde b({\bf q}) \sim T q^{z_\mu}$ and $\tilde
\Delta_\mu({\bf q}) \sim \tilde \Delta_b({\bf q}) \sim T^2 q^{z_\mu}$,
respectively. Hence there is a thermal renormalization (screening) of
these moduli, stronger at small $q$ than for the pure system, that one
can summarize by $\eta_u=\eta'_u=z_\mu > \eta_u^{\rm pure}$.  This
means that the connected or thermal part $\overline{\langle u u
  \rangle}_{\rm conn} \sim T/(\tilde b({\bf q})q^2) \sim
q^{-2-z_\mu}$ becomes independent of $T$ and similarly the
off-diagonal part $\overline{\langle u \rangle \langle u \rangle} \sim
\tilde \Delta_b({\bf q})/(\tilde b({\bf q})^2q^2) \sim
q^{-2-z_\mu}$, also is independent of $T$.
%{\red to make sure this is correct, wee should
%  really work out the phonon correlator with disorder (again!), and
%  quote the formula here} 
So both in-plane displacements and out-of-plane displacements (from
\eqref{outofplane}) are independent of temperature to leading order,
and hence are marginally frozen.  The membrane thus exhibits a glassy
behavior.  Note that the physical interpretation of the condition
$z_\mu > z_\mu^{c1} = \eta_u^{\rm pure}$ for this phase to occur,
corresponds to stress disorder producing in-plane displacements that
are larger than the thermal phonon fluctuations controlled by the pure
fixed point of the homogeneous membrane.  Since for physical membranes
$\zeta=1/2 + z_\mu/4$ in this phase, reaching the value $1$ at
$z_\mu^{c2}=2$ we expect the membrane to crumple for $z_\mu\geq
z_\mu^{c2}$, corresponding to disorder produced by frozen unscreened
disclinations \cite{NRpra}.

%{\red Leo I am unsure about the following statement that you wrote, can we discuss it?
%"Although the out-of-plane
%fluctuations are dynamic, the in-plane phonon fluctuations are
%marginally frozen and the membrane has a glassy behavior"
%I am ok for the marginal glass behavior of phonons, but it
%seems the out of plane are also disorder dominated: in which sense
%could they be "dynamic" ?}
%

Note that the assumption that we are looking for a solution with
$G_{\Delta}({\bf k})=0$, that is $\tilde \Delta_\kappa=0$, is
equivalent to assuming that $\langle h_{\bf k} \rangle =0$ in each
disorder configuration, i.e., no symmetry breaking in each disorder
configuration.  As we see below this parity symmetric subspace is
unstable to introduction of even weak curvature disorder. The local
in-plane strains introduced by the stress disorder can be screened by
local buckling of the membrane. In the absence of interactions the
resulting ground state is highly degenerate with respect to local
buckling of any sign. Simple arguments suggest that to minimize the
curvature energy, the interactions between "puckers" are expected to
be ``antiferromagnetic''\cite{RNpra}
%, although recent Monte Carlo
%simulations find both types of behaviors\cite{MCpuckers}. 
In any case the restriction of vanishing curvature disorder (symmetric
membrane) does not preclude the possibility of a collective
spontaneous buckling transition.
\end{itemize}

\subsection{Non-zero curvature disorder} 

We now study the more general problem of non-zero curvature disorder.
There are three types of solutions which we describe as
disorder-dominated phases, temperature-dominated (dynamic) phases, and
marginal phases, characterized by $\zeta'>\zeta$, $\zeta'<\zeta$,
$\zeta'=\zeta$, respectively. To study these phases we follow an
analysis similar to the previous sections. In each case we first
determine the renormalized quartic coupling constants via
Eqs. \eqref{couplings2} to \eqref{couplings2d}, and then use them
inside the Eq.\eqref{GGeqn}. The resulting phases are displayed in
Fig. \ref{Fig1_lr}.

It is useful to recall that in the case of SR disorder we found only
two fixed point solutions: (i) the "thermal" fixed point of the
homogeneous membrane, which is stable to weak short-range disorder
(although it does exhibit some signature of the presence of disorder,
e.g., $\zeta'>0$) and characterized by two exponents, $\eta=\eta_{\rm
  pure}$ (as in the strict absence of disorder) and $\eta' = 2 \eta= 2
\eta_{\rm pure}$, which describes the weak (irrelevant)
disorder-induced fluctuations, (ii) the SR disorder dominated $T=0$
fixed point with $\eta=\eta'=\eta_{\rm SRdis}$. Note that in this
Section we will denote systematically the exponents at the SR disorder
fixed point with the subscript ${\rm SRdis}$. As we will see a much
greater variety of phases will arise in the present case of LR
disorder.

\subsubsection{Disorder-dominated phases: $\zeta'>\zeta$} 

Let us now search for disorder-dominated phases. From the
classification in Section \ref{sec:4b} it amounts to either studying a
membrane at $T=0$ or a search for solutions of the SCSA equations with
$\eta'<\eta$, which implies $\zeta' >\zeta$ from
\eqref{zetazetaprime}, where temperature is subdominant.  In both
cases we can neglect $T \Pi_{\kappa\kappa}(\q)$ compared to
$\Pi_{\kappa\Delta}(\q)$, as was done to obtain
Eqs.\eqref{MLcoupling}, except that here we keep the possibly relevant
terms $\Delta_\mu(\q), \Delta_b(\q)$ in the numerator of
Eqs. \eqref{couplings2c}, \eqref{couplings2d}.  We find the
renormalized couplings as
\begin{eqnarray}\label{MLcoupling5}
&& \left(\tilde{\mu}(\q)\;,
\;\tilde{\Delta}_\mu(\q)\right) \simeq
\left({1\over 4\Pi\kd(\q)}\;,\;
{\frac{\Delta_\mu(\q)}{2 \mu^2} + \Pi\dd(\q)\over 8\Pi\kd(\q)^2}\right)\;,\\
&&\simeq  (D^2-1) \left({Z\kap Z\Del\over 4 \Pi(\eta,\eta',D)}q^{4-D-\eta-\eta'}\;,\;
{Z\kap^2\Pi(\eta',D)\over 8
\Pi(\eta,\eta',D)^2}q^{4-D-2\eta}
+ \frac{(D^2-1) \Delta_\mu}{16 \mu^2} \frac{(Z_\kappa Z_\Delta)^2}{\Pi(\eta,\eta',D)^2} 
q^{8 - 2 D - 2 \eta -2 \eta' - z_\mu}
\right) \nonumber \;,\\
&& \left(\tilde{b}(\q)\;,
\;\tilde{\Delta}_b(\q)\right) \simeq \frac{2}{D+1} \left(\tilde{\mu}(\q)\;,
\; {\frac{\Delta_b(\q)}{b^2 (D+1)} + \Pi\dd(\q)\over 8\Pi\kd(\q)^2}
\right)\;,\\
&& \simeq  2 (D-1) \left({Z\kap Z\Del\over 4 \Pi(\eta,\eta',D)}q^{4-D-\eta-\eta'}\;,\;
{Z\kap^2\Pi(\eta',D)\over 8
\Pi(\eta,\eta',D)^2}q^{4-D-2\eta}
+ \frac{(D-1) \Delta_b}{8 b^2} \frac{(Z_\kappa Z_\Delta)^2}{\Pi(\eta,\eta',D)^2} 
q^{8 - 2 D - 2 \eta -2 \eta' - z_\mu}\;
\right)\;, \nonumber
\end{eqnarray}
where we used \eqref{Pscale}, and assumed $\eta'<(4-D)/2$ and 
$\eta+\eta'<4-D$.

Substituting the resulting coupling constants into Eqs.\eqref{GGeqn}
we find the analog of Eqs.\eqref{MLggEqns}, generalized to long-range
disorder,
\begin{eqnarray}\label{LReqn01}
Z\kap k^{-\eta}&=&\kappa + {Z\kap k^{-\eta}D (D-1)\over
4d_c}\left[{2\Sigma({\eta+\eta'\over2},\eta',D)\over\Pi(\eta,\eta',D)}
-{\Sigma(\eta,D)\Pi(\eta',D)\over\Pi(\eta,\eta',D)^2}\right]\\
&&\;\;\;- D(D-1) Z\kap A\Del {\Sigma(\eta+\eta'+{z_\mu\over2}+{D\over2}-2,\eta,D)\over
4d_c \Pi(\eta,\eta',D)^2}k^{4-D-\eta-2\eta'-z_\mu}\;, \nonumber \\
{Z\kap^2\over Z\Del}k^{-2\eta+\eta'}&=& \label{LReqn02}
\Delta\kap k^{-z\kap}+{D (D-1) Z\kap^2\over4d_c Z\Del\Pi(\eta,\eta',D)^2}
\left[A\Del\Sigma(\eta+\eta'+{z_\mu\over2}+{D\over2}-2,\eta',D)
k^{4-D-2\eta-\eta'-z_\mu}\right.\\
&&\left.\;\;\;+\Sigma(\eta,\eta',D)\Pi(\eta',D)k^{-2\eta+\eta'}
\right]\;, \nonumber
\end{eqnarray}
where we have defined another explicitly positive universal amplitude,
\begin{equation}\label{Adelta}
A\Del={2 Z\Del^2\Delta\over D b^2}\;,
\end{equation}
and $\Delta$ was defined in \eqref{effDelta}. If we set the LR
disorders $A_\Delta=0$ and $\Delta_\kappa=0$, we recover the equations
\eqref{MLggEqns} for short-range disorder, which were analyzed in
Section \ref{sec:4b}. There it was found that in that case there are
no disorder dominated solution with $\eta'>\eta$, but there is one
disorder-dominated solution at $T=0$ (although marginal with respect
to introduction of temperature, since it is characterized by exponents
$\eta=\eta'=\eta_{\rm SRdis}$).

% old version not exactly correct ** recheck 
%Substituting the resulting
%coupling constants into Eqs.\ref{GGeqn} we find an analog of Eqs.\ref{MLggEqns} generalized to long-range disorder
%%
%\begin{eqnarray}\label{LReqn}
%Z\kap k^{4-\eta}&=&{Z\kap k^{4-\eta}D\over
%4d_c(D+1)}\left[{2\Sigma({\eta+\eta'\over2},\eta')\over\Pi(\eta,\eta')}
%-{\Sigma(\eta)\Pi(\eta')\over\Pi(\eta,\eta')^2}\right]\\
%&&\;\;\;-Z\kap A\Del {\Sigma(\eta+\eta'+{z_\mu\over2}+{D\over2}-2,\eta)\over
%2d_c(D+1)^2\Pi(\eta,\eta')^2}k^{8-D-\eta-2\eta'-z_\mu}\;,\\
%{Z\kap^2\over Z\Del}k^{4-2\eta+\eta'}&=&
%\Delta\kap k^{4-z\kap}+{D Z\kap^2\over4d_c(D+1)^2Z\Del\Pi(\eta,\eta')^2}
%\left[A\Del\Sigma(\eta+\eta'+{z_\mu\over2}+{D\over2}-2,\eta')
%k^{8-D-2\eta-\eta'-z_\mu}\right.\\
%&&\left.\;\;\;+(D+1)\Sigma(\eta,\eta')\Pi(\eta')k^{4-2\eta+\eta'}
%\right]\;,
%\end{eqnarray}
%%
%where we have defined another explicitly positive universal amplitude
%%
%\begin{equation}\label{Adelta}
%A\Del={Z\Del^2\Delta\over4\mu^2}\;.
%\end{equation}
%%

We now describe the relevance of the two types of LR disorder. By comparing the first term on the right-hand side of Eq. \eqref{LReqn02}
to the left-hand side, one sees that the long-range curvature disorder 
is irrelevant %at $T \neq 0$ 
as long as 
\bea \label{curvcond} 
z_\kappa< 2\eta - \eta'  \quad \Leftrightarrow \quad \text{LR curvature disorder irrelevant\;.} 
\eea 
Thus we see that at the $T=0$ SR disorder fixed point, the LR
curvature disorder is irrelevant, as long as
\bea
z_\kappa< \eta_{\rm SRdis} 
\eea 
%{\red I removed some not quite coherent and very confusing text here, it is %commented in the tex file here
%please recheck}

%{\red Not satisfactory, move somewhere else Note that in fact the condition \eqref{curvcond} holds for any $\eta,\eta'$ (it is not restricted to disorder dominated phases) hence we can study it also at the thermal fixed point of the membrane in
%presence of (irrelevant) weak disorder, and we see that there LR curvature disorder 
%is always relevant there. $z_\kappa<2\eta-\eta'$. Therefore, for
%example, at AL fixed point, where $2\eta_{AL}=\eta'_{AL}$, the long-range
%curvature disorder is marginally relevant. On the other hand at
%at ML fixed point, where $\eta_{ML}=\eta'_{ML}$,
%the long-range curvature disorder is relevant, with eigenvalue $\eta_{ML}$.}
%{\red As we mentioned previously, disorder-free AL fixed point is the only 
%stable finite temperature phase when disorder is short-range.
%This can be seen from the fact that there are no 
%disorder dominated solutions ($\eta'<\eta$)
%to Eq.\eqref{LReqn} when both long-range curvature and long-range stress 
%disorders are irrelevant.}

By looking at the last two terms of either equations
Eqs. \eqref{LReqn01}, \eqref{LReqn02} we see that the long-range stress disorder is irrelevant for
\bea
z_\mu<4-D-2\eta'=\eta_u'  \quad \Leftrightarrow \quad \text{LR stress disorder irrelevant\;.}  
\eea 
This last condition is consistent with the one we
found in analyzing $\Delta\kap(\q)=0$ case, except here it is $\eta'$
instead of $\eta$, that controls the roughness of the phase.

We therefore look for long-range (LR) disordered solutions to the above
equations. We find that there are three possible LR phases that we
call Long-Range Stress Glass (LRSG), Long-Range Curvature Glass (LRCG), 
and Long-Range Mixed Glass (LRMG), corresponding to the relevant 
stress, curvature and both disorders, respectively.\\

(1-i) {\bf Long-Range Stress Glass (LRSG)}: \\

This solution is defined by relevance of LR stress disorder and irrelevance of 
the LR curvature disorder term $\Delta_\kappa k^{4-z_\kappa}$ and is therefore
only stable for $z\kap<2\eta-\eta'$. The relevance of LR stress requires
that all terms in \eqref{LReqn01} scale similarly, which implies
\begin{equation}\label{etaPrime}
\eta'=\eta'(z_\mu):=2-D/2-z_\mu/2\;.
\end{equation}
Combining this result with Eqs.\eqref{LReqn01}, \eqref{LReqn02} leads 
to the equation which determines $\eta=\eta(z_\mu)$ (plugging $\eta'=\eta'(z_\mu)$ in the 
first equation)
and the universal amplitude $A\Del(z_\mu)$ (plugging both $\eta=\eta(z_\mu)$ and $\eta'=\eta'(z_\mu)$ in the second equation)
\begin{eqnarray}\label{LRetaA}
&&{D (D-1) \Sigma({\eta+\eta'\over2},\eta',D)\over2d_c\Pi(\eta,\eta',D)} 
-{\Sigma(\eta,D)\over \Sigma(\eta,\eta',D)} =1
\;,\\
&&A\Del(z_\mu)={4d_c\Pi(\eta,\eta',D)^2\over D(D-1)\Sigma(\eta,\eta',D)}  
- \Pi(\eta',D)\;. \nonumber
\end{eqnarray}
We note that the first equation coincides with the equation obtained
from the system \eqref{MLetaEqns} (obtained for short-range disorder)
when eliminating $\Pi(\eta,\eta',D)^2$ from the second equation there and
plugging it into the first. Hence the relation \eqref{LRetaA} between
$\eta$ and $\eta'$ also holds for SR disorder, but the value of the exponent
$\eta'$ here is imposed by the LR disorder as in \eqref{etaPrime}. Also, if
one sets $A\Del(z_\mu)=0$ in the second equation, one recovers the system 
\eqref{MLetaEqns}. 

On general grounds, and given the assumptions leading to \eqref{etaPrime} and 
\eqref{LRetaA}, we expect that this LRSG phase is stable as long as the
following conditions hold
\begin{eqnarray}\label{condStable}
\eta'(z_\mu)&<&\eta(z_\mu)\;,
\;\;{\rm equivalent\; to}\;\;
\zeta'(z_\mu)>\zeta(z_\mu)\;, \quad \text{and} \quad A\Del(z_\mu) >0\;, \\
\eta'(z_\mu)&>&0\;,
\;\;{\rm equivalent\; to}\;\;
\zeta'(z_\mu)<1\;. \label{condStable2}
\end{eqnarray}
The first condition is that it is disorder-dominated and corresponds
to the assumption made in deriving \eqref{LRetaA}. As we discuss below
it corresponds to the condition $z_\mu > z_\mu^{c1} = 2 - 2 \eta_{\rm
  SRpure}$. The second condition states that the membrane is flat and
defines the right boundary of LRSG, with $z_\mu^{c2}=4-D$, beyond which
the membrane crumples, with $\eta'(z_\mu^{c2})=0$. An additional
condition used in the derivation, $\eta'(z_\mu)+\eta(z_\mu)<4-D$,
should also be verified.  Finally one also needs
\bea
z\kap&<&2\eta(z_\mu)-\eta'(z_\mu)\;,
\;\;{\rm equivalent\; to}\;\;
z\kap<2 \eta(z_\mu)+{D\over2}-2+z_\mu/2\;,
\eea
which is the requirement that the LR curvature disorder is
irrelevant. It defines the upper boundary of the LRSG, above which
both LR curvature and LR stress disorders are relevant and the
membrane is described by the LRMG phase.

To analyze the solutions to \eqref{LRetaA} we must start with the SR
fixed point, which is a solution to \eqref{LRetaA} with
$\eta=\eta'=\eta_{\rm SRdis}$ and $A\Del(z_\mu)=0$.  A true LR
solution of \eqref{LRetaA} with $A\Del(z_\mu)>0$ emerges from this SR
fixed point for $z_\mu>z_\mu^{c1}$ where $z_\mu^{c_1}$ is defined by
the condition $\eta'(z_\mu^{c_1}) = \eta_{\rm SRdis}$,
which from \eqref{etaPrime} gives $z_\mu^{c_1} = 2 - 2 \eta_{\rm SRdis}$. \\

Let us now discuss in more details the case of the physical membrane,
with $D=2$ and $d_c=1$. We find
\bea
&& \eta_{\rm SRdis} = \sqrt{6}-2 = 0.44949.. \quad , \quad z_{\mu}^{c1}=6 - 2 \sqrt{6} =1.10102\;, \\
&& \eta'(z_\mu) = \eta_{\rm SRdis} - \frac{z_\mu-z_\mu^{c1}}{2}\;, \\
&& \eta(z_\mu) = \eta_{\rm SRdis} - \frac{23-8\sqrt{6}}{58} (z_\mu-z_{\mu}^{c1})
- 0.0112487 (z_\mu-z_{\mu}^{c1})^2 + ...\;, \\
&& A\Del(z_\mu)= 0.0494198 (z_\mu-z_{\mu}^{c1}) - 0.0548815 (z_\mu-z_{\mu}^{c1})^2 + ...\;, 
\eea 
where $\frac{23-8\sqrt{6}}{58}=0.0586911..$. One thus verifies that
the solution obeys the above conditions \eqref{condStable}, as well as
$\eta(z_\mu)+ \eta'(z_\mu)<2$. Note that $T$ does not appears and hence for
$z_\mu > z_\mu^{c1}$ this phase is stable at any temperature $T$. At
the lower boundary $z_\mu=z_\mu^{c1}$ the marginal phase, characterized
by $\eta'=\eta$ becomes the ground state (itself marginally unstable
to temperature).  Below $z_\mu^{c1}$ the LRSG phase becomes unstable
to the marginal LRSM phase, which is analyzed below (see
Fig. \ref{Fig1_lr}). This lower boundary is a straight vertical line
because in LRSG phase the exponents are independent of $z_\kappa$.  At
the upper boundary $z_\mu=z_\mu^{c2}=2$ the flat membrane is unstable,
and we find that $\eta(z_\mu^{c2})\approx0.3877$
with $A\Del(z_\mu)=0.0209$.\\

%for $D=2,d_c=1$. 
%The last condition gives the requirement that
%the phase is dominated by disorder at long scales and defines the lower
%critical $z_\mu^{c1}$ at which $\eta'(z_\mu^{c1})=\eta(z_\mu^{c1})$ and
%marginal phase characterized by $\eta'=\eta$ becomes the ground state.
%Using this condition we find that Eq.\ref{LRetaA{a}} reduces to 
%Eq.\ref{MLfinalEqn}. Therefore on this lower phase boundary we have
%$\eta'(z_\mu^{c1})=\eta(z_\mu^{c1})=\eta_{ML}\approx0.449$, for $D=2,\ d_c=1$. 
%Combining this with the expression for $\eta'(z_\mu)$ Eq.\ref{etaPrime} we find 
%that $z_\mu^{c1}=\eta_u^{ML}\approx 1.102$,
%
%below which LRSG becomes unstable to the marginal LRSM phase to be
%analyzed below. This lower boundary is a straight vertical line because in
%LRSG phase the exponents are independent of $z_\kappa$.
%It is satisfying to observe that the universal
%intrinsically positive amplitude $A\Del(z_\mu^{c1})$ of Eq.\ref{LRetaA{b}}
%is positive everywhere inside the domain of stability of LRSG, and becomes
%negative precisely at $z_\mu^{c1}$ where the LRSG solution is no longer valid.

(1-ii) {\bf Long-Range Curvature Glass (LRCG)}: \\

In this phase the LR stress disorder is irrelevant, i.e., $z_\mu < 4-D
- 2\eta'$ and the LR curvature disorder is relevant.  At large scales
we can therefore drop terms in Eqs.\eqref{LReqn01}, \eqref{LReqn02},
that involve stress disorder (those involving the amplitude $A\Del$).
Furthermore, for LR curvature disorder to be relevant we must balance
the powers of $k$ of the term $\Delta\kap k^{-z_\mu}$ with the
remaining terms of \eqref{LReqn02}, which leads to the relation
\begin{equation}\label{zKappa}
2 \eta(z\kap) - \eta'(z\kap)=z_\kappa\;,
\end{equation}
where we denote $\eta(z\kap)$ and $\eta'(z\kap)$ the (as yet unknown)
exponents in the LRCG phase. Dropping the irrelevant stress disorder
term, Eq. \eqref{LReqn01} then gives the condition
\begin{equation}\label{CGetaeqn}
{D(D-1)\over4d_c}
\left[{2\Sigma({\eta+\eta'\over2},\eta',D)\over\Pi(\eta,\eta',D)}
-{\Sigma(\eta,D)\Pi(\eta',D)\over\Pi(\eta,\eta',D)^2}\right]=1\;.
\end{equation}
These equations together determine the anomalous exponents $\eta(z\kap)$
and $\eta'(z\kap)$ in this phase. In addition one can define another
intrinsically positive amplitude, 
\be
A\kd := {\Delta\kap Z\Del\over Z\kap^2}\;.
\ee
Dropping the irrelevant stress-disorder term in Eq.\eqref{LReqn02} 
allows to determine the value of this amplitude as
\begin{eqnarray}\label{Akd}
A\kd = A\kd(z\kap) := 1-{D (D-1) \Sigma(\eta,\eta')\Pi(\eta')\over4d_c \Pi(\eta,\eta')^2}\;,
\end{eqnarray}
where one must plug in the right-hand side the values of the exponents
$\eta=\eta(z_\mu)$ and $\eta'=\eta'(z_\mu)$ of the LRCG phase. 
It is thus a universal amplitude (i.e., depending only on $z_\kappa$)
characteristic of the LRCG phase. 

From the above derivation we can see that the region of stability of the
LRCG ground state is defined by the following boundaries
\begin{eqnarray}\label{condStableCG}
z_\mu&<&4-D-2\eta(z\kap)=\eta_u(z\kap)\;,\\
\eta'(z\kap)&<&\eta(z\kap)\;,
\;\;{\rm equivalent\; to}\;\;
\zeta'(z\kap)>\zeta(z\kap)\;, \label{condStableCG2}\\
\eta'(z\kap)&>&0\;,
\;\;{\rm equivalent\; to}\;\;
\zeta'(z\kap)<1\;, \label{condStableCG3} 
\end{eqnarray}
where the first condition enforces the irrelevance of the 
LR stress disorder and determines the lower-right
boundary of the LRCG phase towards the LRMG phase, in which both LR
curvature and LR stress disorders are simultaneously relevant, 
as illustrated in Fig. \ref{Fig1_lr}.
The conditions \eqref{condStableCG2}, \eqref{condStableCG3} are
identical to Eqs. \eqref{condStable}, \eqref{condStable2},
except that in LRCG the exponents are only
functions of $z_\kappa$, which is why the lower- and upper-boundaries are
straight horizontal lines.

To analyze the system of equations \eqref{zKappa} and \eqref{CGetaeqn}
we must again start from the SR disorder fixed point. That determines
the lower boundary $z\kap =z\kap^{c1}$, such that
$\eta'(z\kap^{c1})=\eta(z\kap^{c1})=\eta_{SRdis}$.  From
\eqref{zKappa} it then gives $z\kap^{c1}=\eta_{SRdis}$. On this
boundary the amplitude $A\kd(z\kap^{c1})=0$. For $z\kap > z\kap^{c1}$
there is a nontrivial solution for the LRCG with
$A\kd(z\kap^{c1})>0$. At the upper boundary, $z\kap = z\kap^{c2}$,
defined by the equation \eqref{condStableCG3}, the flat phase is
unstable and the membrane crumples. Let us now specialize to the
physical membrane $D=2, d_c=1$. Then one finds, for $z_\kappa$, above,
but close to the lower-boundary $z\kap^{c1}$
\bea
&&  z_{\kappa}^{c1}= \eta_{\rm SRdis} = \sqrt{6}-2 = 0.44949..\;, \\
&& \eta'(z_\kappa) = \eta_{\rm SRdis} + a' (z_\kappa-z_{\kappa}^{c1}) + .. \quad , \quad a'=2 a-1= -0.46305..\;, \\
&& \eta(z_\kappa) = \eta_{\rm SRdis}  + a (z_\kappa-z_{\kappa}^{c1})  + .. \quad , \quad a= \frac{2}{19} \left(5-\sqrt{6}\right)=0.268475..\;, \\
&& A\Del(z_\kappa)= A (z_\mu-z_{\mu}^{c1})  + .. \quad , \quad A=
\frac{1}{95} \left(108+43 \sqrt{6}\right)= 2.24556..\;,
\eea 
%with $a= \frac{2}{19} \left(5-\sqrt{6}\right)=0.268475..$, $a'=2 a-1= -0.46305$ 
%and $A= \frac{1}{95} \left(108+43 \sqrt{6}\right)= 2.24556$. 
which satisfies all the required stability conditions. More generally, 
solving Eqs. \eqref{zKappa} and \eqref{CGetaeqn} in the whole
LRCG phase we confirm that $\eta'(z\kap)$
decreases and $\eta(z\kap)$ increases as $z\kap$ increases. 
For the upper boundary defined by equation \eqref{condStableCG3} 
at which the membrane crumples, we find $z\kap^{c2}\approx 1.19695$, 
with $\eta'(z\kap^{c2})=0$ and $\eta(z\kap^{c2})=z\kap^{c2}/2 \approx 0.598473$.\\

%
%for $D=2,d_c=1$. As the lower boundary $z\kap^{c1}$ is approached from above
%$\eta'(z\kap)$ approaches $\eta(z\kap)$ from below, and the universal
%amplitude $A\kd(z\kap)$, which is positive everywhere inside LRCG, vanishes,
%$A\kd(z\kap^{c1})=0$. On this boundary to the marginal phases, 
%$\eta'(z\kap^{c1})=\eta(z\kap^{c1})$, and the defining equation \ref{CGetaeqn}
%reduces to Eq.\ref{MLfinalEqn}. Using Eq.\ref{zKappa} we therefore find 
%$z\kap^{c1}=\eta_{ML}\approx0.449$, for 

(1-iii) {\bf Long-Range Mixed Glass (LRMG)}: \\

A new glassy ground state and phase results when both types of LR
disorder are relevant.  The LRMG phase borders the LRSG, LRCG and the
Crumpled Glass phases, with the boundaries already defined in the
discussion of LRSG and LRCG, above.  The relevance of both types of
disorders requires us to balance the powers of $k$ in both
Eqs.\eqref{LReqn01},\eqref{LReqn02} and therefore the exponents are
completely determined to be
\begin{eqnarray}\label{LRMGeta}
\eta(z\kap,z_\mu)&=&{z_\kappa\over2}-{z_\mu\over4}+1-{D\over4}\;,\\
\eta'(z_\mu)&=&2-{D\over2}-{z_\mu\over2}\;.
\end{eqnarray}
We expect these expressions to be exact, since they are a result
of dimensional analysis.
The two universal amplitudes defined in the previous sections are also determined by
the Eqs.\eqref{LReqn01},\eqref{LReqn02}  and are given by
\begin{eqnarray}\label{AdAkd}
A\Del = A\Del(z\kap,z_\mu)&=&{1\over\Sigma(\eta,D)}
\left[2\Pi(\eta,\eta',D)\Sigma({\eta+\eta'\over2},\eta',D)
-\Sigma(\eta,D)\Pi(\eta',D)\right]
-{4d_c\Pi(\eta,\eta',D)^2\over D(D-1) \Sigma(\eta,D)}\;,\\
A\kd = A\kd(z\kap)&=&1+{\Sigma(\eta,\eta',D)\over\Sigma(\eta,D)}
-{D(D-1)\Sigma(\eta,\eta',D)\Pi(\eta',D)\over4d_c\Pi(\eta,\eta',D)^2}
\\
&& +{D(D-1)\Sigma(\eta,\eta',D)\over4d_c\Sigma(\eta,D)}
\left[{\Sigma(\eta,D)\Pi(\eta',D)\over\Pi(\eta,\eta',D)^2}
-{2\Sigma({\eta+\eta'\over2},\eta',D)\over\Pi(\eta,\eta',D)}\right]\;.
\end{eqnarray}
The consistency of these amplitudes with the results of the previous section
can easily be checked. We summarize by recalling the equations of the phase boundaries
\bea
&& z_\kappa = 2 \eta_{LRSG}(z_\mu) + \frac{D}{2}-2 + \frac{z_\mu}{2} \quad , \quad \text{LRSG-LRMG boundary,} \\
&& z_\mu = 4 - D - 2 \eta'_{LRCG}(z_\kappa) \quad , \quad \text{LRCG-LRMG boundary,} \\
&& z_\mu=2 \quad \text{or} \quad z_\kappa = z_\kappa^{(c_2)} \quad , \quad \text{boundary to crumpled phase.}
\eea 
Note that the condition that the LRMG is a disorder-dominated phase is
\bea
\eta'(z_\mu) < \eta(z\kap,z_\mu) \quad \Leftrightarrow \quad 2 z_\kappa + z_\mu > 4 - D\;.
\label{cond4} 
\eea 
As can be seen in Fig. \ref{Fig1_lr} this condition is always satisfied
in the LRMG phase: the triple point LRCG-LRMG-LRSG, where the
three phases has coordinates given by $z_\mu^{c_1}=2 - 2 \eta_{\rm SRdis}$ and
$z_\kappa = \eta_{\rm SRdis}$ is located precisely on the line where \eqref{cond4}
holds as an equality. 

%%
%\begin{eqnarray}\label{AdAkd}
%A\Del(z\kap,z_\mu)&=&{D(D+1)\over2\Sigma(\eta)}
%\left[2\Pi(\eta,\eta')\Sigma({\eta+\eta'\over2},\eta')
%-\Sigma(\eta)\Pi(\eta')\right]
%-{2d_c(D+1)^2\Pi(\eta,\eta')^2\over\Sigma(\eta)}\;,\\
%A\kd(z\kap)&=&1+{D\Sigma(\eta,\eta')\over2\Sigma(\eta)}
%-{D\Sigma(\eta,\eta')\Pi(\eta')\over4d_c(D+1)\Pi(\eta,\eta')^2}
%+{D^2\Sigma(\eta,\eta')\over8d_c(D+1)\Sigma(\eta)}
%\left[{\Sigma(\eta)\Pi(\eta')\over\Pi(\eta,\eta')^2}
%-{2\Sigma({\eta+\eta'\over2},\eta')\over\Pi(\eta,\eta')}\right]\;.
%\end{eqnarray}
%%

\subsubsection{Temperature-dominated phases, $\zeta'<\zeta$}

We now examine solutions to the SCSA equations \eqref{GGeqn},
\eqref{GGeqn2}, \eqref{couplings2}-\eqref{couplings2d},
\eqref{Pijexplicit} of the heterogeneous membrane, for which the
roughness of the membrane is dominated by (dynamic) thermal
fluctuations, i.e., solutions with $\eta' > \eta$.  The thermal fixed
point of the homogeneous membrane, that we have already examined is an
example of this type of a solution, at which disorder is completely
irrelevant. This solution falls on the $z_\mu$-axis of
Fig. \ref{Fig1_lr} for $0<z_\mu<0.358$. We have already seen that there
are no short-range disordered solutions and we therefore examine LR
disordered fixed points of temperature dominated type.  As we will see
below, there turns out to be three solutions of this type which we
name Long-Range Curvature (LRC), Long-Range Stress (LRS), and
Long-Range Mixed (LRM), in analogy with the glass solutions we have
studied in the previous section.

For all temperature dominated solutions $\eta'>\eta$ and $T>0$, at
long wavelengths we have, from \eqref{Gscale2}, $G\Del(\kb)\ll
G\kap(\kb)$ and therefore $\Pi\kd(\q)\ll T\Pi\kk(\q)$ from
\eqref{Pscale}, i.e., thermal screening dominates.  Hence the term
$\Pi\kd(\q)$ can be neglected in the numerators of the renormalized
couplings \eqref{couplings2}, as was done to derive \eqref{ALcoupling2}
in the case of SR disorder. Here, however we do not drop
$\Delta_\mu(\q), \Delta_b(\q)$ in the numerators of
Eqs.\eqref{couplings2c},\eqref{couplings2d}, because they might turn
out to be relevant.

Let us assume that $0< \eta' < \frac{4-D}{2}$, then $\Pi\dd(\q)$
diverges at small $q$, and one finds the LR generalization of
\eqref{ALcoupling} or, equivalently the generalization of
\eqref{DeltaKapVanish} in presence of curvature disorder,
\begin{eqnarray}\label{ALcouplingLR}
&& \left(\tilde{\mu}(\q)\;,
\;\tilde{\Delta}_\mu(\q)\right) \simeq
\left({1\over 2 T \Pi\kk(\q)}\;,\;{\frac{\Delta_\mu(\q)}{2 \mu^2} 
+ \Pi\dd(\q)\over 2 T^2 \Pi\kk(\q)^2}\right)\;,\\
&&\simeq  (D^2-1) \left({Z\kap^2\over 2 T \Pi(\eta,D)}q^{4-D-2\eta}\;,\;
{Z\kap^4\Pi(\eta',D)\over 2 T^2
Z\Del^2\Pi(\eta,D)^2}q^{4-D+2\eta'-4\eta} + \frac{(D^2-1) \Delta_\mu Z_\kappa^4}{
4 T^2 \mu^2 \Pi(\eta,D)^2} q^{8-2 D  - 4 \eta - z_\mu} \right)\;,\nonumber \\
&& \left(\tilde{b}(\q)\;,
\;\tilde{\Delta}_b(\q)\right)\simeq 
\frac{2}{D+1} \left(\tilde{\mu}(\q)\;, {\frac{\Delta_b(\q)}{(D+1)b^2} 
+ \Pi\dd(\q)\over 2 T^2 \Pi\kk(\q)^2} \right)\;, \nonumber \\
&& \simeq 2(D-1)
\left({Z\kap^2\over 2 T \Pi(\eta,D)}q^{4-D-2\eta}\;,
\;
{Z\kap^4\Pi(\eta',D)\over 2 T^2
Z\Del^2\Pi(\eta,D)^2}q^{4-D+2\eta'-4\eta} + \frac{(D-1) \Delta_b Z_\kappa^4}{
2 T^2 b^2 \Pi(\eta,D)^2} q^{8-2 D  - 4 \eta - z_\mu} \right)\;. \nonumber 
%\left(\tilde{b}(\q)\;,
%\;\tilde{\Delta}_b(\q)\right)&\simeq& (D^2-1)
%\left({1\over T (D+1)\Pi\kk(\q)}\;,\;
%{\Pi\dd(\q)\over T^2 (D+1)\Pi\kk(\q)^2}\right)\;\\
%&=&\left({Z\kap^2\over T (D+1)\Pi(\eta,D)}q^{4-D-2\eta}\;,\;
%{Z\kap^4\Pi(\eta',D)\over T^2 (D+1)
%Z\Del^2\Pi(\eta,D)^2}q^{4-D+2\eta'-4\eta}\right)\;,
\end{eqnarray}
Plugging these formula into \eqref{GGeqn} we obtain
\begin{eqnarray}\label{EtallEtaP}
k^{-\eta}&=&{D(D-1) \Sigma(\eta,D)\over
d_c \Pi(\eta,D)} k^{-\eta}
%-{D\Pi(\eta')\Sigma(2\eta-\eta',\eta) \over d_c(D+1)\Pi(\eta)^2}
%{Z\kap^3\over Z\Del^2}k^{4-3\eta+2\eta'} 
+ O(k^{\eta'-2 \eta}) 
- O(k^{\min(2 \eta',4-D)-3 \eta}) \\
&&\;\;\;- A\kap{2 \Sigma(2 \eta+z_\mu/2+D/2-2,\eta,D)
\over{d_c \Pi(\eta,D)^2}}k^{4-D-3\eta-z_\mu}\;, \nonumber \\
k^{-2\eta+\eta'}&=&A_{\kappa \Delta} k^{-z\kap} \nonumber 
 +
O(k^{\min(2 \eta',4-D) + \eta'-4\eta})
%{D\over d_c(D+1)}{Z\kap^4\Pi(\eta')\over Z\Del^3\Pi(\eta)^2}\Sigma(2\eta-\eta',\eta')
%k^{4+3\eta'-4\eta}
\\
&&\;\;\;+ A\kap{2 \Sigma(2 \eta+z_\mu/2+D/2-2,\eta',D)
\over{d_c \Pi(\eta,D)^2}}k^{4-D-4\eta+\eta'-z_\mu}\;, \nonumber
\end{eqnarray}
where we divided the first line by $Z_\kappa$ and the second by
$Z_\kappa^2/Z_\Delta$.  Note that there are terms which we did not
write explicitly as they are subdominant when $\eta' >\eta$ and do not
involve LR disorder. Their analysis is exactly as in \eqref{ALggEqns}
and they can be dropped. We recall that we defined earlier the two
dimensionless amplitudes
\be \label{defAmp} 
A_\kappa = {(D-1) Z\kap^2\Delta\over b^2 T^2} \quad , \quad 
A_{\kappa \Delta}=\frac{\Delta_\kappa Z_\Delta}{Z_\kappa^2}\;.
\ee 
%We observe that for the temperature dominated solutions that we are 
%looking for here the second terms in Eqs.\ref{EtallEtaP{a,b}}
%are irrelevant with respect to the left hand sides of respective equations.
%This is true because $3\eta-2\eta'<\eta$ and  $4\eta-3\eta'<2\eta-\eta'$ for
%solutions with $\eta<\eta'$. At long wavelengths we therefore drop these terms
%in both equations obtaining a simplified set of equations
%
%\begin{eqnarray}\label{EtaSimple}
%k^{4-\eta}&=&{D\Sigma(\eta)\over
%d_c(D+1)\Pi(\eta)}k^{4-\eta}
%-A\kap{2 \Sigma(2 \eta+z_\mu/2+D/2-2,\eta)
%\over{d_c (D+1)^2\Pi(\eta)^2}}k^{8-D-3\eta-z_\mu}\;,\\
%k^{4-2\eta+\eta'}&=&A\kd k^{4-z\kap}
%+A\kap{2 \Sigma(2 \eta+z_\mu/2+D/2-2,\eta')
%\over{d_c (D+1)^2\Pi(\eta)^2}}k^{8-D-4\eta+\eta'-z_\mu}\;,
%\end{eqnarray}
%%
%which we analyze next.
The relevance of the stress disorder in these equations is determined
by the value of $z_\mu$.  We further observe that if we discard
irrelevant terms the first equation in the system \eqref{EtallEtaP}
does not involve curvature disorder, neither $\eta'$ or $z\kap$, and
therefore is identical to Eq.\eqref{dkvEqn}. Hence, in all the
temperature-dominated phases studied in this section, all quantities
related to the (connected) thermal correlations, i.e., $\eta(z_\mu)$,
$A\kap(z_\mu)$, the small $q$ behavior of $G_\kappa(\q)$ and the
thermal roughness exponent $\zeta$, are identical to those found in
section \ref{sec:stressonly}, where stress-only disorder was
considered (i.e., with $\Delta\kap(\q)=0$). In these thermal phases
with LR disorder, the thermal (connected) and disorder (off-diagonal)
correlations effectively mutually decouple at large scale. Let us now
study these phases in more details. \\

(2-i) {\bf long-range curvature disorder (LRC)}: \\

As the name implies, the LRC flat phase is characterized by relevant
LR curvature disorder and irrelevant LR stress disorder. Thus it must
satisfy the condition of irrelevance of LR stress disorder, which is
\be z_\mu < 4 - D - 2 \eta\;, \label{irrstress} \ee in which case one
can neglect in both equations \eqref{EtallEtaP} at small $k$ the terms
proportional to $A_\kappa$. Consider first the case $z\kap=0$: then
the situation is very similar to the case of short-range disorder,
studied in Section \ref{sec:4b} with both types of disorder.  Indeed
there, the stress disorder is also irrelevant, and balancing the two
first terms in the second equation in \eqref{EtallEtaP} we obtain
$\eta'=2\eta_{\rm pure}$, where $\eta=\eta_{\rm pure}$ is the exponent
of the homogeneous membrane. For physical membranes $D=2,d_c=1$ this
leads to $\eta' \approx1.66$.  Now when LR curvature disorder is
included, i.e., $z_\kappa>0$, balancing again the two first terms in
the second equation in \eqref{EtallEtaP} gives $\eta'(z\kap)$ and the
universal amplitude $A\kd$ in the LRC phase
\begin{eqnarray}\label{LRCeta}
\eta&=&\eta_{\rm pure}\;,\\
\eta'(z\kap)&=&2\eta_{\rm pure}-z\kap\;, \nonumber \\
A\kd&=&1\;. \nonumber 
\end{eqnarray}

This phase is only stable inside the region $z_\mu<z_\mu^{c3}$, 
outside of which the LR stress disorder becomes relevant, and a new thermal 
phase which we call LRM appears. From \eqref{irrstress} 
the boundary separating LRC and LRM phases is attained at 
\begin{equation}\label{zmucLRC}
z_\mu = z_\mu^{c3} =4-D-2\eta_{\rm pure}=\eta_u^{\rm pure}\;.
\end{equation}
The upper boundary of LRC, $z\kap^{c3}$, at which LRCG becomes more stable at 
all temperatures and all strengths of disorder is determined by 
the condition $\eta'(z\kap^{c3})=\eta(z\kap^{c3})=\eta_{\rm pure}$, giving from 
Eq. \eqref{LRCeta}
\begin{equation}\label{zkapcLRC}
z\kap^{c3}=\eta^{\rm pure}\;.
\end{equation}
\\

(2-ii) {\bf long-range mixed disorder (LRM)}: \\

From discussion above and from Eqs. \eqref{EtallEtaP} we see that a
new marginal, temperature-dominated ground state LRM emerges when
$z_\kappa>0$ and $z_\mu>4-D-2\eta_{\rm pure}=\eta^{\rm pure}_u
(\approx0.358$ for physical membranes) and both LR curvature and LR
stress disorders are relevant. A balance between all the terms in
equations \eqref{EtallEtaP} completely determines both exponents,
\begin{eqnarray}\label{LRMeta}
\eta(z_\mu)&=&2-{D\over2}-{z_\mu\over2}\;,\\
\eta'(z\kap,z_\mu)&=&2\eta(z_\mu)-z\kap=4-D-z_\mu-z\kap\;.
\end{eqnarray}
We expect these expressions to be exact, since they are a result of
dimensional analysis. The two universal amplitudes
$A\kap(z_\mu),A\kd(z\kap,z_\mu)$ are also determined by
\eqref{EtallEtaP} and are given by
\begin{eqnarray}\label{AkAkd}
A\kap(z_\mu)&=&\half D(D-1)\Pi(\eta,D)
-{d_c \Pi(\eta,D)^2\over2\Sigma(\eta,D)}\;,\\
A\kd(z\kap,z_\mu)&=&1+{\Sigma(\eta,\eta',D)\over\Sigma(\eta,D)}
-{D(D-1) \Sigma(\eta,\eta',D)\over d_c\Pi(\eta,D)}\;, \nonumber
\end{eqnarray}
where one inserts $\eta=\eta(z_\mu)$ and $\eta'=\eta'(z_\kappa,z_\mu)$
using \eqref{LRMeta}. We observe that $A\kap(z_\mu)$ is identical to that of Eq.\eqref{univA},
and that it vanishes on the vertical line $z_\mu=\eta_u^{\rm pure}$ ($=0.358..$, 
for $D=2,d_c=1$) where the LRM phase
borders the LRC phase (see Fig. \ref{Fig1_lr}).
The LRM phase is also bounded from above by the 
line 
\begin{equation}\label{LRMlb}
z\kap=2-{D\over2}-{z_\mu\over2}\;,
\end{equation}
on which $\eta=\eta'$: there the behavior is no longer temperature dominated,
and the disorder-dominated LRCG phase takes over.
In other words it is stable only when $2 z_\kappa + z_\mu < 4 - D$
which is the exact opposite of the condition \eqref{cond4} 
of stability of the LRMG phase. Finally in Fig. \ref{Fig1_lr}
we see that LRM phase is also bounded from below and
on the right by the curve L on which the amplitude $A\kd(z\kap,z_\mu)$ 
vanishes, signifying the irrelevance of LR curvature disorder. 
The boundary L in the $(z_\mu,z_\kappa)$ plane is thus given by
\begin{equation}\label{boundaryL}
{D (D-1) \Sigma(\eta,\eta',D)\over d_c \Pi(\eta,D)}
-{\Sigma(\eta,\eta',D)\over\Sigma(\eta,D)}=1\;,
\end{equation}
where one inserts $\eta=\eta(z_\mu)$ and $\eta'=\eta'(z_\kappa,z_\mu)$
using \eqref{LRMeta}.\\

(2-iii) {\bf long-range stress disorder (LRS)}: \\

As $z_\mu$ is increased and $z\kap$ is lowered past the boundary L, the LR
curvature disorder becomes irrelevant, while the LR stress disorder remains
important and we enter a new, temperature-dominated LRS phase.
Since the LR stress is still relevant in the LRS phase, $\eta(z_\mu)$ is still
given by Eq.\eqref{LRMeta} and the associated amplitude $A_\kappa$
is the same as in the
LRM phase, given by Eq. \eqref{AkAkd}. The
irrelevance of the LR curvature disorder in this phase allows us to drop the
curvature term in \eqref{EtallEtaP}, resulting in the equation 
\begin{equation}\label{LRSetaP}
{D (D-1) \Sigma(\eta(z_\mu),\eta')\over d_c\Pi(\eta(z_\mu))}
-{\Sigma(\eta(z_\mu),\eta')\over\Sigma(\eta(z_\mu))}=1\;,
\end{equation}
that, together with the value of $\eta(z_\mu)$ given in
Eq.\eqref{LRMeta} determines the $\eta'(z_\mu)$ exponent.  We note
that unlike in the LRM phase, in the LRS phase the values of the two
exponents $\eta'(z_\mu)$ and $\eta(z_\mu)$ are functions only of
$z_\mu$.

We observe that the above equation \eqref{LRSetaP} is identical to the
second equation in \eqref{AkAkd} with the amplitude $A\kd$ set to $0$
or equivalently to Eq.\eqref{boundaryL}. Hence we find that starting
from the LRM phase, where the curvature disorder is relevant and has a
finite amplitude given by the second line of Eq.\eqref{AkAkd}, the
amplitude first decreases and finally vanishes on the boundary
L. Beyond this boundary, inside the LRS phase, the curvature disorder
is no longer relevant.

The boundary L separating the LRM and the LRS phases can also be obtained 
from the second of Eqs. \eqref{EtallEtaP}
%\eqref{EtaSimple{b}} 
by noting where the LR curvature disorder term becomes irrelevant with
respect to the left hand side or with respect to LR stress disorder
term (which in the LRC phase have the same scaling with $k$). Looking
at the powers of $k$ we find that L is determined by
$z\kap=2\eta(z_\mu)-\eta'(z_\mu)$, which together with the first 
Eq.\eqref{LRMeta} becomes
\begin{equation}\label{boundaryLagain}
z\kap=4-D-z_\mu-\eta'(z_\mu)\;.
\end{equation}
In the above equation it is important to remember to use
$\eta'(z_\mu)$ inside the LRS phase, given by Eq.\eqref{LRSetaP}.  The
two determining equations of the L boundary
\eqref{LRSetaP},\eqref{boundaryLagain} are identical to
Eqs.\eqref{LRMeta},\eqref{boundaryL} used in the determination of the L
boundary from requirement of vanishing of $A\kd(z\kap,z_\mu)$, but
have a very different interpretation.  It is satisfying to observe
that both analysis gives an identical curve $z\kap=z^L\kap(z_\mu)$ for
the L boundary.

Finally it is easy to see that the LRS phase is also bound by a 
vertical line separating it from a marginal phase LRSM. On this boundary the
temperature roughness no long dominates over disorder and 
$\eta(z_\mu)=\eta'(z_\mu)$. Using this condition inside Eq.\eqref{LRSetaP} we
find that on this boundary, in addition to the first equation in \eqref{LRMeta},
i.e., $\eta(z_\mu)=2-{D\over2}-{z_\mu\over2}$, 
$\eta(z_\mu)$ is also given by
\begin{equation}\label{etaB}
{D (D-1) \Sigma(\eta)\over2d_c \Pi(\eta)}=1\;,
\end{equation}
which together with Eqs.\eqref{eqSCSApure}, \eqref{Eta} gives 
\begin{eqnarray}\label{etaBii}
\eta&=&\eta_{\rm pure}(2d_c)\;,\\
&=&{2\over3},\;\;{\rm for}\; D=2,d_c=1\;.
\end{eqnarray}
Combining with the above other determination of $\eta(z_\mu)$ from
Eq.\eqref{LRMeta} we obtain the right boundary of the LRS phase, 
\begin{eqnarray}\label{boundaryR}
z_\mu&=& 4-D-2\eta_{\rm pure}(2d_c)=\eta_u^{\rm pure}(2d_c)\;,\\
&=&{2\over3},\;\;{\rm for}\; D=2,d_c=1\;,
\end{eqnarray}
as represented in the Fig.\ref{Fig1_lr}.

\subsubsection{Marginal phases, $\zeta'=\zeta$}

Finally we examine solutions to the SCSA equations \eqref{GGeqn},
\eqref{GGeqn2}, \eqref{couplings2}-\eqref{couplings2d},
\eqref{Pijexplicit} of the heterogeneous membrane in LR disorder that
are marginal, i.e., characterized by $\eta=\eta'$. As we will see
below there are two such solutions, which we call Long-Range Stress
Marginal (LRSM) and Long-Range Curvature Marginal (LRCM). As the names
imply, in the LRCM phase the LR curvature disorder is relevant and LR
stress disorder is irrelevant, while in the LRSM phase the LR stress
disorder is relevant and LR curvature disorder is
irrelevant. \cite{MMphase} In the phase diagram of Fig. \ref{Fig1_lr},
the LRCM phase is the shaded region that overlaps with the temperature
dominated phases LRC, LRM, LRS, as well as the disorder-dominated
phase LRCG and the second marginal phase LRSM. It is likely, that,
although all these phases are stable in the same region of $z\kap$ and
$z_\mu$ the degeneracy will be lifted by the other physical parameters
of the model, such as temperature $T$ and disorder strengths
$\Delta\kap, \Delta_\mu,\Delta_\lambda$. Varying these additional
parameters will induce real thermodynamic phase transitions between
the overlapping phases, that can be detected experimentally or
numerically. We have not discuss the nature of these phase transitions
and leave it for future investigation.  Physically we expect that for
membranes with $z\kap$ and $z_\mu$, in the shaded region we expect that
at low temperatures and strong disorder LRCG will be the stable
phase. In the opposite limit of high temperatures and weak disorder,
temperature dominated phases LRC, LRM, LRS will characterize the
disorder membrane.  The marginal LRSM, LRCM phases will exist at
intermediate values of temperature and disorder strength.

For the marginal phases with $\eta=\eta'$ at long wavelengths
$\Pi\kk(\q)$ and $\Pi\kd(\q)$ will have the same scaling behavior with
$q$. We extend the calculation of Section \ref{sec:search} to include
LR disorder.

From Eqs. \eqref{couplings2} and \eqref{Pscale} we find at small $q$,
\begin{eqnarray}\label{MLcoupling22}
&& \left(\tilde{\mu}(\q)\;,
\;\tilde{\Delta}_\mu(\q)\right) \simeq
\left({1\over 2 ( T \Pi\kk(\q) + 2\Pi\kd(\q))}\;,\;{\frac{\Delta_\mu(\q)}{2 \mu^2} + \Pi\dd(\q) \over 2 ( T \Pi\kk(\q) + 2\Pi\kd(\q))^2}\right)\;,\\
&& \simeq \frac{ (D^2-1)q^{4-D-2\eta}}{2 \Pi(\eta,D)} \left(\frac{1}{T Z\kap^{-2} + 2 Z\Del^{-1} Z\kap^{-1} } \;,\;
{Z\Del^{-2} + (D^2-1) \frac{\Delta_\mu}{2 \mu^2 \Pi(\eta,D)} q^{4-D-2 \eta -z_\mu}  \over (T Z\kap^{-2} + 2 Z\Del^{-1} Z\kap^{-1})^2 }
\right)\;,\\
&& \left(\tilde{b}(\q)\;,
\;\tilde{\Delta}_b(\q)\right) \simeq \frac{2}{D+1} \left(\tilde{\mu}(\q)\;,
\; {\frac{\Delta_b(\q)}{(D+1) b^2} + \Pi\dd(\q) \over 2 ( T \Pi\kk(\q) + 2\Pi\kd(\q))^2} \right)\;, \\
&&  \simeq \frac{ (D-1)q^{4-D-2\eta}}{ \Pi(\eta,D)} \left(\frac{1}{T Z\kap^{-2} + 2 Z\Del^{-1} Z\kap^{-1} } \;,\;
{Z\Del^{-2} + (D-1) \frac{\Delta_b}{b^2 \Pi(\eta,D)} q^{4-D-2 \eta -z_\mu}  \over (T Z\kap^{-2} + 2 Z\Del^{-1} Z\kap^{-1})^2 }
\right)\;.
\end{eqnarray}
We now introduce the dimensionless ratio 
\begin{equation}\label{rRatio}
r={Z_\kappa\over T Z_\Delta}\;
\end{equation}
to characterize the ratio of the disorder amplitude to the thermal
amplitude.  It can also be written $r = a_\Delta/a_T$ in the notations
of Section \eqref{sec:search}.
Substituting the expressions \eqref{MLcoupling22} into Eqs.\eqref{GGeqn},
\eqref{GGeqn2}, we obtain the self-consistency equations
%\begin{eqnarray}\label{MLggEqnsT}
%&& Z\kap k^{-\eta} = {D (D-1) \over
%d_c} \left[ {T Z_\kappa^{-1} + Z_{\Delta}^{-1}  \over T Z\kap^{-2} + 2 Z\Del^{-1} Z\kap^{-1}}
% - {Z\Del^{-2} Z_\kappa^{-1}\over (T Z\kap^{-2} + 2 Z\Del^{-1} Z\kap^{-1})^2 }
% \right]   \frac{\Sigma(\eta,D)}{\Pi(\eta,D)} k^{-\eta} \\
% && - \frac{2 T^2 A_\kappa}{d_c Z_\kappa^3} 
%  {1\over (T Z\kap^{-2} + 2 Z\Del^{-1} Z\kap^{-1})^2} \frac{\Sigma(2 \eta+z_\mu/2+D/2-2,\eta,D)}{\Pi(\eta,D)^2} 
%   k^{4-D- 3\eta-z_\mu}
%\;,\;\;\;\;\;\;\\
%&& {Z\kap^2\over Z\Del}k^{-\eta}= \Delta\kap k^{-z_\kappa} +{D(D-1)\over
%d_c} {Z\Del^{-3} \over (T Z\kap^{-2} + 2 Z\Del^{-1} Z\kap^{-1})^2 }
%\frac{\Sigma(\eta,D)}{\Pi(\eta,D)}
%k^{-\eta}  \\ \label{MLggEqns2T} 
%&&  + \frac{2 T^2 A_\kappa}{d_c Z_\kappa^2 Z_\Delta} 
%  {1\over (T Z\kap^{-2} + 2 Z\Del^{-1} Z\kap^{-1})^2} \frac{\Sigma(2 \eta+z_\mu/2+D/2-2,\eta,D)}{\Pi(\eta,D)^2} 
%  k^{4-D-3 \eta-z_\mu}
%\end{eqnarray}
\begin{eqnarray}\label{MLggEqnsT1}
&& 1= {D (D-1) \over
d_c} \left[ \frac{1+r}{1+2 r} - \frac{r^2}{(1+2 r)^2} 
 \right]   \frac{\Sigma(\eta,D)}{\Pi(\eta,D)} - \frac{A_\kappa}{(1+2 r)^2} 
  \frac{2 \Sigma(2 \eta+z_\mu/2+D/2-2,\eta,D)}{d_c \Pi(\eta,D)^2}  k^{4-D-2 \eta-z_\mu}
\;,\;\;\;\;\;\;\\
&& 1= A_{\kappa \Delta} k^{\eta-z_\kappa} +{D(D-1)\over
d_c} \frac{r^2}{(1+2 r)^2}
\frac{\Sigma(\eta,D)}{\Pi(\eta,D)}   + \frac{A_\kappa}{(1+ 2 r)^2} 
   \frac{2 \Sigma(2 \eta+z_\mu/2+D/2-2,\eta,D)}{d_c \Pi(\eta,D)^2} 
   k^{4-D-2 \eta-z_\mu}\;, \nonumber
\end{eqnarray}
where in the first line we have divided by $Z_\kappa k^{-\eta}$ and in
the second line by $Z_\kappa^2 k^{-\eta}/Z_\Delta$. The definitions of
the LR disorder dimensionless amplitudes $A_\kappa$ and $A_{\kappa
  \Delta}$ are recalled in \eqref{defAmp}.  We note that as expected
from the definition of $r$ these equations reduce to disorder- and
temperature-dominated ones for $r=\infty$ and $r=0$, respectively.

%Applying general equations \ref{ReduceMU,ReduceB} to this case we obtain
%%
%\begin{eqnarray}\label{MarginalCouplings}
%\left(\tilde{\mu}(\q)\;\;\tilde{\Delta}_\mu(\q)\right)
%&=&\left({Z\kap^2/2\over1+2r}\;,\;
%{\Delta_\mu Z\kap^2/4\mu^2\over(1+2r)^2\Pi(\eta)}q^{4-D-2\eta-z_\mu}
%+{r^2Z\kap^2/2\over1+2r}\right){q^{4-D-2\eta}\over\Pi(\eta)}\;,\\
%\left(\tilde{b}(\q)\;\;\tilde{\Delta}_b(\q)\right)
%&=&\left({Z\kap^2/(D+1)\over1+2r}\;,\;
%{\Delta_b Z\kap^2/(d+1)^2b^2\over(1+2r)^2\Pi(\eta)}q^{4-D-2\eta-z_\mu}
%+{r^2Z\kap^2/(D+1)\over1+2r}\right){q^{4-D-2\eta}\over\Pi(\eta)}\;.
%\end{eqnarray}
%%
%We note that as expected from the definition of $r$ these equations reduce to 
%disorder and temperature dominated ones for $r=\infty$ and $r=0$,
%respectively. Substituting these renormalized coupling constants inside
%Eq.\ref{GGeqn}, we obtain
%%
%\begin{eqnarray}\label{EtaMarginal}
%1&=&{(1+r)D\Sigma(\eta)\over(1+2r)d_c(D+1)\Pi(\eta)}
%-A\kap{2 \Sigma(2 \eta+z_\mu/2+D/2-2,\eta)
%\over{(1+2r)^2 d_c (D+1)^2\Pi(\eta)^2}}k^{4-D-2\eta-z_\mu}
%-{r^2 D\Sigma(\eta)\over(1+2r)^2 d_c(D+1)\Pi(\eta)}\;,\\
%1&=&A\kd k^{\eta-z\kap}
%+A\kap{2 \Sigma(2 \eta+z_\mu/2+D/2-2,\eta)
%\over{(1+2r)^2 d_c (D+1)^2\Pi(\eta)^2}}k^{4-D-2\eta-z_\mu}
%+{r^2 D\Sigma(\eta)\over(1+2r)^2 d_c(D+1)\Pi(\eta)}\;,
%\end{eqnarray}
%%
We now use these equation to study the two marginal flat phase of a 
disordered polymerized membrane.\\

(3-i) {\bf long-range stress marginal disorder (LRSM):} \\

In the LRSM phase the LR curvature disorder 
is irrelevant and the $A\kd k^{\eta-z\kap}$ term in the second equation of 
\eqref{MLggEqnsT1} can
be dropped at long wavelengths.  The relevance of LR stress disorder then
demands that 
\begin{equation}\label{etaM}
\eta(z_\mu)=\eta'(z_\mu)=2-D/2-z_\mu/2\;.
\end{equation}
The sum of the two equations \eqref{MLggEqnsT1} gives the equation
\begin{equation}
\label{limitEqn}
{(1+r)D(D-1) \Sigma(\eta,D)\over(1+2r)d_c\Pi(\eta,D)}=1\;,
\end{equation}
which determines $r$ as a function of $z_\mu$ by plugging in
$\eta=\eta(z_\mu)$ (see below).  For $r=0$ this equation is identical
to Eq.\eqref{etaB} for $\eta(z_\mu)=\eta'(z_\mu)=\eta_{\rm
  pure}(2d_c)$ ($=2/3$ for $D=2,d_c=1$) on the boundary of LRSM phase
with the LRS phase.  For $r=\infty$ the above equation correctly
reduces to Eq.\eqref{MLfinalEqn}, which determines
$\eta(z_\mu)=\eta'(z_\mu)=\eta_{\rm pure}(4d_c)=\eta_{\rm SRdis}$
($=0.449..$ for $D=2,d_c=1$) i.e., the exponents of the $T=0$ SR
disorder fixed point, which is also the exponent on the boundary of
LRSM phase with the LRSG phase.

Combining these considerations with Eq.\eqref{etaM} we conclude that
the LRSM phase exists in the range $\eta_u^{\rm
  pure}(2d_c)<z_\mu<\eta_u^{\rm pure}(4d_c)$ ($2/3<z_\mu<2-2 \eta_{\rm
  SRdis} \approx 1.102$, for $D=2,d_c=1$), corresponding (in the
language of the renormalization group) to a stable $T>0$ fixed point
moving down, as $z_\mu$ is increased, until $T=0$ is reached for
$r=\infty$.  Hence as $z_\mu$ varies in this range, this marginal
phase continuously interpolates between the temperature dominated LRS
and the disorder dominated LRSG phases, by having the universal
amplitude ratio vary in the range $0<r(z_\mu)<\infty$.

As can be seen from second of Eqs.\eqref{MLggEqnsT1} the LRSM phase is
also bounded from above by the line $z_\kappa=2-D/2-z_\mu/2$, i.e.,
the line same as the upper boundary of LRM phase \eqref{LRMlb}, where
LR curvature disorder becomes relevant.

Equation \eqref{limitEqn} together with \eqref{etaM}
also determine the $z_\mu$ dependence of the amplitude ratio $r(z_\mu)$
and, taking the difference between the two equations in \eqref{MLggEqnsT1}, also determines 
the universal amplitude $A\kap(z_\mu)$, as 
\begin{eqnarray}\label{univAmps}
r(z_\mu)&=& \frac{D(D-1) \Sigma(\eta(z_\mu),D) - d_c \Pi(\eta(z_\mu),D) }{2 d_c \Pi(\eta(z_\mu),D)  
- D(D-1) \Sigma(\eta(z_\mu),D) }\;, \\
%{D\Sigma(\eta(z_\mu))-2d_c(D+1)\Pi(\eta(z_\mu))
%\over4d_c(D+1)\Pi(\eta(z_\mu))-D\Sigma(\eta(z_\mu))}\;,\\
A\kap(z_\mu)&=&{1\over4}(1+3 r(z_\mu))D(D-1)\Pi(\eta(z_\mu))\;.
\end{eqnarray}

(3-ii) {\bf long-range curvature marginal disorder (LRCM):} \\

Finally there is a solution to the SCSA equations \eqref{MLggEqnsT1}
for which the LR curvature is relevant and the LR stress disorder is
irrelevant. Balancing the first two terms in the second equation of
\eqref{MLggEqnsT1} we obtain
\begin{equation}\label{LRCMeta}
\eta'=\eta=\eta(z_\kappa)=z_\kappa\;.
\end{equation}
Neglecting the LR stress disorder in the first equation of
\eqref{MLggEqnsT1} we immediately obtain
\begin{equation}\label{Ieqn}
{(1+2r)^2\over(1+3r+r^2)}={D (D-1) \Sigma(\eta,D)\over d_c \Pi(\eta,D)}\;.
\end{equation}
Solving above equation for $r(z\kap)$ we find
\be \label{univAmpsii}
r(z\kap) = {3I(z\kap)-4+\sqrt{(5I(z\kap)^2-4I(z\kap))}\over2(4-I(z\kap))}
\quad , \quad 
I(z\kap):={D(D-1) \Sigma(z\kap,D)\over d_c\Pi(z\kap,D)}\;,
\ee
which leads, for the physical membrane, $D=2$, $d_c=1$, to the
explicit expression
\bea \label{formula} 
r(z\kap) =
\frac{-z_{\kappa } \left(5 z_{\kappa }+4\right)+\sqrt{2} \sqrt{\left(z_{\kappa }-1\right)
   \left(z_{\kappa }+1\right) \left(z_{\kappa } \left(9 z_{\kappa
   }+4\right)-10\right)}+6}{2 \left(z_{\kappa } \left(z_{\kappa }+4\right)-2\right)}\;.
\eea
Using Eq. \eqref{MLggEqnsT1} to compute $A\kd(z\kap)$ leads to
\begin{eqnarray}\label{univAmpsii2}
A\kd(z\kap)&=&{1+3r(z\kap)\over 1+3r(z\kap)+r(z\kap)^2}\;,
\end{eqnarray}
The physical range of $r(z\kap)$ is $0<r(z\kap)<\infty$. From the
above equation and by comparing with Eqs.\eqref{MLfinalEqn} and
\eqref{ALeta} we find that at $r=0$
$\eta(z\kap)=\eta'(z\kap)=\eta_{\rm pure}(d_c)$ ($=0.820852..$ for
$D=2$, $d_c=1$) and at $r=\infty$ $\eta(z\kap)=\eta'(z\kap)=\eta^{\rm
  pure}(4d_c)=\eta^{\rm SRdis}(d_c)$ ($=0.449..$ for
$D=2,d_c=1$). Plotting the explicit formula \eqref{formula} fully
confirms these predictions, with $r>0$ in the range. Since $\eta_{\rm
  pure}(d_c)>\eta_{\rm SRdis}(d_c)=\eta_{\rm pure}(4d_c)$ we conclude
that at the upper horizontal part of the boundary of the LRCM phase,
Fig. \ref{Fig1_lr}, $r=0$ and $z\kap=\eta=\eta'=\eta_{\rm pure}(d_c)$,
while at the lower boundary $r=\infty$ and $z\kap=\eta=\eta'=\eta_{\rm
  SRdis}(d_c)$.

The condition that the LR stress disorder is irrelevant implies that
other part of the upper boundary separating LRCM and LRCG phases is
given by $z_\kappa<2-D/2-z_\mu/2$, same as the boundary of the LRM,
LRS and LRSM phases with the LRCG. Thus the LRCM solution exists only
in the shaded area illustrated in Fig.\ref{Fig1_lr}. In this shaded
region the LRCM solution ``coexists'' with the disorder-dominated LRCG
phase and one of the temperature-dominated solutions (LRC, LRM, LRS
phases), as well as with the other marginal LRSM phase. We expect that
a thermodynamic phase transition between these phases can be
controlled by varying temperature and the strength (amplitude) of
disorder The present marginal fixed point interpolates continuously
between the temperature-dominated ($r=0$) and the LRCG behavior
($r=\infty$) as $z_\kappa$ increases within the range $\eta^{\rm
  SRdis}(d_c)<z\kap<\eta^{\rm pure}(d_c)$.
\section{Conclusions}

Motivated by a number of physical realizations, particularly freely
suspended graphene, we presented a detailed study of tensionless
elastic sheets in the presence of thermal fluctuations and/or local
heterogeneities due to quenched random internal disorder. We developed
a general continuum elastic theory and used it to study statistical
mechanics of the crumpled, flat, and a rich variety of glassy wrinkled
phases, that are driven by anomalously strong effects of thermal
fluctuations and short- and long-ranged quenched internal disorder. We
also presented a detailed analysis of some of the associated phase
transitions, with a particular focus on the flat-to-crumpled
transition in a ``phantom'' (i.e., neglecting self-avoiding
interaction) membrane. 

Throughout, we utilized a powerful field-theoretic method of the
Self-Consistent Screening Approximation that complements the
renormalization group methods, together with the expansion in
intrinsic and embedding dimensionalities. As discussed in detail, the
advantage of the SCSA is its ability to reproduce and interpolate
between exact results in three complementary limits. It is thus highly
constrained and is expected to give both qualitative and quantitative
predictions, as has been verified in a number of numerical studies.

\acknowledgments

%*On leave from Laboratoire de Physique de l'Ecole Normale Superieure,
%24 rue Lhomond, Paris 75231 Cedex 05, Laboratoire Propre du CNRS
%Associe A L'Ens et a L'Universite Paris Sud.  

We thank D. R. Nelson for numerous interactions over the years.  For
the period 1990-1993, when this work was completed, we also take the
opportunity to thank D. Bensimon, D. Bowick, M. Kardar, G. Grest,
T. Lubensky, M. Mezard, D. Morse, and J. Toner for stimulating
discussions.  For the more recent period of finishing the writing of
the present manuscript we specially thank F. Guinea for useful
comments. We also acknowledge discussion with D. Mouhanna.  The
research part was done while PLD was on leave from LPTENS at the
Harvard Physics Department. We thank LPTENS and University of Colorado
at Boulder for hospitality while this manuscript was completed.
%Radzihovsky acknowledges support by the Hertz Fellowship,
%by NSF Grant No. DMR91-15491 and through the Harvard Materials
%Research Laboratory. 
LR also acknowledges support by the NSF grants DMR-1001240, MRSEC
DMR-1420736, PHY-1125915, Simons Investigator Fellowship, and thanks
KITP and \'Ecole Normale Sup\'erieure for hospitality during his
sabbatical stay, when part of this work was completed.

%PLD acknowledges support from NSF grant
%DMS-9100383.

\appendix

%\section{Algebra of matrices $M_{\alpha \beta, \gamma \delta}(q)$ and 
%$N_{\alpha \beta, \gamma \delta}(q)$. }

\section{Useful integrals and identities}

In this appendix we present calculational details for a certain class
of integrals that are ubiquitous in calculations with massless field
theories. For completeness we derive integrals that are more general
than necessary for the SCSA calculations of the main text. The
motivation is that these more general integrals are necessary for
higher order perturbative loop calculations, e.g., as needed to assess
the accuracy of the SCSA (not performed here). For this appendix to
function as a useful future reference we attempt to derive most
relations from first principles.

\subsection{Area of a $D$-dimensional Sphere}

Let $S_{D-1}$ denote an ``area'' of a $D-1$-dimensional sphere.  Then
for an arbitrary function $f(x)$ spherical symmetry of the integrand
gives
\begin{equation}\label{SD}
\int d^D p f(p^2)=S_{D-1}\int dp p^{2D-1} f(p^2)\;.
\end{equation}
Although $S_{D-1}$ can be calculated directly by for example by going to
$D$-dimensional polar coordinates and integrating over the polar angles, we
will use a short cut. 

Since $S_{D-1}$ is independent of function $f(x)$, let us pick a 
convenient one and compute the above integral. Let us take $f(x)=e^{-x}$, 
and calculate the left and right hand sides of Eq.\eqref{SD}.
The left hands side can be easily integrated in rectangular 
coordinates
\begin{eqnarray}\label{SDleft}
\int d^D p e^{-p^2}
&=&\prod_i^D \int dp_i e^{-p_i^2}\;,\;\;
=\left(\int_{-\infty}^{\infty}dp_1 e^{-p_1^2}\right)^D\;,\;\;
=(\pi)^{D/2}\;.
\end{eqnarray}
The right hand side is easily computed by making a change of variables
$p^2\rightarrow x$
\begin{eqnarray}\label{SDright}
S_{D-1}\int_0^\infty dp p^{D-1} e^{-p^2}
&=&\half S_{D-1}\int_0^\infty dx x^{D/2-1} e^{-x}\;,\\
&=&\half S_{D-1}\g(D/2)\;.
\end{eqnarray}
Equating the two results and solving for $S_D$ we find
\begin{equation}\label{SD2}
S_{D-1}={2(\pi)^{D/2}\over\g(D/2)}\;.
\end{equation}

\subsection{A Class of Spherically Symmetric Integrals}

\begin{equation}\label{SymmInt}
\int_p{p^{2s}\over(p^2+r)^t}=r^{s-t-D/2}
{\g(s+D/2)\g(t-s+D/2)\over(4\pi)^{D/2}\g(D/2)\g(t)}\;,
\end{equation}
where we use notation $\int_p\equiv\int{d^D p\over (2\pi)^D}\;.$
%
%\begin{equation}\label{notation}
%\int_p\equiv\int{d^D p\over (2\pi)^D}\;.
%\end{equation}
%

Proof:

\begin{eqnarray}\label{proof}
I_1&=&\int_p {p^{2s}\over(p^2+r)^t}\;,\\
&=&{1\over\g(t)}\int_0^\infty dy y^{t-1}\int p^{2s} e^{-y(p^2+r)}\;,\cr
&=&{S_{D-1}\over(2\pi)^D\g(t)}\int_0^\infty dy y^{t-1}e^{-r y}
  \int_0^\infty dp p^{2s+D-1} e^{-y p^2}\;,\\
&=&{1\over(4\pi)^{D/2}\g(D/2)\g(t)}\int_0^\infty dy y^{t-1}e^{-r y}
  \int_0^\infty dx x^{s+D/2-1} e^{-y x}\;,\\
&=&{\g(s+D/2)\over(4\pi)^{D/2}\g(D/2)\g(t)}
  \int_0^\infty dy y^{t-s-D/2-1}e^{-r y}\;,\\
&=&r^{s-t+D/2}{\g(s+D/2)\g(t-s-D/2)\over(4\pi)^{D/2}\g(D/2)\g(t)}\;.
\end{eqnarray}

\subsection{Spherical Average of $\hat{p}_{\alpha_1}\hat{p}_{\alpha_2}
\ldots\hat{p}_{\alpha_m}$}

We now extend above result to an average of a product of unit vectors
over a sphere. This will be needed to derive some of the integral
identities in the subsequent sections. We will have to compute
integrals of the form
\begin{equation}\label{IntegralForm}
\int_p\hat{p}_{\alpha_1}\hat{p}_{\alpha_2}
\ldots\hat{p}_{\alpha_m} f(p^2)\;,
\end{equation}
where $f(x)$ is an arbitrary function. Because of the spherical
symmetry of $f(p^2)$, the angular part of the integral can be done and
amounts to averaging the fully symmetric tensor
$\hat{p}_{\alpha_1}\hat{p}_{\alpha_2}\ldots\hat{p}_{\alpha_m}$ over
the $D-1$-dimensional sphere. We denote this average by angular
brackets. Since $\delta\ab$ is the only spherically symmetric tensor
and carries two indices, we immediately conclude that spherical
average over a product of odd unit vectors (odd $m$) vanishes. For $m$
even rotational invariance and unit norm of $\hat{p}_{\alpha_1}$ gives
\begin{eqnarray}\label{AngleAve}
\langle\hat{p}_{\alpha_1}\hat{p}_{\alpha_2}\ldots\hat{p}_{\alpha_{2n}}\rangle
&=&{1\over C(D,n)}\left[\delta_{\alpha_1 \alpha_2}\delta_{\alpha_3 \alpha_4}
\cdots\delta_{\alpha_{2n-1} \alpha_{2n}}+{\rm pairings}\right]\;,\\
&=&{1\over C(D,n)} S^{(n)}_{\alpha_1\alpha_2\ldots\alpha_{2n}}\;,
\end{eqnarray}
where the fully symmetric tensor
$S^{(n)}_{\alpha_1\alpha_2\ldots\alpha_{2n}}$ consists of distinct
pairings, i.e., permutations that include all distinct rearrangements
of $2n$ indices among the $\delta\ab$, but do not include permutations
such as $\delta_{\alpha_1 \alpha_2} \rightarrow\delta_{\alpha_2
  \alpha_1}$, nor permutations of pairs of indices such as
$\delta_{\alpha_1 \alpha_2}\delta_{\alpha_3 \alpha_4}
\rightarrow\delta_{\alpha_3 \alpha_4}\delta_{\alpha_2 \alpha_1}$.
For instance 
\bea S^{(1)}_{\alpha_1 \alpha_2} =
  \delta_{\alpha_1 \alpha_2}\;, \quad S^{(2)}_{\alpha_1 \alpha_2
    \alpha_3 \alpha_3} = \delta_{\alpha_1 \alpha_2} \delta_{\alpha_3
    \alpha_4} + \delta_{\alpha_1 \alpha_3} \delta_{\alpha_2 \alpha_4}
  + \delta_{\alpha_1 \alpha_4} \delta_{\alpha_2 \alpha_3}\;,
\eea
and so on, $S^{(n)}$ has $(2n-1)!!$ terms (see below).

We now prove that the constant $C(D,n)$ for arbitrary space
dimensionality $D$ and arbitrary number $2n$ of unit vectors is given by
\begin{equation}\label{CDanswer}
C(D,n)=D(D+2)(D+4)\cdots(D+2(n-1)) = 2^{n} \left(\frac{D}{2}\right)_{n}\;,
\end{equation}
where $(x)_n=x(x+1)..(x+(n-1))=\Gamma(x+n)/\Gamma(x)$ is the Pochhammer symbol.

We first note that for $D=1$ the left hand side of the Eq.\eqref{AngleAve} is
$1$. Since in this case $\delta\ab=1$, the expression in the square
brackets gives the total number of permutations of $2n$ indices not
including the interchanges within the pair nor pair interchanges. We therefore
find that 
\begin{eqnarray}\label{CDspecial}
C(1,n)&=&{(2n)!\over (2!)^n n!} = (2n-1)!!\;,
\end{eqnarray}
which agrees with the expression for $C(D,n)$ in Eq.\eqref{CDanswer} 
to be proved.

To derive $C(D,n)$ let us contract all the indices in
Eq.\eqref{AngleAve} in one particular way, say, by letting
$\alpha_1=\alpha_2, \alpha_3=\alpha_4, \ldots,
\alpha_{2n-1}=\alpha_{2n}$, and sum over repeated indices.  The left
hand side is then identically $1$. We now must compute the resulting
number in the square brackets on right hand side of
Eq.\eqref{AngleAve}.  To make things clear let us first work out an
example for $n=3$.  Then fully symmetric tensor
$S^{(3)}_{\alpha_1\alpha_2\ldots\alpha_6}$ is a sum of
$(2\times3-1)!!=15$ permutation terms
\begin{eqnarray}\label{Example}
&&\left[\delta_{\alpha_1\alpha_2}\delta_{\alpha_3\alpha_4}
      \delta_{\alpha_5\alpha_6}\right]\\
&&+\left[\delta_{\alpha_1 \alpha_2}
  \left(\delta_{\alpha_6 \alpha_3}\delta_{\alpha_4 \alpha_5}+
       \delta_{\alpha_6 \alpha_4}\delta_{\alpha_3 \alpha_5}\right)
+\delta_{\alpha_3 \alpha_4}
  \left(\delta_{\alpha_6 \alpha_1}\delta_{\alpha_2 \alpha_5}+
   \delta_{\alpha_6 \alpha_2}\delta_{\alpha_1\alpha_5}\right)
+\delta_{\alpha_5 \alpha_6}
  \left(\delta_{\alpha_2 \alpha_3}\delta_{\alpha_4 \alpha_1}+
   \delta_{\alpha_1 \alpha_4}\delta_{\alpha_3 \alpha_2}\right)\right]\nonumber\\
&&+\left[\delta_{\alpha_6 \alpha_1}
  \delta_{\alpha_2 \alpha_3}\delta_{\alpha_4 \alpha_5}
  +{\rm 7\; permutations}\right]\;\nonumber
\end{eqnarray}
where the $7$ permutations in the last term include only those
rearrangements of indices that like the first term of these $8$
permutations do not allow the contractible pairs of indices
$(\alpha_1,\alpha_2), (\alpha_3,\alpha_4),(\alpha_5,\alpha_6)$ to sit
on the same $\delta\ab$. In above expression we purposely organized
the terms into groups inside square brackets with the pattern that
will generalize for arbitrary $n$.  The first group (containing one
term) has all pairs of indices to be contracted sitting on the same
$\delta\ab$. This group (term) will obviously lead to a $D^3$ term
upon contraction, since each $\delta\aas=D$. The second group consists
of $3\times2$ terms that have only one $\delta\ab$ that has
contractible indices on the same $\delta\ab$, contributing a factor of
$D$. This $\delta\ab$ is multiplied by the terms in parenthesis that
do not have any pairs of indices that are contractible sitting on the
same $\delta\ab$, and therefore these products of two $\delta\ab$ will
collapse into one $\delta\aas$ upon contraction, contributing only a
single power of $D$. The total contribution of group two is therefore
proportional to $D^2$.  The third group does not have any $\delta\ab$
that contains contractible indices and each of the products of three
$\delta\ab$ in this group collapses to a single power of $D$, much
like in group two the products in parenthesis did.  Upon a contraction
we obtain $C(D,3)$ which is a $3$rd order polynomial in $D$
\begin{eqnarray}\label{CDpExample}
C(D,3)&=&D^3+6D^2+8D\;,\;\; =D(D+2)(D+4)\;.
\end{eqnarray}

It is obvious that for an arbitrary $n$ this structure will
generalize.  $C(D,n)$ will be a polynomial of order $D^n$, with
monomial $D^{n-k}$ for $k=0,\ldots, n-1$ coming from a $k$th group
that we described above. As in the example the zeroth group that will
contribute $D^n$ will come from the $n$-product of $\delta\ab$, with
all contractible pairs of indices belonging to the same
$\delta\ab$. It is obvious that there is only one such term and
therefore the coefficient of $D^n$ is $1$.  The $D^{n-1}$ will come
from group one. $n-1$ powers of $D$ can be broken up into two
contributions. One is a factor of $D^{n-2}$ due to $n-2$ tensors
$\delta\ab$ that have all $2(n-2)$ contractible indices sitting on the
same $\delta\ab$. The additional factor of $D$ comes from the
remaining product of two $\delta\ab$, that has all the contractible
indices located on different $\delta\ab$ and therefore fully
collapsible to $\delta\aas=D$. A $k$th term proportional to $D^{n-k}$
will come from the group that contains exactly $n-k-1$ contractible
pairs of indices belonging to the same $\delta\ab$ times an additional
factor of $D$ from the fully collapsible $k+1$-product of $\delta\ab$.

We must now compute the coefficients of these monomials. For the $D^{n-k}$ 
term we first count the number of ways of selecting which of the $n-k-1$ pairs
of indices will sit on the same $\delta\ab$. This equals to the number of
ways of choosing $n-k-1$ out of $n$ objects giving a multiplicative
contribution 
\begin{equation}\label{Fa}
{n!\over (n-k-1)! (k+1)!}
\end{equation}
to the coefficient of $D^{n-k}$ (contributing $n-k-1$ of the powers upon
the contraction). For each of these choices of $n-k-1$ pairs of indices, there
are $(2k)!!$ permutations of the remaining $k+1$ pairs that will fully
collapse to a single $\delta\aas$, giving an additional factor of 
$D$ after contraction. Hence the general $C(D,n)$ polynomial has a 
following form
\begin{eqnarray}\label{CDpGeneral}
C(D,n)&=&D^n+{n!\over(n-2)! 2!} (2)!!D^{n-1}
+{n!\over(n-3)! 3!} (4)!!D^{n-2}+\ldots\;\nonumber\\
&&+{n!\over(n-(k+1))!(k+1)!} (2 k)!!D^{n-k}+\ldots +(2 (n-1))!!D\;.
\end{eqnarray}
A derivation of above polynomial can be expressed diagrammatically. We
denote each of the $\delta_{\alpha_1\alpha_2}$ by a vertex with two
legs carrying indices $\alpha_1,\alpha_2$. In this representation the
fully symmetric tensor $S^{(n)}_{\alpha_1\alpha_2\ldots\alpha_{2n}}$
in Eq.\eqref{AngleAve} is given by a sum over string of $n$ vertices,
with $2n$ indexed legs.  A full contraction is then represented by
pairwise connection of all the $2n$ legs on each $n$-vertex. As is the
case for standard Feynman diagrammatics, each of the resulting loops
contributes a factor of $D$.  The $k$th term of above polynomial
$D^{n-k}$ is generated by a sum over all possible leg connections that
end up with $n-k$ loops. Diagrammatic combinatorics then leads to a
polynomial identical to Eq.\eqref{CDpGeneral}.  Although it is not
obvious, this polynomial can be factorized into the form of
Eq.\eqref{CDanswer}, proving the result that we will use below.

\subsection{Feynman Parameters Integrals}

Often to perform integrals we need to combine denominators of the
integrand. This can be done using Feynman identity
\begin{equation}\label{Combine}
{1\over\prod_1^n A_i^{a_i}}=
{\g\big(\sum_1^n a_i\big)\over\prod_1^n\g(a_i)}
\prod_1^n\left(\int_0^\infty dx_i x_i^{a_i-1}\right)
{\delta\big(1-\sum_1^n x_i\big)\over\big(\sum_1^n x_i
A_i\big)^{\sum_1^n a_i}}\;.
\end{equation}

Proof:
\begin{eqnarray}\label{Feynman}
{1\over\prod_1^n A_i^{a_i}}
&=&{1\over\g(a_1)\ldots\g(a_n)}\int_0^\infty dx_1\ldots dx_n
x_1^{a_1-1}\ldots x_n^{a_n-1} e^{-A_1 x_1-A_2 x_2-\ldots-A_n x_n}\;,
\;\;\;\;\;\;\;\;\;\;\;\;\;\;\;\;\;\;\;\\
&=&{1\over\prod_1^n\g(a_i)}\prod_1^n\left(\int_0^\infty dx_i x_i^{a_i-1}\right)
e^{-\sum_1^n A_i x_i}\int_0^\infty ds\delta\big(\sum_1^n x_i -s\big)\;,\\
&=&{1\over\prod_1^n\g(a_i)}\prod_1^n\left(\int_0^\infty dx_i x_i^{a_i-1}\right) 
\delta\big(\sum_1^n x_i - 1\big)
\int_0^\infty ds s^{\sum_1^n a_i - 1} e^{-\big(\sum_1^n A_i x_i\big)s},
\;\;\;\;\;\;\;\;\;\;\;\;\;\;\;\;\;\\
&=&{\g\big(\sum_1^n a_i\big)\over\prod_1^n\g(a_i)}
\prod_1^n\left(\int_0^\infty dx_i x_i^{a_i-1}\right) 
{\delta\big(\sum_1^n x_i - 1\big)
\over\big(\sum_1^n x_i A_i\big)^{\sum_1^n a_i}}\;,
\end{eqnarray}
where in above we introduced unity in a form of a $\delta$-function and
made a change of variables $x_i\rightarrow x_i s$, followed by
integration over $s$.

\subsection{Two Propagator Integrals of Products
$q_{\alpha_1}q_{\alpha_2}\ldots q_{\alpha_m}$}

We now use results of previous subsections to derive integrals that are
necessary for evaluating expressions that arise in the SCSA of the main text.
We will derive somewhat more general integrals required for 
higher order computations in theory of membranes. 

We are interested in the integrals of the form
\begin{equation}\label{wantIntegral}
I_{\alpha_1 \alpha_2\ldots \alpha_m}(a,b,{\bf p})=
\int_q{q_{\alpha_1}q_{\alpha_2}\ldots q_{\alpha_m}\over(\p+\q)^{2a}
  q^{2b}}\;.
\end{equation}
We will begin with the simplest case of $m=0$, slowly increasing the level of
complexity,
\begin{equation}\label{Io}
I(a,b,{\bf p}))=\int_q{1\over(\p+\q)^{2a} q^{2b}}\;.
\end{equation}
Using Eq.\eqref{Combine} with $n=2$, $A_1=(\p+\q)^2$, $A_2=q^2$,
$a_1=a$, and $a_2=b$ to combine the denominators we find
\begin{eqnarray}\label{DeriveI}
I(a,b)&=&\int_q{1\over(\p+\q)^{2a} q^{2b}}\;,\\
&=&{\g(a+b)\over\g(a)\g(b)}\int_q\int_0^1 dx {x^{a-1} (1-x)^{b-1}
  \over\left[x(\p+\q)^2+(1-x)q^2\right]^{a+b}}\;,\\
&=&{\g(a+b)\over\g(a)\g(b)}\int_0^1 dx x^{a-1} (1-x)^{b-1}
  \int_q{1\over\left(q^2+r\right)^{a+b}}\;,
\end{eqnarray}
where we first integrated over one of the Feynman's
parameters using the $\delta$-function, shifted the $q$ integration variable
by $\q\rightarrow\q-x\p$ and defined $r=x(1-x)p^2$. Noting that the remaining
rotationally invariant integral over $\q$ is of the form of
Eq.\eqref{SymmInt} we obtain
\begin{eqnarray}\label{DeriveIii}
I(a,b)&=&p^{D-2a-2b}{\g(a+b-D/2)\over\fpi\g(a)\g(b)}\int_0^1 dx x^{a-1} (1-x)^{b-1}\left[x(1-x)\right]^{D/2-a-b}\;,
\;\;\;\;\;\;\;\;\;\;\;\;\;\;\;\;\;\;\;\\
&=&p^{D-2a-2b}{\g(a+b-D/2)\over\fpi\g(a)\g(b)}\int_0^1 dx 
  x^{D/2-b-1} (1-x)^{D/2-a-1}\;,\\
&=&p^{D-2a-2b}{\g(D/2-a)\g(D/2-b)\g(a+b-D/2)\over\fpi\g(a)\g(b)\g(D-a-b)}\;.
\label{DeriveIiic}
\end{eqnarray}
Performing similar steps for $m=1$ we obtain
\begin{eqnarray}\label{Ia}
I_{\alpha_1}(a,b)
&=&\int_q{q_{\alpha_1}\over(\p+\q)^{2a} q^{2b}}\;,\\
&=&{\g(a+b)\over\g(a)\g(b)}\int_q\int_0^1 dx {x^{a-1} (1-x)^{b-1}q_{\alpha_1}
  \over\left[x(\p+\q)^2+(1-x)q^2\right]^{a+b}}\;,\\
&=&{\g(a+b)\over\g(a)\g(b)}\int_0^1 dx x^{a-1} (1-x)^{b-1}
  \int_q{q_{\alpha_1}-x p_{\alpha_1}\over\left[q^2+r\right]^{a+b}}\;,\label{I1c}\\
&=&-p_{\alpha_1}p^{D-2a-2b}{\g(a+b-D/2)
  \over\fpi\g(a)\g(b)}\int_0^1 dx x^{a} (1-x)^{b-1}
  \left[x(1-x)\right]^{D/2-a-b},\;\;\;\;\;\;\;\;\;\;\;\;\;\;\;\\
&=&-p_{\alpha_1}p^{D-2a-2b}{\g(a+b-D/2)\over\fpi\g(a)\g(b)}
   \int_0^1 dx x^{D/2-b} (1-x)^{D/2-a-1}\;,\\
&=&-p_{\alpha_1}p^{D-2a-2b}{\g(D/2-a)\g(D/2-b+1)\g(a+b-D/2)
  \over\fpi\g(a)\g(b)\g(D-a-b+1)}\;,
\end{eqnarray}
where in \eqref{I1c} the integral of the part proportional
$q_{\alpha_1}$ obviously vanishes by rotational symmetry. Above result
can also be simply obtained by noticing that
\begin{eqnarray}\label{recursionIa}
I_{\alpha_1}(a,b)&=&{1\over2(b-1)}{\pt\over\pt p_{\alpha_1}}
  \int_q{1\over(\p+\q)^{2(b-1)} q^{2a}}\;,\\
&=&{1\over2(b-1)}{\pt\over\pt p_{\alpha_1}}I(b-1,a)\;,\label{I1b2}\\
&=&-p_{\alpha_1}p^{D-2a-2b}{\g(D/2-a)\g(D/2-b+1)\g(a+b-D/2)
  \over\fpi\g(a)\g(b)\g(D-a-b+1)}\;,\label{I1c2}
\end{eqnarray}
where in going from \eqref{I1b2} to \eqref{I1c2} we used
Eq.\eqref{DeriveIiic} and the recursion formula for Gamma-functions,
$\g(a)=a\g(a-1)$.

For $m=2$ we have
\begin{eqnarray}\label{Iab}
I_{\alpha_1 \alpha_2}(a,b)
&=&\int_q{q_{\alpha_1}q_{\alpha_2}\over(\p+\q)^{2a} q^{2b}}\;,\\
&=&{\g(a+b)\over\g(a)\g(b)}\int_0^1 dx x^{a-1} (1-x)^{b-1}
  \int_q{q_{\alpha_1}q_{\alpha_2}+x^2 p_{\alpha_1}p_{\alpha_2}
  \over\left(q^2+r\right)^{a+b}}\;,\label{I2ab_b}\\
&=&{\g(a+b)\over\g(a)\g(b)}\int_0^1 dx x^{a-1} (1-x)^{b-1}
  \int_q\left[{\delta_{\alpha_1 \alpha_2}\over D}
  {q^2\over\left(q^2+r\right)^{a+b}}+
  {x^2
    p_{\alpha_1}p_{\alpha_2}\over\left(q^2+r\right)^{a+b}}\right]\;,
\label{I2ab_c}\\
&=&{p^{2(D/2-a-b+1)}\over\fpi\g(a)\g(b)}\int_0^1 dx x^{a-1} (1-x)^{b-1}
  \left[\delta_{\alpha_1 \alpha_2}
  \g(a+b-D/2-1)(x(1-x))^{D/2-a-b+1}\right.\nonumber\\
&&\;\;\;\;\;\;\;\;\;+\left.\hat{p}_{\alpha_1}\hat{p}_{\alpha_2}
  \g(a+b-D/2)x^2(x(1-x))^{D/2-a-b}\right]\;,\\
&=&p^{2(D/2-a-b+1)}\left[\delta_{\alpha_1\alpha_2}
  {\g(a+b-D/2-1)\g(D/2-a+1)\g(D/2-b+1)
  \over2\fpi\g(a)\g(b)\g(D-a-b+2)}\right.\nonumber\\
&&\;\;\;\;\;\;\;\;\;+\left.\hat{p}_{\alpha_1}\hat{p}_{\alpha_2}
  {\g(a+b-D/2)\g(D/2-a)\g(D/2-b+2)
  \over\fpi\g(a)\g(b)\g(D-a-b+2)}\right],
\end{eqnarray}
where in \eqref{I2ab_b} again by symmetry the terms with odd number of
$q\as$ vanished and to get to \eqref{I2ab_c} we used
Eq.\eqref{AngleAve} for $n=1$. We also introduced a unit vector
$\hat{\pb}=\pb/p$. Another way to obtain above result is to note the
recursion relation
\begin{equation}\label{recursionIab}
I_{\alpha_1 \alpha_2}(a,b)={1\over 4(b-1)(b-2)}
{\pt\over\pt p_{\alpha_1}\pt p_{\alpha_2}}I(b-2,a)
+{1\over 2(b-1)}\delta_{\alpha_1\alpha_2}I(b-1,a)\;,
\end{equation}
and use Eq.\eqref{DeriveIii}.

By repeating the integration steps of
Eqs.\eqref{DeriveI},\eqref{DeriveIii},\eqref{Ia},\eqref{Iab} for
higher values of $m$ we observe a general structure that emerges. For
arbitrary $m$ in making the shift in the $q$ integration we will
obtain in the numerator $m$ factors
$(q_{\alpha_i}+xp_{\alpha_i})$. Upon expanding this product all the
terms containing an odd number $q_{\alpha_i}$ vanish by rotational
symmetry.  For $m=2n$ even, we organize all the remaining terms into
groups by the number of pairs of $q_{\alpha_i}$ contained in the term,
because via Eq.\eqref{AngleAve} the $\q$ integration for each member
of a group is of the same form.  Each member of a group of $k$ pairs
of $q_{\alpha_i}$ is multiplied by a one of the possible products
$2(n-k)$ vectors $x\hat{p}_{\alpha_i}$, carrying the remaining
$2(n-k)$ indices. After the $\q$ integration a $k$th group becomes
proportional to a tensor $S^{(k,2n)}_{\alpha_1
  \alpha_2\ldots\alpha_{2n}}(\hat{\p})$ that is a sum of tensors of
the form
\begin{equation}\label{tensorForm}
S^{(k)}_{\alpha_1\alpha_2\ldots\alpha_{2k}}
\hat{p}_{\alpha_{2k+1}}\hat{p}_{\alpha_{2k+2}}\ldots \hat{p}_{\alpha_{2n}}\;,
\end{equation}
with the permutations ranging over all possible ways of dividing up
the $2n$ indices between the
$S^{(k)}_{\alpha_1\alpha_2\ldots\alpha_{2k}}$ and
$\hat{p}_{\alpha_{2k+1}}\hat{p}_{\alpha_{2k+2}}\ldots\hat{p}_{\alpha_{2n}}$
tensors.

For odd $m=2n+1$ the structure is very similar to that of $m=2n$, but
with an additional factor of $p_{\alpha_i}$ multiplying each tensor,
symmetrized over all indices.  We now summarize the results of
integration for several values of $m$.

$m=3$:
\begin{eqnarray}\label{Iabc}
I_{\alpha_1 \alpha_2 \alpha_3}(a,b)
&=&-p^{D-2a-2b+3}\left[S^{(1,3)}_{\alpha_1\alpha_2\alpha_3}
  {\g(a+b-D/2-1)\g(D/2-a+1)\g(D/2-b+2)
  \over2\fpi\g(a)\g(b)\g(D-a-b+3)}\right.\\
&&\left.+S^{(0,3)}_{\alpha_1\alpha_2\alpha_3}
  {\g(a+b-D/2)\g(D/2-a)\g(D/2-b+3)
  \over\fpi\g(a)\g(b)\g(D-a-b+3)}\right]\;,\nonumber
\end{eqnarray}
where as explained above
\begin{eqnarray}\label{tensorsIabc}
S^{(1,3)}_{\alpha_1\alpha_2\alpha_3}
&=&\hat{p}_{\alpha_1}S^{(1)}_{\alpha_2\alpha_3}+
  \hat{p}_{\alpha_2}S^{(1)}_{\alpha_3\alpha_1}+
  \hat{p}_{\alpha_3}S^{(1)}_{\alpha_1\alpha_2}\;,\\
S^{(0,3)}_{\alpha_1\alpha_2\alpha_3}
&=&\hat{p}_{\alpha_1}\hat{p}_{\alpha_2}\hat{p}_{\alpha_3}\;,
\end{eqnarray}

$m=4$:
\begin{eqnarray}\label{Iabcd}
I_{\alpha_1\alpha_2\alpha_3\alpha_4}(a,b)
&=&p^{D-2a-2b+4}\left[S^{(2,4)}_{\alpha_1\alpha_2\alpha_3\alpha_4}
  {\g(a+b-D/2-2)\g(D/2-a+2)\g(D/2-b+2)
  \over4\fpi\g(a)\g(b)\g(D-a-b+4)}\right.\\
&&\left.+S^{(1,4)}_{\alpha_1\alpha_2\alpha_3\alpha_4}
  {\g(a+b-D/2-1)\g(D/2-a+1)\g(D/2-b+3)
  \over2\fpi\g(a)\g(b)\g(D-a-b+4)}\right.\nonumber\\
&&\left.+S^{(0,4)}_{\alpha_1\alpha_2\alpha_3\alpha_4}
  {\g(a+b-D/2)\g(D/2-a)\g(D/2-b+4)
  \over\fpi\g(a)\g(b)\g(D-a-b+4)}\right]\;,\nonumber
\end{eqnarray}
where the tensors are
\begin{eqnarray}\label{tensorsIabcd}
S^{(2,4)}_{\alpha_1\alpha_2\alpha_3}
&=&S^{(2)}_{\alpha_1\alpha_2\alpha_3\alpha_4}\;,\\
S^{(1,4)}_{\alpha_1\alpha_2\alpha_3\alpha_4}
&=&\hat{p}_{\alpha_1}\hat{p}_{\alpha_2}S^{(1)}_{\alpha_3\alpha_4}+
  \hat{p}_{\alpha_1}\hat{p}_{\alpha_3}S^{(1)}_{\alpha_2\alpha_4}+
  \hat{p}_{\alpha_1}\hat{p}_{\alpha_4}S^{(1)}_{\alpha_2\alpha_3}+\\
&&\hat{p}_{\alpha_2}\hat{p}_{\alpha_3}S^{(1)}_{\alpha_1\alpha_4}+
  \hat{p}_{\alpha_2}\hat{p}_{\alpha_4}S^{(1)}_{\alpha_1\alpha_3}+
  \hat{p}_{\alpha_3}\hat{p}_{\alpha_4}S^{(1)}_{\alpha_1\alpha_2}\;,\nonumber\\
S^{(0,4)}_{\alpha_1\alpha_2\alpha_3\alpha_4}
&=&\hat{p}_{\alpha_1}\hat{p}_{\alpha_2}\hat{p}_{\alpha_3}\hat{p}_{\alpha_4}\;.
\end{eqnarray}

$m=5$:
\begin{eqnarray}\label{Iabcde}
I_{\alpha_1\alpha_2\alpha_3\alpha_4\alpha_5}(a,b)
&=&-p^{D-2a-2b+5}\left[S^{(2,5)}_{\alpha_1\alpha_2\alpha_3\alpha_4\alpha_5}
  {\g(a+b-D/2-2)\g(D/2-a+2)\g(D/2-b+3)
  \over4\fpi\g(a)\g(b)\g(D-a-b+5)}\right.\\
&&\left.+S^{(1,5)}_{\alpha_1\alpha_2\alpha_3\alpha_4\alpha_5}
  {\g(a+b-D/2-1)\g(D/2-a+1)\g(D/2-b+4)
  \over2\fpi\g(a)\g(b)\g(D-a-b+5)}\right.\nonumber\\
&&\left.+S^{(0,5)}_{\alpha_1\alpha_2\alpha_3\alpha_4\alpha_5}
  {\g(a+b-D/2)\g(D/2-a)\g(D/2-b+5)
  \over\fpi\g(a)\g(b)\g(D-a-b+5)}\right]\;,\nonumber
\end{eqnarray}
where the tensors are
\begin{eqnarray}\label{tensorsIabcde}
S^{(2,5)}_{\alpha_1\alpha_2\alpha_3\alpha_4\alpha_5}
&=&\hat{p}_{\alpha_1}S^{(2)}_{\alpha_1\alpha_2\alpha_3\alpha_4}
  + 4{\rm\; other\;\; permutations}\;,\label{tensorsIabcde25}\\
S^{(1,5)}_{\alpha_1\alpha_2\alpha_3\alpha_4\alpha_5}
&=&\hat{p}_{\alpha_1}\hat{p}_{\alpha_2}\hat{p}_{\alpha_3}
  S^{(1)}_{\alpha_4\alpha_5}
  + 9{\rm\; other\;\; permutations}\;,\\
S^{(0,5)}_{\alpha_1\alpha_2\alpha_3\alpha_4\alpha_5}
&=&\hat{p}_{\alpha_1}\hat{p}_{\alpha_2}\hat{p}_{\alpha_3}
  \hat{p}_{\alpha_4}\hat{p}_{\alpha_5}\;.
\end{eqnarray}
In the expression for tensor
$S^{(1,5)}_{\alpha_1\alpha_2\alpha_3\alpha_4\alpha_5}$ the
permutations are the $5!/2!/3!=10$ ways of choosing a set of $3$
indices that will sit on the $\hat{\p}$ vectors. Similarly, in
\eqref{tensorsIabcde25} the $5!/1!/4!=5$ permutations correspond to
different ways of choosing which one index will sit on
$\hat{\p}$. From above expressions, it is quite obvious how to
construct the solution to the integrals of this form for an arbitrary
$m$.

\subsection{Summary}

Let us define 
\bea
H_{n,m}(a,b) =  {\g(a+b-D/2-n)\g(D/2-a+n)\g(D/2-b+m-n)
  \over2^n \fpi\g(a)\g(b)\g(D-a-b+m)}
\eea 
Then we have shown
\bea
&& I_{\alpha_1}(a,b) = -  p^{D-2a-2b+1 } S^{(0,1)}_{\alpha_1} H_{0,1}(a,b) \\
&&
I_{\alpha_1 \alpha_2}(a,b) = 
p^{D-2a-2b+2}\left[S^{(1,2)}_{\alpha_1 \alpha_2} H_{1,2}(a,b) +
 S^{(0,2)}_{\alpha_1 \alpha_2} H_{0,2}(a,b) \right] \nonumber  \\
&& I_{\alpha_1 \alpha_2 \alpha_3}(a,b) =
-p^{D-2a-2b+3}\left[S^{(1,3)}_{\alpha_1\alpha_2\alpha_3} H_{1,3}(a,b)
+ S^{(0,3)}_{\alpha_1\alpha_2\alpha_3} H_{0,3}(a,b) \right] \nonumber  \\
&& I_{\alpha_1\alpha_2\alpha_3\alpha_4}(a,b)
= p^{D-2a-2b+4}\left[S^{(2,4)}_{\alpha_1\alpha_2\alpha_3\alpha_4} H_{2,4}(a,b)
 +S^{(1,4)}_{\alpha_1\alpha_2\alpha_3\alpha_4} H_{1,4}(a,b)
 +S^{(0,4)}_{\alpha_1\alpha_2\alpha_3\alpha_4} H_{0,4}(a,b) \right] \nonumber  \\
&& I_{\alpha_1\alpha_2\alpha_3\alpha_4\alpha_5}(a,b) = 
-p^{D-2a-2b+5}\left[S^{(2,5)}_{\alpha_1\alpha_2\alpha_3\alpha_4\alpha_5} H_{2,5}(a,b)
 +S^{(1,5)}_{\alpha_1\alpha_2\alpha_3\alpha_4\alpha_5} H_{1,5}(a,b)
 +S^{(0,5)}_{\alpha_1\alpha_2\alpha_3\alpha_4\alpha_5} H_{0,5}(a,b) \right] \nonumber 
\end{eqnarray}
Hence each term $H_{n,m}(a,b)$ is paired with the symmetric tensor
$S^{(n,m)}$ with $m$ indices and $n$ Kronecker delta, defined above,
i.e.,
\bea
&& S^{(0,1)}_{\alpha_1} = \hat p_{\alpha_1}\;,\quad S^{(1,2)}_{\alpha_1 \alpha_2} =  \delta_{\alpha_1\alpha_2}\;,\quad
S^{(0,2)}_{\alpha_1 \alpha_2} = \hat{p}_{\alpha_1}\hat{p}_{\alpha_2}\;,\ldots\;,
\eea 
i.e., $m-2 n$ is the number of $\hat p$'s appearing in the tensor
$(n,m)$. We also note useful recursion relations,
\bea
&& H_{n,m+1}(a,b) = \frac{D-2 b+2 m-2 n}{2 (D-a-b+m)} H_{n,m}(a,b)\;,\\
&& H_{n-1,m}(a,b) = \frac{(2 a+2 b-D-2 n) (-2 b+D+2 m-2 n)}{-2 a+D+2 n-2} H_{n,m}(a,b)\;. 
\eea 

\section{Results for the SCSA integrals}
\label{FinalResults}

Here we give the results for the integrals needed in the text.
From \eqref{Iabcd} we have
\bea
 \Pi(\eta,\eta',D) &= & \int_p {(p_\alpha P_{\alpha \beta}^T(\hat {\bf q}) p_{\beta})^2
\over|\pb|^{4-\eta}|\pb+\hat \q|^{4-\eta'}}\; 
=  P^T_{\alpha_1 \alpha_2}(\hat {\bf q}) 
P^T_{\alpha_3 \alpha_4}(\hat {\bf q}) 
I_{\alpha_1\alpha_2\alpha_3\alpha_4}(2-\frac{\eta}{2},2-\frac{\eta'}{2},\hat {\bf q})\;, \\
& =& (D^2-1)
 {\g(2-\frac{\eta+\eta'}{2}-\frac{D}{2})\g(\frac{D}{2}+\frac{\eta}{2})\g(\frac{D}{2}+\frac{\eta'}{2})
  \over4\fpi\g(2- \frac{\eta}{2})\g(2- \frac{\eta'}{2})\g(D+\frac{\eta+\eta'}{2} )} \label{Pieteeta2new} 
  \eea 
which leads to the integral %$A(D,\eta,\eta')$ is 
given in the text in \eqref{Aresult}. Note that the other terms
vanish because of the transverse projectors and we used that
\bea
P^T_{\alpha_1 \alpha_2}({\bf q}) P^T_{\alpha_3 \alpha_4}({\bf q}) S^{(2,4)}_{\alpha_1\alpha_2\alpha_3}
= D^2-1\;.
\eea 

Let us consider now the second integral. We have
\bea
&&\Sigma(\eta,\eta',D) = \int_q {(\hat k_\alpha P_{\alpha \beta}^T(\hat {\bf q}) \hat k_{\beta})^2
|\q|^{4-D - 2\eta} 
\over |\hat \ks+ \q|^{4-\eta'}}\;,\\ 
%\hat k_{\alpha_1} \hat k_{\alpha_2} \hat k_{\alpha_3} \hat k_{\alpha_4} 
%\big[ \delta_{\alpha_1,\alpha_2} \delta_{\alpha_3,\alpha_4} I(2 - \frac{\eta'}{2} , \frac{D}{2} + \eta'-2, \hat {\bf k}) 
%\\
%&& - \delta_{\alpha_1,\alpha_2} 
%I_{\alpha_3,\alpha_4}(2 - \frac{\eta'}{2} , \frac{D}{2} + \eta'-1, \hat {\bf k}) 
%- \delta_{\alpha_3,\alpha_4} 
%I_{\alpha_1,\alpha_2}(2 - \frac{\eta'}{2} , \frac{D}{2} + \eta'-1, \hat {\bf k}) 
%+ I_{\alpha_1,\alpha_2,\alpha_3,\alpha_4}(2 - \frac{\eta'}{2} , \frac{D}{2} + \eta', \hat {\bf k}) \big] \nonumber \\
&& =  I(2 - \frac{\eta'}{2} , \frac{D}{2} + \eta-2, \hat {\bf k})  
 - 2 \hat k_{\alpha_1} \hat k_{\alpha_2} I_{\alpha_1,\alpha_2}(2 - \frac{\eta'}{2} , \frac{D}{2} + \eta-1, \hat {\bf k}) 
 + \hat k_{\alpha_1} \hat k_{\alpha_2} \hat k_{\alpha_3} \hat k_{\alpha_4}  
I_{\alpha_1,\alpha_2,\alpha_3,\alpha_4}(2 - \frac{\eta'}{2} , \frac{D}{2} + \eta, \hat {\bf k})\;.\nonumber
\eea 
Using the above expressions \eqref{DeriveIii}, \eqref{Iab}, \eqref{Iabcd}, \eqref{tensorsIabcd}
and performing the tensor contractions we obtain
\bea
&&\Sigma(\eta,\eta',D) = \frac{\left(D^2-1\right) \Gamma
   (2-\eta) \Gamma
   \left(\frac{D}{2}+\frac{\eta'}{2}\right
   ) \Gamma
   \left(\eta-\frac{{\eta}'}{2}\right
   )}{4 (4 \pi)^{D/2} \Gamma
   \left(2-\frac{{\eta}'}{2}\right
   ) \Gamma
   \left(\frac{D}{2}+\eta\right) \Gamma
   \left(\frac{D}{2}-
   \eta+\frac{{\eta}'}{2}+2 \right)}\;.
\eea 
We notice that the results for $\Pi$ and $\Sigma$ can be rewritten
in a more compact unified way 
\bea
&&\Pi(\eta,\eta',D) = \frac{D^2-1}{4 (4 \pi)^{D/2}} 
F_D(\frac{D}{2} + \frac{\eta}{2},  \frac{D}{2} + \frac{\eta'}{2})\;, \\
&& \Sigma(\eta,\eta',D) = \frac{D^2-1}{4 (4 \pi)^{D/2}} 
F_D(2-\eta, \frac{D}{2} + \frac{\eta'}{2})\;,
\eea 
with
\bea
F_D(a,b) = \frac{ \Gamma(a) \Gamma(b) \Gamma(2 + \frac{D}{2} - a - b)}{\Gamma(2 + \frac{D}{2}-a)
\Gamma(2 + \frac{D}{2}-b) \Gamma(a+b)}\;.
\eea 

%\newpage
%
%\bea
%B(\eta,D)
%%k^{-4+\eta}\int_q |\q|^{\eta_u} |\ks-\q|^{-(4-\eta)} 
%%\left(k\as P^T\ab(q) k\bs\right)^2\;,\\
%&=&{(D^2-1)\Gamma(\eta/2)\Gamma(D/2+\eta/2)\Gamma(2-\eta)
%\over4(4\pi)^{D/2}\Gamma(2-\eta/2)\Gamma(D/2+\eta)\Gamma(D/2+2-\eta/2)} 
%\;.
%\end{eqnarray}
%
%
%\newpage
%
%\appendix{C}{Details of the $\epsilon^2$ calculations.}
%
%\appendix{D}{Matrix algebra of the crumpling transition calculations.}
%
%\appendix{E}{Long-range disorder calculational details.}
% 

\section{Crumpling transition integrals}
\label{app:crumpling} 

Here we work out integrals necessary for the analysis of the crumpling
transition. To utilize the results of the previous sections, let us
rewrite \eqref{(3.9)} in a generalized form, as
\be
\Pi_{\alpha \beta, \gamma \delta}(\p)
= {1 \over 4} \int_q (q_{\alpha} (p_\beta+q_{\beta}) + q_{\beta} (p_\alpha+q_{\alpha}))
(q_{\gamma} (p_\delta+q_{\delta}) + q_{\delta} (p_\gamma+q_{\gamma}))
\frac{1}{(\pb + \q)^{2 a}  q^{2b}}\;, 
\ee
where in the case of interest $2a = 2 b = 4 -\eta$. Then in terms of 
previous derived integrals, it can be rewritten as
\bea
\Pi_{\alpha \beta, \gamma \delta}(\p)=
{\rm sym}_{\alpha,\beta} {\rm sym}_{\gamma,\delta} [
p_\beta p_\delta I_{\alpha \gamma}(a,b,\pb) 
+ p_\delta I_{\alpha \beta \gamma}(a,b,\pb)
+ p_\beta I_{\alpha \gamma \delta}(a,b,\pb)] + I_{\alpha \beta \gamma \delta}(a,b,\pb)\;,
\eea
where ${\rm sym}_{\alpha,\beta}$ is a symmetrization over
$\alpha,\beta$ indices.

To calculate the polarization integrals, $\pi_i(\p)$, we will use that 
\bea
(D-1) \pi_3(\p) &=& P^T_{\alpha \beta}(\p) \Pi_{\alpha \beta, \gamma \delta}(\p) P^T_{\gamma \delta}(\p)\;,\\
\pi_5(\p) &=& P^L_{\alpha \beta}(\p) \Pi_{\alpha \beta, \gamma \delta}(\p) P^L_{\gamma \delta}(\p)\;,\\
\sqrt{D-1}\pi_4(\p)&=& 
P^L_{\alpha \beta}(\p) \Pi_{\alpha \beta, \gamma \delta}(\p) P^T_{\gamma \delta}(\p)
=
P^T_{\alpha \beta}(\p) \Pi_{\alpha \beta, \gamma \delta}(\p) P^L_{\gamma \delta}(\p)\;.
\eea 
It is also convenient to define
\bea
S^{(n,4)}_{AB}  = P^{A}_{\alpha \beta}(\p) S^{(n,4)}_{\alpha \beta \gamma \delta}  P^{B}_{\gamma \delta}(\p)\;.
\eea 
Thus we have, using that $B_{\alpha \beta} {\rm sym}_{\alpha,\beta}
A_{\alpha \beta} = B_{\alpha \beta} A_{\alpha \beta}$ for any symmetric tensor $B$, 
\bea
&& (D-1) \pi_3(\p) = P^T_{\alpha \beta}(\p) I_{\alpha \beta \gamma \delta}(a,b,\pb)  P^T_{\gamma \delta}(\p)  = S^{(2,4)}_{TT} H_{2,4}(a,b)
= (D^2-1)   H_{2,4}(a,b) p^{D-2 a- 2b+4}\;,
\eea 
where we used $S^{(2,4)}_{TT} = S^{(2)}_{TT}=D^2-1$ and
$S^{(1,4)}_{TT} =S^{(0,4)}_{TT} =0$. Thus, we obtain
\bea
\pi_3(\p) = (D+1) H_{2,4}(a,b) p^{D-2 a- 2b+4} =|_{a=b=2-\eta/2} \frac{\Pi(\eta,D)}{D-1} p^{D-4+2 \eta}\;.
\eea 

Next we have 
\bea
\sqrt{D-1} \pi_4(\p) &=& 
P^L_{\alpha \beta}(\p) ( I_{\alpha \beta \gamma \delta}(a,b,\pb) + p_\beta I_{\alpha \gamma \delta}(a,b,\pb) )  P^T_{\gamma \delta}(\p)\;, \\
&=& p^{D-2a-2b+4} ( S^{(2,4)}_{LT} H_{2,4}(a,b) + S^{(1,4)}_{LT} H_{1,4}(a,b) 
- P^L_{\alpha \beta}(\p) \hat p_\beta S^{(1,3)}_{\alpha \gamma \delta} 
P^T_{\gamma \delta}(\p) H_{1,3}(a,b) )\;,  \nn \\
&=& (D-1)  ( H_{2,4}(a,b) + H_{1,4}(a,b) - H_{1,3}(a,b) )\;,\nn\\ 
&=& (D-1) (5 - 2 a - 2 b + D) H_{2,4}(a,b)  p^{D-2a-2b+4}\;,
\eea
using that $S^{(0,4)}_{LT} =0$ and $P^L_{\alpha \beta}(\p) \hat
p_\beta S^{(0,3)}_{\alpha \gamma \delta} P^T_{\gamma
  \delta}(\p)=0$. In the last line we used that
$S^{(2,4)}_{LT}=S^{(1,4)}_{LT}=P^L_{\alpha \beta}(\p) \hat p_\beta
S^{(1,3)}_{\alpha \gamma \delta} P^T_{\gamma \delta}(\p)=D-1$. Hence
we have
\bea
\frac{ \pi_4(\p)}{ \pi_3(\p) } = \frac{\sqrt{D-1}}{D+1} (5 - 2 a - 2 b + D) =|_{a=b=2-\eta/2} \frac{\sqrt{D-1}}{D+1} (D + 2 \eta - 3)\;. 
\eea
Then we have,
\bea
\pi_5(\p) &=& 
%P^L_{\alpha \beta}(\p) (p_\beta p_\delta I_{\alpha \gamma}(a,b,\pb) 
%+ p_\delta I_{\alpha \beta \gamma}(a,b,\pb)
%+ p_\beta I_{\alpha \gamma \delta}(a,b,\pb ) + I_{\alpha \beta \gamma \delta}(a,b,\pb) )P^L_{\gamma \delta}(\p) \\
%&& = 
\hat p_\alpha \hat p_\gamma p^2 I_{\alpha \gamma}(a,b,\pb) 
+ 2 \hat p_{\alpha} \hat p_{\gamma} \hat p_{\delta} p I_{\alpha \gamma \delta}(a,b,\pb )  
+ \hat p_{\alpha} \hat p_\beta \hat p_{\gamma} \hat p_{\delta}  I_{\alpha \beta \gamma \delta}(a,b,\pb)\;,\\
&& 
= p^{D-2a-2b+4} \left[H_{1,2}(a,b) + H_{0,2}(a,b) - 6 H_{1,3}(a,b)
- 2 H_{0,3}(a,b) 
+3  H_{2,4}(a,b)
 + 6  H_{1,4}(a,b)
 + H_{0,4}(a,b) 
\right]\;. \nn \\
\eea

With this, we obtain
\bea
\frac{ \pi_5(\p)}{ \pi_3(\p) } &=& \frac{1}{D+1} 
(4 a^2-4 D (a+b-2)+\frac{4 (b-1) (a-b+1)}{-2 a+D+2}-\frac{4 (a-1) (a-b-1)}{-2 b+D+2}\\
&& +8 a
   b-16 a+4 b^2-16 b+D^2+15)\;, \nn \\
&=&|_{a=b=2-\eta/2}  \frac{(-22+31D-10D^2+D^3+43\eta-32D\eta+5 D^2\eta-24\eta^2+8D\eta^2+4 \eta^3)}{(D+1)(D-2+\eta)}\;.
\eea

For the remaining integrals we use Mathematica. We find
\bea
\pi_2(\p) &=& \frac{1}{D-1} (W_2)_{\alpha \beta,\gamma\delta}  \Pi_{\alpha \beta, \gamma \delta}(\p)\;, \\
%&&
%= \frac{1}{D-1}  (W_2)_{\alpha \beta,\gamma\delta} 
%{\rm sym}_{\alpha,\beta} {\rm sym}_{\gamma,\delta} [
%p_\beta p_\delta I_{\alpha \gamma}(a,b,\pb) 
%+ p_\delta I_{\alpha \beta \gamma}(a,b,\pb)
%+ p_\beta I_{\alpha \gamma \delta}(a,b,\pb)] + I_{\alpha \beta \gamma \delta}(a,b,\pb)  \\
& = &
( 2 H_{2,4}(a,b) + 2 H_{1,4}(a,b) + \frac{1}{2} H_{1,2}(a,b)
- 2 H_{1,3}(a,b)) p^{D-2a-2b+4} \;,
\eea
where $(W_2)_{\alpha \beta,\alpha \beta} = D-1$. This leads to
\bea
&& \frac{ \pi_5(\p)}{ \pi_3(\p) } = \frac{1}{D+1}  (\frac{2 (b-1) (a-b+1)}{D+2-2 a} + \frac{2 (a-1) (b-a+1)}{D+2-2 b}
) =|_{a=b=2-\eta/2} 
\frac{1}{D+1}  \frac{2 (2-\eta)}{D+\eta-2}\;.
\eea 

Finally,
\bea
\pi_1(\p) &=& \frac{2}{(D-2) (D+1)} (W_2)_{\alpha \beta,\gamma\delta}  \Pi_{\alpha \beta, \gamma \delta}(\p)\;, \\
%&&
%= \frac{1}{D-1}  (W_2)_{\alpha \beta,\gamma\delta} 
%{\rm sym}_{\alpha,\beta} {\rm sym}_{\gamma,\delta} [
%p_\beta p_\delta I_{\alpha \gamma}(a,b,\pb) 
%+ p_\delta I_{\alpha \beta \gamma}(a,b,\pb)
%+ p_\beta I_{\alpha \gamma \delta}(a,b,\pb)] + I_{\alpha \beta \gamma \delta}(a,b,\pb)  \\
& = &
 2 H_{2,4}(a,b) p^{D-2a-2b+4}\;, 
\eea
where $(W_1)_{\alpha \beta,\alpha \beta} = (D-2) (D+1)/2$, 
which leads to (for any $a,b$)
\bea
\frac{ \pi_1(\p)}{ \pi_3(\p) } &=&\frac{2}{D+1}\;.
\eea 

We now calculate the height-field self-energy integrals, 
$b_i(a,b,D)$, defined by (up to a small 
change in notations and the change $\q \to - p \q$),
\bea
b_i(a,b,D) = \int_q q^{-2 b} |\hat \p + \q|^{-2 a}
\hat p_{\alpha}(\hat p_\beta+q_{\beta}) (W_i)_{\alpha\beta,\gamma\delta}(\q) \hat p_{\gamma} (\hat p_\delta+q_{\delta})\;,
\label{(3.15app)}
\eea
which we will need in the text in \eqref{(3.15)} for $a = 2 -\eta/2$ and $b=D/2-2+ \eta$. 

Let us start with $b_3$. From the definition 
\bea
(D-1) b_3 &=& \int_q q^{-2 b} |\hat \p + \q|^{-2 a} 
(\hat p_{\alpha} P_{\alpha\beta}^T(\q) \hat p_{\beta})^2
= \Sigma(2+b-D/2,4- 2 a)\;,\\
&=&
 \frac{D^2-1}{4 (4 \pi)^{D/2}}  \frac{ \Gamma(D/2-b) \Gamma(\frac{D}{2} + 2-a) \Gamma(a+b- \frac{D}{2})}{
\Gamma(2 + b)
\Gamma(a) \Gamma(D+2-a-b)} = (D^2-1) H_{2,4}(a,b+2)\;,
\eea 
which for $a = 2 -\eta/2$ and $b=D/2-2+ \eta$ leads to
\bea
b_3=(D+1) \frac{\Sigma(\eta,D)}{D^2-1}\;.
\eea

Using Mathematica we find next,
\bea
\sqrt{D-1} b_4 &=& 2 (D-1) (1+ 2 b - D) H_{2,4}(a,b+2)\;,
\eea
which for $a = 2 -\eta/2$ and $b=D/2-2+ \eta$ gives 
\bea
b_4=2 \sqrt{D-1} (2 \eta-3)  \frac{\Sigma(\eta,D)}{D^2-1}\;.
\eea
We also find
\bea
b_5 &=&\left( -\frac{4 (b+1) (-a+b+1)}{-2 a+D+2}-\frac{4 (a-1) (b+1)}{D-2 (a+b-1)}+4 b^2-4 (b+1) D+8 b+D^2+3\right) H_{2,4}(a,b+2)\;,
\eea
which for $a = 2 -\eta/2$ and $b=D/2-2+ \eta$ leads to
\bea
b_5= -\frac{2 D^2-4 D \eta ^2+16 D \eta -15 D-4 \eta ^3+24 \eta ^2-43 \eta +22}{D+\eta -2}\times \frac{\Sigma(\eta,D)}{D^2-1}\;.
\eea
Similarly, we obtain
\bea
\hspace{-1cm}b_2 &=& \frac{2 (D-1) \left(a^2 (8 b-4 D+4)+4 a (b-D-2) (2 b-D+1)+(b-D-3) (b-D) (2 b-D)+4
   (b+1)\right)}{(2 a-D-2) (2 (a+b-1)-D)}
 H_{2,4}(a,b+2).\;\;\;\;\;\;\;\;\; \nn\\
\eea
which for $a = 2 -\eta/2$ and $b=D/2-2+ \eta$ reduces to
\bea
b_2= -\frac{(D-1) \left(D^2+2 \eta -4\right)}{D+\eta -2}
 \times \frac{\Sigma(\eta,D)}{D^2-1}\;.
\eea
Finally, we find
\bea
&& b_1 = (D-2)(D+1) 
 H_{2,4}(a,b+2)
\eea
which for $a = 2 -\eta/2$ and $b=D/2-2+ \eta$ gives
\bea
b_1= (D-2)(D+1)
 \times \frac{\Sigma(\eta,D)}{D^2-1}\;.
\eea


\begin{thebibliography}{99}
\bibitem{GaussCurv} The other geometrical invariant is the intrinsic
  Gaussian curvature. However the integral of the Gaussian curvature
  can be shown to be a topological invariant proportional to the Euler
  character or the genus of the manifold and therefore will not be
  considered in this paper.
\bibitem{ModeCount} In the isotropic description the theory is
  $O(d)\times O(D)$ invariant and there are a total of $Dd$
  independent deformational modes.  In the spontaneously broken flat
  phase the symmetry is reduced to $O(d_c)\times O(D)\times O(D)$. The
  general principles require that when a group G is broken down to a
  group H, the number of massless Goldstone modes is equal to the
  dimension of the coset space G/H, here $O(d)/O(D)/O(d_c)$. We
  therefore expect $Dd_c$ massless modes and $D^2$ massive ones
  corresponding to $\pt\as\h$ Goldstone modes and $\pt\as u\bs$
  elastic deformations, respectively.
\bibitem{CommentDist} As usual at long wavelengths the higher moments
  are expected to be irrelevant in renormalization group sense.
\bibitem{comment4dc} It is not clear whether there is an underlying
  physical reason for this $d_c\rightarrow 4d_c$ property or that it
  is just an accident special to the SCSA.
\bibitem{MMphase} Actually there is an additional mixed marginal phase
  in which both LR curvature and LR stress disorders are relevant,
  analogous to the mixed LRM and LRMG phases and the exponents are
  completely determined with $\eta=\eta'=z\kap$.  However, this phase
  turns out to only exist on a special line $z\kap=2-D/2-z_\mu/2$ and
  is therefore not likely to be experimentally significant.  For this
  reason we do not describe in any further detail.
\end{thebibliography}
\end{document}